\documentclass[sigconf,prologue,table,xcdraw]{acmart}
\usepackage{graphicx}
\usepackage{subcaption}
\usepackage{booktabs}
\usepackage{diagbox}
\usepackage{adjustbox}
\usepackage{makecell}
\usepackage{enumitem}
\usepackage{hyperref}
\usepackage{color}
\usepackage{colortbl}
\colorlet{shadecolor}{gray!15}
\colorlet{taskonecolor}{blue!5!white}   
\colorlet{tasktwocolor}{green!5!white}  
\colorlet{sectionborder}{blue!50!black}

\usepackage{soul}

\usepackage{multirow}
\usepackage{algorithm}
\usepackage{algorithmic}

\AtBeginDocument{%
  }

\setcopyright{acmlicensed}
\copyrightyear{2018}
\acmYear{2018}
\acmDOI{XXXXXXX.XXXXXXX}
\acmConference[Conference acronym 'XX]{Make sure to enter the correct
  conference title from your rights confirmation email}{June 03--05,
  2018}{Woodstock, NY}
\acmISBN{978-1-4503-XXXX-X/2018/06}

\begin{document}

\title{Heterogeneous Graph Backdoor Attack}

\author{Jiawei Chen}
\email{jchen015@odu.edu}
\affiliation{%
  \institution{Old Dominion University}
  \city{Norfolk}
  \state{Virginia}
  \country{USA}
}

\author{Lusi Li}
\email{lusili@cs.odu.edu}
\affiliation{%
  \institution{Old Dominion University}
  \city{Norfolk}
  \state{Virginia}
  \country{USA}
}

\author{Daniel Takabi}
\email{takabi@odu.edu}
\affiliation{%
  \institution{Old Dominion University}
  \city{Norfolk}
  \state{Virginia}
  \country{USA}
}

\author{Masha Sosonkina}
\email{msosonki@odu.edu}
\affiliation{%
  \institution{Old Dominion University}
  \city{Norfolk}
  \state{Virginia}
  \country{USA}
}

\author{Rui Ning}
\email{rning@cs.odu.edu}
\affiliation{%
  \institution{Old Dominion University}
  \city{Norfolk}
  \state{Virginia}
  \country{USA}
}

\renewcommand{\shortauthors}{Chen et al.}

\begin{abstract}

Heterogeneous Graph Neural Networks (HGNNs) excel in modeling complex, multi-typed relationships across diverse domains, yet their vulnerability to backdoor attacks remains unexplored. To address this gap, we conduct the first investigation into the susceptibility of HGNNs to existing graph backdoor attacks, revealing three critical issues: (1) high attack budget required for effective backdoor injection, (2) inefficient and unreliable backdoor activation, and (3) inaccurate attack effectiveness evaluation. To tackle these issues, we propose the \textbf{Heterogeneous Graph Backdoor Attack (HGBA)}, the first backdoor attack specifically designed for HGNNs, introducing a novel relation-based trigger mechanism that establishes specific connections between a strategically selected trigger node and poisoned nodes via the backdoor metapath. HGBA achieves efficient and stealthy backdoor injection with minimal structural modifications and supports easy backdoor activation through two flexible strategies: Self-Node Attack and Indiscriminate Attack. Additionally, we improve the ASR measurement protocol, enabling a more accurate assessment of attack effectiveness. Extensive experiments demonstrate that HGBA far surpasses multiple state-of-the-art graph backdoor attacks in black-box settings, efficiently attacking HGNNs with low attack budgets. Ablation studies show that the strength of HBGA benefits from our trigger node selection method and backdoor metapath selection strategy. In addition, HGBA shows superior robustness against node feature perturbations and multiple types of existing graph backdoor defense mechanisms. Finally, extension experiments demonstrate that the relation-based trigger mechanism can effectively extend to tasks in homogeneous graph scenarios, thereby posing severe threats to broader security-critical domains.

\end{abstract}



\maketitle

\section{Introduction}
\label{sec:Intro}

Recently, graphs have emerged as a critical data structure for modeling complex, non-Euclidean relationships across domains including social networks \cite{myers2014information,tang2010graph,newman2002random,majeed2020graph}, molecular chemistry \cite{trinajstic2018chemical,balaban1985applications,janezic2015graph}, and recommendation systems \cite{wu2022graph,wang2021graph,gao2023survey,gao2022graph}. Graph Neural Networks (GNNs) have achieved remarkable success in processing graph-structured data by employing recursive message-passing mechanisms that aggregate information from neighboring nodes. 
This approach has made GNNs particularly effective for node classification \cite{xiao2022graph,zhao2024disambiguated, zhao2021graphsmote, wu2019net}, 
graph classification \cite{errica2019fair,yao2025mecon, zhao2024disambiguated}, and link prediction \cite{zhang2018link, arrar2024comprehensive, mei2024dynamic}.
Despite these advances, multiple studies \cite{xi2021graph, zhang2023graph, yang2022transferable, dai2023unnoticeable, zhang2024rethinking} have revealed that GNNs are highly vulnerable to backdoor attacks, where attackers can hijack model's behavior using the trigger, raising significant security concerns for their real-world deployments \cite{wang2021review, zhang2021graph, he2021overview}.

A critical limitation in current research is that most existing studies, if not all, exclusively construct graph backdoor attacks on homogeneous graphs that contain only a single type of node and edge. This focus fails to address the reality that most real-world graphs are inherently heterogeneous, featuring diverse node types interconnected through complex relational structures.
It is worth mentioning that although a range of processing solutions, such as Heterogeneous Graph Neural Networks (HGNNs)~\cite{chen2023heterogeneous, fan2019metapath, salamat2021heterographrec}, have been introduced to handle heterogeneous graphs, their vulnerabilities to backdoor attacks remain underexplored. As HGNNs gain traction in practical applications, it becomes essential to understand whether existing graph backdoor attack methodologies remain effective in these more complex, heterogeneous contexts.


\noindent\fcolorbox{white}{shadecolor}{%
\begin{minipage}{0.978\linewidth}
\textbf{In this work,}  we \textbf{1)} systematically investigate the vulnerabilities of HGNNs 
to existing graph backdoor attacks (GBAs), and \textbf{2)} \textcolor{black}{propose the first backdoor attack against HGNNs based on a novel trigger mechanism, addressing the limitations of existing approaches.}

\end{minipage}}

\vspace{1mm}
\noindent\textbf{(1) Threat Investigation to Reveal Challenges: } To systematically investigate the susceptibility of HGNNs, we evaluated five state-of-the-art GBAs on three widely used heterogeneous graph datasets using six HGNNs (three tailored HGNNs and three retrofitted GNNs). Our investigation focused on semi-supervised node classification (SSNC), a predominant task that aims to classify nodes in a heterogeneous graph using the class labels of only a small subset of nodes. Our preliminary analysis reveals two critical limitations of existing attacks on heterogeneous graphs and uncovers methodological flaws in how attack success rates (ASR) have been calculated on SSNC tasks in previous studies \cite{zhang2024rethinking,ding2025spear}. 


\underline{\textit{Prior-Attack Limitation:}}\textit{ High attack budget for backdoor injection}. Existing GBAs rely on subgraph-based triggers, where each poisoned node must be connected to a fixed or adaptively generated subgraph. These methods typically demand a substantial attack budget, measured by the number of added nodes and edges to ensure effective backdoor injection. The challenge is further exacerbated in heterogeneous graphs, where preserving the graph's heterogeneous properties necessitates incorporating multiple node types and edge relationships during trigger construction, thereby significantly increasing the attack budget.

\underline{\textit{Post-Attack Limitation:}}\textit{ Inefficient and unreliable backdoor activation}. Despite the high attack budget required for successful backdoor injection, activating the backdoor in real-world 
heterogeneous graph scenarios remains a significant challenge. Existing attacks using subgraph-based triggers require complex operations, including creating multiple new nodes of varying types, assigning specific features to each node, and establishing multiple inter-node relationships. These operations are not only inefficient but also particularly fragile in dynamic graph environments, where structural and attribute changes can easily disrupt the trigger’s effectiveness. 
Our experimental results confirm this limitation, showing that certain variations in graph evolution can substantially reduce the ASR. 

\underline{\textit{Inaccurate ASR Measurement}}. In addition to the aforementioned limitations, our investigation uncovers a fundamental methodological issue in how existing studies evaluate attack effectiveness using ASR on SSNC tasks. In graph-structured data, nodes are inherently interdependent, with each node’s prediction heavily influenced by its neighbors. Prior studies \cite{zhang2024rethinking,ding2025spear} evaluate ASR by simultaneously attaching triggers to multiple test nodes, thus inadvertently creating interference effects to the prediction of all nodes. Therefore, this cross-node interference artificially inflates or deflates ASR, resulting in unreliable assessments.


\vspace{1mm}
\noindent\textbf{(2) Our Attack - HGBA:} To tackle those issues, we propose \textbf{Heterogeneous Graph Backdoor Attack (HGBA)}, the first backdoor attack specifically designed to target HGNNs. In contrast to existing attacks \cite{xi2021graph,zhang2023graph,yang2022transferable,dai2023unnoticeable,zhang2024rethinking}, HGBA departs in significant ways.

\underline{\textit{Advance Relation-based Triggers}}. Unlike existing approaches that use subgraphs as triggers, HGBA introduces a fundamentally different design with novel relation-based triggers. Its benefits are three-fold: \textit{1) High Attack Effectiveness with Low Attack Budget}: HGBA dramatically reduces the attack budget because it only requires adding only a single edge between the preselected trigger node and the poisoned node to set the trigger, eliminating the need for injecting complex subgraph with new nodes. Despite this, HGBA still maintains high attack effectiveness, achieved through our carefully designed strategies for optimal trigger node selection and backdoor metapath identification. \textit{2) Easy and Flexible Backdoor Activation}: HGBA enables easier backdoor activation by establishing just a single connection between the trigger node and the target node, which drastically reduces the complexity and time overhead required for backdoor activation compared to subgraph-based methods. Additionally, HGBA supports flexible activation strategies tailored to different attack scenarios, such as Self-Node Attacks for situations where attackers target only their own created nodes to be misclassified as the target label, and Indiscriminate Attacks, where attackers can trigger the backdoor on any node, whether their own or others. \textit{3) Stealthiness}: HGBA constructs poisoned datasets with minimal structural modifications and no additional node injections, making detection highly challenging. The backdoored model maintains performance parity on clean samples, eliminating suspicion of performance degradation. Most significantly, backdoor activation is also stealthy by leveraging natural graph evolution patterns without creating synthetic nodes.

\underline{\textit{Downstream-Model-Agnostic}}. 
HGBA achieves high attack effectiveness and stealthiness in black-box settings where attackers have minimal information about the victim model. This effectiveness stems from leveraging HGNN's powerful ability to learn diverse relationships between different nodes, utilizing specific node relationships as triggers rather than targeting model-specific vulnerabilities. This approach allows the backdoor to propagate through any downstream model that learns from the poisoned data, independent of any internal knowledge.

\underline{\textit{Superior Robustness}}. HGBA achieves higher stability by designing triggers that depend solely on the trigger node and backdoor metapath without requiring any feature information from the target node or its neighborhood. This feature-independent approach makes our attack inherently robust against dynamic changes in target node characteristics that commonly occur in evolving real-world graphs. Furthermore, HGBA still maintains high attack performance even under multiple kinds of potential graph backdoor defenses.

\underline{\textit{Attack-Extensible}}. Although HGBA was initially designed to attack HGNNs for node classifications, our experiments have further demonstrated that its innovative trigger design can be effectively extended to both node and graph classification tasks within homogeneous graph scenarios, thereby constituting severe threats for more security-critical domains (e.g., toxic chemical classification \cite{Mehta2023BenchmarkingTM,chen2021chemical,jiang2021ggl}, cybersecurity detection \cite{yan2023graph,sozol2024anomaly}).


\vspace{1mm}
\noindent\fcolorbox{white}{shadecolor}{%
\begin{minipage}{0.978\linewidth}
\textbf{Contributions: }We summarize our main contributions as below:
\end{minipage}}


\vspace{1mm}
\begin{itemize}[leftmargin=*]
    \item We conduct the first systematic evaluation of state-of-the-art GBAs on HGNNs, revealing two key limitations of existing GBAs and an issue with inaccurate ASR measurements in prior studies assessing attack effectiveness, offering key insights for advancing graph backdoor attack research.

    \item We propose the Heterogeneous Graph Backdoor Attack (HGBA), the first efficient backdoor attack specifically targeting HGNNs, which introduces a novel relation-based trigger that overcomes key limitations of existing graph backdoor attacks in real-world settings.



    \item Extensive experiments demonstrate HGBA’s effectiveness and practicality under a black-box setting. Additionally, we reveal that existing mainstream graph backdoor defenses fail to counter HGBA, underscoring the demand for more refined defenses.


    \item We also improve the ASR measurement strategy to isolate the influence between triggers, enabling accurate assessment of individual trigger impact on node classification and enhancing the reliability of evaluation methods for graph backdoor attacks.
    
\end{itemize}

\section{Background and Related Works}
\label{sec:related_work}
\subsection{Graph Neural Networks}
\label{sec:gnns}

Graph Neural Networks (GNNs) are designed to learn node representations in homogeneous graphs, which only consist of single-type nodes and edges. Based on mechanisms for processing graph data, GNNs are categorized as spectral and non-spectral approaches. 

For spectral approaches, Kipf et al.\cite{kipf2016semi} proposed the Graph Convolutional Network (GCN), which effectively propagates information across graphs using a localized approximation of spectral convolutions. Non-spectral approaches instead rely on message-passing mechanisms. Velickovic et al.~\cite{velivckovic2017graph} introduced the Graph Attention Network (GAT), which uses attention weights to focus on the most relevant neighbors during aggregation. Hamilton et al.~\cite{hamilton2017inductive} developed GraphSAGE, which samples fixed-size neighborhoods rather than using the full graph structure, enabling scalability to large graphs and inductive learning for unseen nodes.

Although GNNs are originally designed and evaluated on homogeneous graphs, the Heterogeneous Graph Benchmark (HGB) by Lv et al. \cite{lv2021we} has shown that GNNs can indirectly handle heterogeneous graphs by extracting homogeneous subgraphs based on metapath, achieving competitive performance on heterogeneous graph tasks and sometimes even matching or surpassing specialized HGNNs. This makes GNNs an effective solution for handling heterogeneous graphs. Therefore, our work also considers retrofitted GNNs as target models to be attacked.

\subsection{Heterogeneous Graph Neural Networks}
\label{sec:hgnns}

\vspace{1mm}
\noindent\textbf{Heterogeneous Graph.} In practice, real-world networks typically involve multiple types of entities and relationships, forming heterogeneous graphs. 

\begin{figure}[h]
    \centering
    \includegraphics[width=1\linewidth]{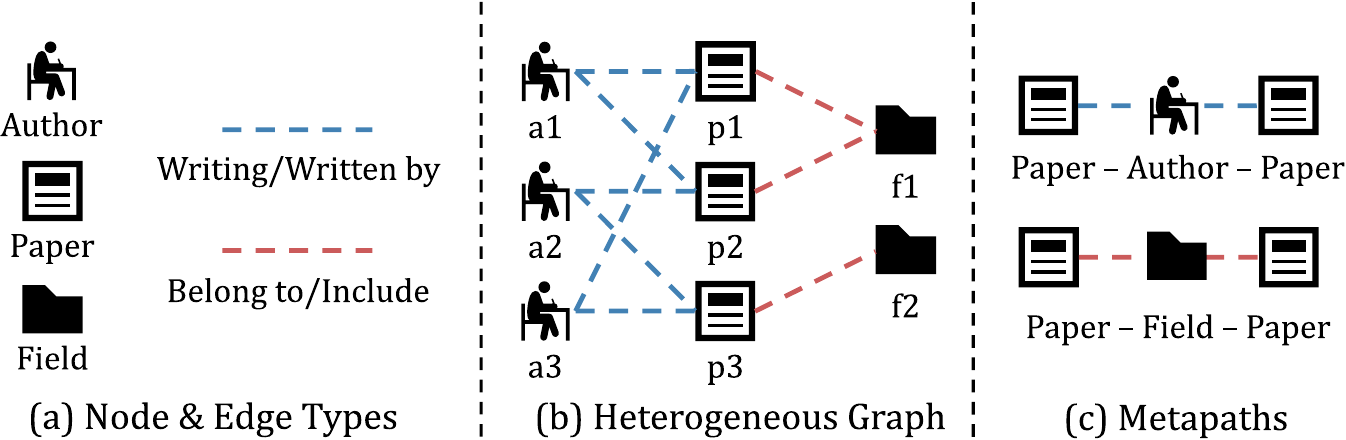}
    \caption{An illustrative example of a heterogeneous graph. (a) Three types of nodes and two types of edges. (b) A heterogeneous graph (a paper citation network) consists of three types of nodes and two types of edges. (c) Two types of meta paths are involved in (b).}
    \label{fig:heterogeneous_graph_example}
\end{figure}

A heterogeneous graph is formally defined as $G = (V, E)$ with a node type mapping function $\text{type}(v): V \rightarrow \mathcal{A}$ and an edge type mapping function $\text{type}(e): E \rightarrow \mathcal{R}$, where $\mathcal{A}$ and $\mathcal{R}$ represent the sets of node and edge types, respectively. The condition $|\mathcal{A}| + |\mathcal{R}| > 2$ distinguishes heterogeneous graphs from homogeneous ones, ensuring diversity in node and edge types. For example, a paper citation network (as shown in Fig.~\ref{fig:heterogeneous_graph_example} b) may contain multiple node types (Author, Paper, Field) connected through various relation types ("Writing", "Belong to", etc.).

\noindent\textbf{Metapath}. In heterogeneous graphs, metapaths describe structured sequences of node and edge types that capture semantic relationships. Formally, a metapath \( P \) is defined as a path in the form of \( A_1 \stackrel{R_1}{\rightarrow} A_2 \stackrel{R_2}{\rightarrow} \dots \stackrel{R_l}{\rightarrow} A_{l+1} \) (abbreviated as \( A_1A_2\dots A_{l+1} \)). This represents a composite relation \( R = R_1 \circ R_2 \circ \dots \circ R_l \) between nodes of type \( A_1 \) and type \( A_{l+1} \), where \( \circ \) denotes the composition operator on relations. Metapaths provide a powerful mechanism for characterizing the complex semantic relationships that exist between nodes in heterogeneous graphs. For example, Figure~\ref{fig:heterogeneous_graph_example}(c) shows how metapaths like Paper-Author-Paper (PAP) and Paper-Field-Paper (PFP) connect papers in a paper citation network, capturing co-authorship and thematic similarities, respectively, to reveal diverse semantic relationships.


To capture the complex relationships in heterogeneous graphs, researchers have developed specialized Heterogeneous Graph Neural Networks (HGNNs). Schlichtkrull et al.~\cite{schlichtkrull2018modeling} introduced the Relational Graph Convolutional Networks (RGCN), which extend the traditional GCN by incorporating relation-specific transformations to handle multiple edge types. Wang et al.~\cite{wang2019heterogeneous} proposed the Heterogeneous Graph Attention Network (HAN), which employs hierarchical attention mechanisms to capture the importance of different node and relation types. Zhang et al.~\cite{zhang2019heterogeneous} developed HetGNN, which samples heterogeneous neighbors via random walks and processes them by node type, using a two-module architecture for content embedding and neighbor aggregation. More recently, Hu et al.~\cite{hu2020heterogeneous} introduced the Heterogeneous Graph Transformer (HGT), which incorporates type-specific parameters to model heterogeneous attention mechanisms and uses relative temporal encoding to capture dynamic structural dependencies while employing an efficient sampling algorithm for scalable training on large graphs.

\subsection{Graph Backdoor Attacks} 
\label{sec:gbas}
Graph backdoor attacks aim to compromise GNNs by embedding hidden vulnerabilities that can be maliciously triggered. Unlike adversarial attacks that target the inference phase, backdoor attacks manipulate the training process to create a model that performs normally on clean inputs but exhibits targeted misclassification when a specific trigger pattern is present. In the context of GNNs, where the goal is to learn node representations on a graph, backdoor attacks exploit both the graph structure and node features to inject malicious patterns that create this backdoor vulnerability.

\begin{figure}[h]
    \centering
    \includegraphics[width=\linewidth]{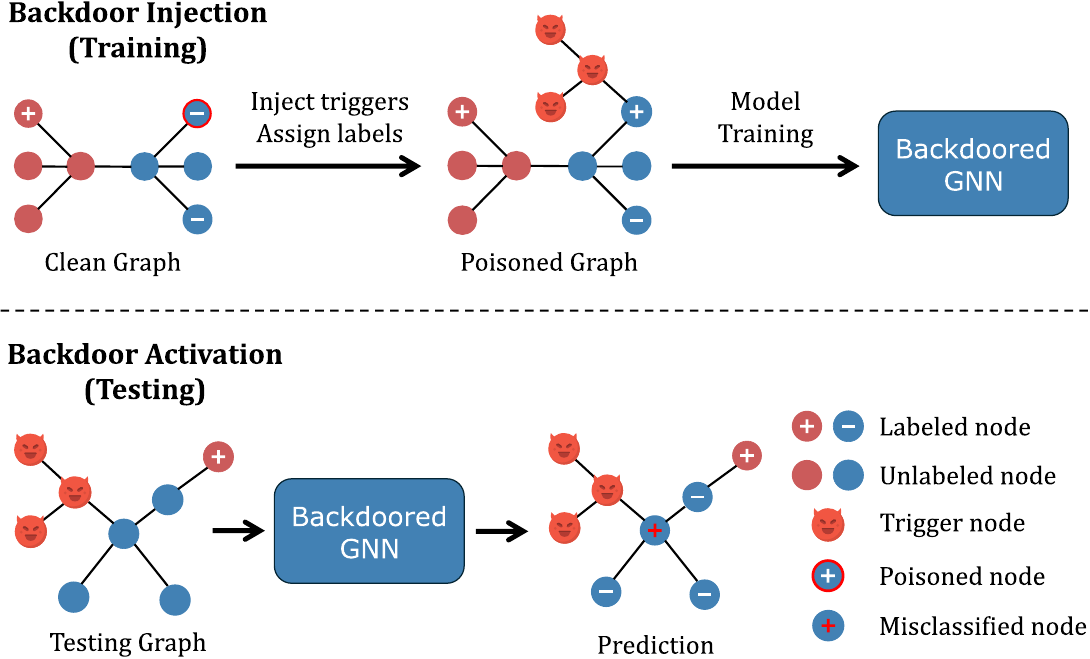}
    \caption{General Framework of Current Graph Backdoor Attacks on SSNC Tasks on Homogeneous Graphs.}
    \label{fig:general_framework_gba}
\end{figure}

The general process of current graph backdoor attacks on SSNC tasks on homogeneous graphs is illustrated in Figure~\ref{fig:general_framework_gba}. In the backdoor injection phase (training), for a clean graph \(G = (V, E)\), attackers select a subset of nodes \( V_p \subset V \) as poisoned nodes (e.g., blue nodes with red borders), attach the subgraph-based trigger \( T \) to each node in \( V_p \), and modify their labels to the target label \( y_t \in Y \) (the positive class in this case), thereby generating a poisoned graph \(G_{poisoned}\). The GNN model \( f \), trained on \(G_{poisoned}\), establishes a spurious correlation between the presence of triggers \( T \) and the target label \( y_t \). This malicious association causes the backdoored GNN \( f_b \) to misclassify nodes with triggers \( T \) as \( y_t \), achieving the backdoor injection objective. During the backdoor activation phase (testing), attackers target the specific node by attaching the trigger \( T \). The backdoored GNN model \( f_b \) misclassifies the target node as \( y_t \), while preserving performance on clean nodes without triggers.

\ul{\textit{While considerable progress has been made in graph backdoor attacks, it is important to note that existing approaches primarily target homogeneous graphs.}} For instance, Zhang et al.~\cite{zhang2021backdoor} utilized random subgraphs as triggers for attacking graph classification tasks. Almost simultaneously, Xi et al.~\cite{xi2021graph} proposed Graph Trojan Attack (GTA), generating adaptive subgraph triggers based on neighborhood subgraphs of target nodes, which has been demonstrated to be extendable to node classification tasks in their work. After extending the works of Zhang et al. and the GTA to SSNC tasks, Dai et al.~\cite{dai2023unnoticeable} identified the limitations of requiring numerous poisoned nodes and producing easily detectable triggers, which led them to develop the Unnoticeable Graph Backdoor Attack (UGBA) with strategic node selection and stealthy trigger generation.
More recently, Zhang et al.~\cite{zhang2024rethinking} observed that triggers from previous methods often fall outside the data distribution and introduced the Distribution Preserving Graph Backdoor Attack (DPGBA) to generate in-distribution trigger features.


\noindent\fcolorbox{white}{taskonecolor}{%
  \begin{minipage}{\dimexpr\linewidth-2\fboxsep-2\fboxrule\relax}
    \noindent\textbf{Research Task 1 (RT1): Threat Investigation.} Despite these advancements in graph backdoor attacks designed for attacking GNNs, the security of HGNNs against backdoors remains largely unexplored. To tackle this, we investigate to reveal that the complexity of heterogeneous graphs necessitates significantly higher attack budgets for effectively injecting backdoors into HGNNs compared to GNNs. Furthermore, even when a backdoor is successfully implanted using a high attack budget, current graph backdoor attacks struggle to activate the backdoor in real-world scenarios. Additionally, we identified an issue with attack success rate (ASR) calculation in some previous research caused by cross-node dependence of graph data.
  \end{minipage}%
}

\noindent\fcolorbox{white}{tasktwocolor}{%
  \begin{minipage}{\dimexpr\linewidth-2\fboxsep-2\fboxrule\relax}
    \noindent\textbf{Research Task 2 (RT2): HGBA.} The learned lessons motivate our work on Heterogeneous Graph Backdoor Attack (HGBA), with the aim of designing novel graph backdoor triggers by leveraging the characteristics of graph data and HGNNs, directly addressing the key limitations identified in RT1. 
  \end{minipage}%
}

\vspace{5mm}
\stepcounter{section} 
\noindent\fcolorbox{sectionborder}{taskonecolor}{%
  \parbox{\dimexpr\linewidth-2\fboxsep-2\fboxrule\relax}{%
    \Large\textbf{\thesection \hspace{3mm} Research Tast 1: Threat Investigation}%
  }%
}

\vspace{-3mm}
\subsection{Threat Model}

\noindent\textbf{Attacker's Goal.} Our threat model considers an adversary aiming to compromise HGNNs through backdoor attacks. The attacker's objective is to poison the heterogeneous graph dataset, injecting a backdoor into the trained HGNNs that allows them to hijack model behavior, causing targeted misclassification when specific triggers are activated while maintaining normal model performance on clean inputs. 


\vspace{1mm}
\noindent\textbf{Attacker's Capability during Training.} The attacker operates as a malicious data provider, a realistic scenario in contemporary machine-learning ecosystems where models are frequently trained on aggregated data from multiple sources. In this role, we adopt a challenging \textit{black-box setting} where the attacker lacks knowledge of the victim model’s architecture, hyperparameters, or training procedure. The attacker can only contribute either a portion or the entirety of the heterogeneous graph dataset. Simultaneously, we reasonably assume that to remain undetected, the attacker can only modify the dataset under budget constraints, measured by the number of added nodes and edges, which is particularly
critical in heterogeneous graphs due to their structural complexity.


\vspace{1mm}
\noindent\textbf{Attacker's Capability during Inference.} During inference, the attacker activates the backdoor with capabilities equivalent to regular users: creating accounts (nodes), modifying profile information (node features), and establishing connections (edges), without privileged abilities to directly manipulate arbitrary nodes other than those they created or the inference process.  

\vspace{1mm}
\noindent\textbf{Victim's Capabilities.} We assume victims can employ existing GNN backdoor defenses before or during the training phase to counter potential backdoor attacks, either by examining and sanitizing the heterogeneous graph dataset beforehand or by using robust models during training to enhance resilience.

\subsection{Investigation} 
\label{sec:3.3}

\noindent\fcolorbox{white}{taskonecolor}{%
  \begin{minipage}{\dimexpr\linewidth-2\fboxsep-2\fboxrule\relax}
  \noindent\textbf{Investigation Objectives (IO):}
    \begin{itemize}[leftmargin=*]
        \item \ul{\textit{IO1: Launching Backdoor.}} Are GBAs originally designed for homogeneous graphs (HoG) also applicable to heterogeneous graphs (HeGs) with reasonable attack budgets? 
        \item \ul{\textit{IO2: Exploiting Backdoor.}} After backdoor injection with a high attack budget, can these backdoors be effectively activated using triggers in real-world scenarios?
        \item \ul{\textit{IO3: Evaluating Backdoor.}} Is the attack effectiveness of GBAs being accurately measured in HeGs?
    \end{itemize}
  \end{minipage}%
}

\vspace{1mm}
\noindent\textbf{IO1: Launching Backdoor - Challenges of Injecting Backdoors in HGNNs.}
\label{sec:io1} Injecting backdoors into HGNNs presents unique challenges compared to GNNs, which is the significantly higher attack budget required for effective backdoor implementation in HGNNs. While traditional graph backdoor attacks on HoGs involve introducing subgraphs composed of multiple nodes and edges as triggers to poisoned nodes, adapting these approaches to HeGs requires substantially more additional nodes and edges to preserve graph heterogeneity.

\begin{figure}[h]
    \centering
    \includegraphics[width=0.8\linewidth]{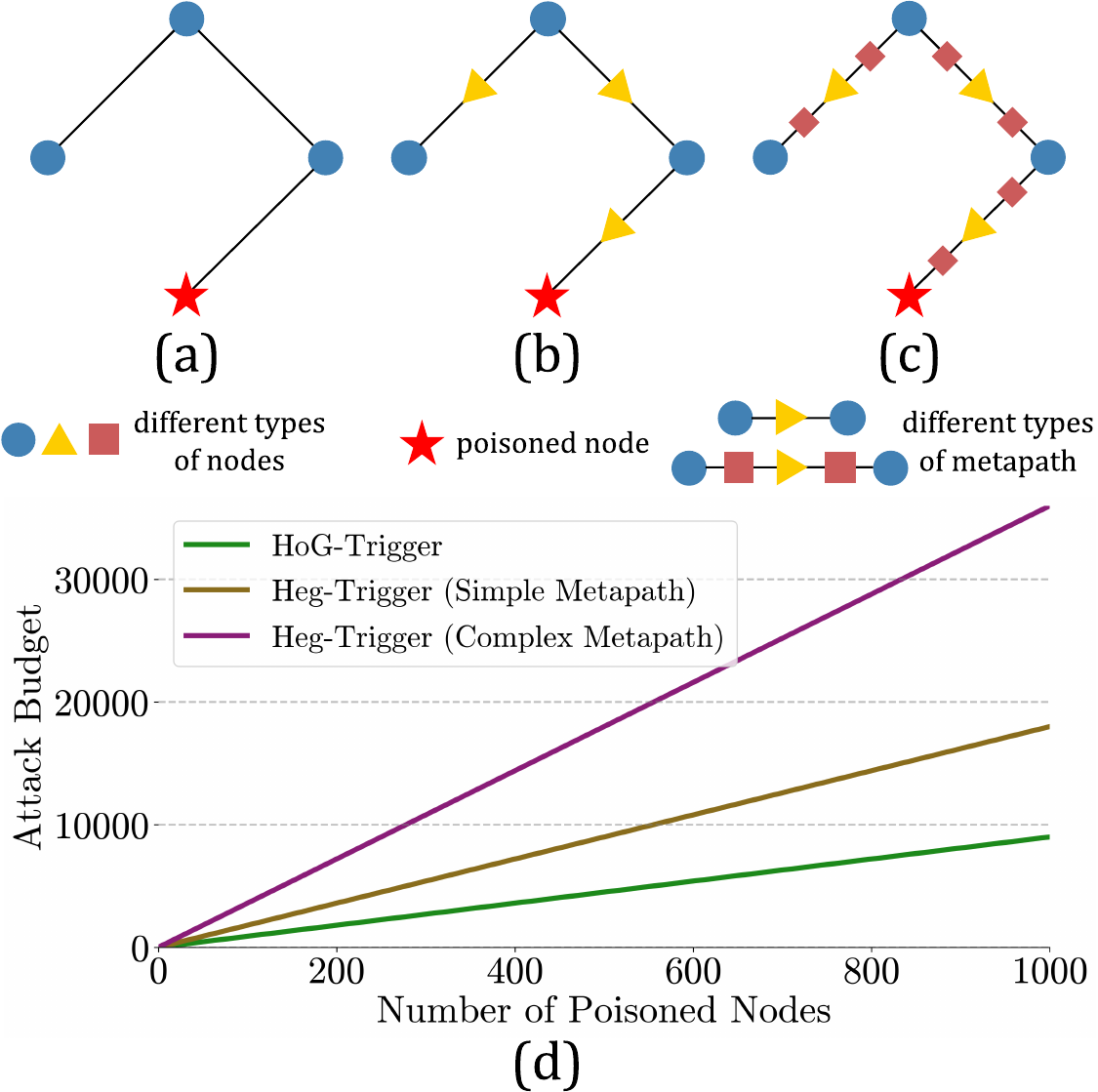}
   \caption{Impact of Heterogeneity on Attack Budget in Graph Backdoor Attacks. (a) Trigger generated by current graph backdoor attacks for HoGs. (b) Augmented triggers for HeGs with added nodes and edges to maintain heterogeneity. (c) Augmented triggers for HeGs incorporating more diverse node types due to complex heterogeneity. (d) Quantified comparison of attack budgets related to graph structural complexity.}
    \label{fig:io1} 
\end{figure}

To stealthily inject backdoor triggers into HeGs, attackers must augment homogeneous triggers generated by current graph backdoor attacks with numerous intermediate nodes and additional edges based on the metapaths inherent to the target heterogeneous graph. This transforms the homogeneous trigger into a heterogeneous subgraph that seamlessly integrates with the original graph structure. In this extended subgraph trigger, the original nodes actively contribute to embedding and activating the backdoor, while the newly added intermediate nodes serve as structural camouflage to maintain compatibility with the graph's heterogeneous nature. Therefore, the structural complexity of HeGs necessitates a higher attack budget for poisoning individual nodes compared to homogeneous graphs as illustrated in Figure~\ref{fig:io1}. Figures~\ref{fig:io1}(a) shows the trigger generated by current graph backdoor attacks for HoGs, while Figures~\ref{fig:io1}(b) demonstrates how these triggers be augmented for HeGs by inserting additional nodes of diverse types to maintain heterogeneity. For HeGs with more complex metapaths, even more node types must be incorporated, as depicted in Figures~\ref{fig:io1}(c). This fundamental structural difference leads to a consistently higher attack budget for HeGs compared to HoGs, as quantified in Figure~\ref{fig:io1}(d).

Furthermore, the inherent complexity of HeGs suggests that attackers need to poison a larger proportion of nodes to achieve high Attack Success Rates (ASRs), further amplifying the required attack budget. To validate the hypothesis that injecting backdoors into HGNNs requires a higher attack budget, we define the attack budget \(B_a\) as the total number of newly added nodes and edges, as shown in Equation~\ref{eq:attack_budget}, where \(N_{\text{new}}\) is the number of newly added nodes and \(E_{\text{new}}\) is the number of newly added edges.

\begin{equation}
\label{eq:attack_budget}
B_a = N_{\text{new}} + E_{\text{new}}
\end{equation}

We conducted experiments with attack budgets set at 1\%, 3\%, 5\%, and 10\% of the number of nodes and edges in the training set. Firstly, we employed state-of-the-art GBAs (DPGBA, UGBA, GTA, SBA-SAMPLE and its variant SBA-GEN) across three wildly used homogeneous graph datasets (Cora, Pubmed, and Flickr), targeting three GNNs (GCN, GAT, and GraphSAGE). Additionally, we performed the same attacks on three popular heterogeneous graph datasets (ACM, DBLP, and IMDB), targeting six HGNNs (three tailored HGNNs and three retrofitted GNNs). The details of the target models are provided in Section \ref{sec:gnns} and Section \ref{sec:hgnns}, with the datasets and experimental settings detailed in Appendix \ref{app:datasets} and Appendix \ref{app:experimental_details}. Average results across all datasets and models for each attack budget are presented in Table~\ref{tab:io1}, with more detailed results available in Appendix ~\ref{app:io1}. The results show that as the $B_a$ increases, Attack Success Rates (ASRs) for both GNNs and HGNNs rise, while clean metrics decline but remain within acceptable limits. However, under the same $B_a$, GNNs experience higher ASRs than HGNNs, indicating that HGNNs are more resistant to current GBAs and require larger $B_a$ to compromise successfully.


\noindent\fcolorbox{white}{shadecolor}{%
\begin{minipage}{0.978\linewidth}
\textbf{Takeaways:} While current GBAs can achieve high performance on GNNs, they are significantly less effective on HGNNs unless afforded much larger attack budgets, underscoring the need for more efficient and targeted strategies in heterogeneous graph settings.


\end{minipage}}

\begin{table}[htbp]
    \renewcommand{\arraystretch}{1} 
    \centering
    \begin{tabular}{ccccccc}
        \toprule
        & \multicolumn{2}{c}{GNNs} & & \multicolumn{3}{c}{HGNNs} \\
        \cmidrule(r){2-3} \cmidrule(r){5-7}
        & Clean Acc & ASR & & C Mic & C Mac & ASR \\
        \midrule
        Clean Graph & 71.43 & - & & 79.67 & 79.41 & - \\
        1\%         & 71.03 & \textbf{66.08} & & 78.81 & 78.20 & \textbf{43.77}\text{\textdownarrow} \\
        3\%         & 70.58 & \textbf{87.33} & & 78.45 & 77.94 & \textbf{52.37}\text{\textdownarrow} \\
        5\%         & 70.13 & \textbf{90.16} & & 78.09 & 77.41 & \textbf{56.75}\text{\textdownarrow} \\
        10\%        & 68.84 & \textbf{92.73} & & 76.41 & 75.32 & \textbf{61.16}\text{\textdownarrow} \\
        \bottomrule
    \end{tabular}
    \caption{Effect of Attack Budget $B_a$ on Current Graph Backdoor Attacks for GNNs and HGNNs. To evaluate the stealthiness, Clean Accuracy is used for GNNs, while Clean Micro-F1 and Clean Macro-F1 are used for HGNNs, which are commonly used metrics in their respective fields. For assessing attack effectiveness, ASR is employed on both GNNs and HGNNs, measuring the accuracy of samples with triggers being classified into the target class. Only clean metrics are reported for clean graphs.}
    \label{tab:io1}
\end{table}

\begin{figure*}
    \centering
    \includegraphics[width=0.9\linewidth]{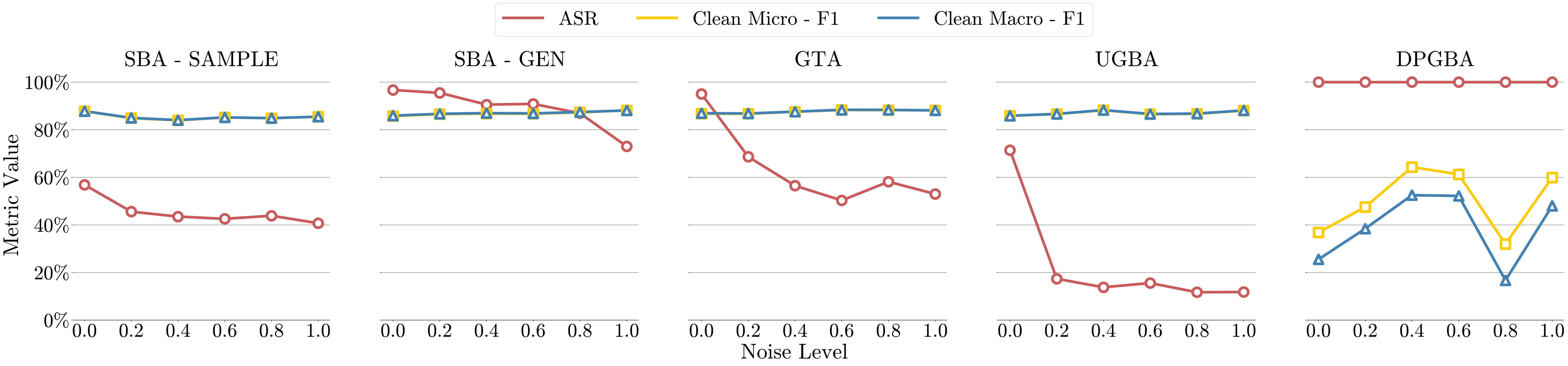}
    \caption{Impact of Node Perturbations on Backdoor Activation for Existing Graph Backdoor Attacks on HGNNs. The representative results on the ACM dataset, using HetGNN as the attacked model.}
    \label{fig:io2}
\end{figure*}

\vspace{1mm}
\noindent\textbf{IO2: Exploiting Backdoor - Challenges in Backdoor Activation in Real-world HeG Scenarios.}
\label{sec:io2} Successfully exploiting backdoors in real-world heterogeneous graph (HeG) scenarios presents significant challenges, even after successful backdoor injection using a significantly high attack budget. Our investigation of IO2 reveals that effectively activating and exploiting backdoors using current subgraph-based triggers faces two major obstacles: operational inefficiency and activation fragility.

In the threat model assumed, attackers operate with capabilities similar to regular users, which is consistent with practical heterogeneous graph environments. Exploiting backdoors through subgraph-based trigger attacks requires executing a complex sequence of operations across different node types and relation patterns inherent to HeGs. These operations include creating new accounts (adding nodes of appropriate types), modifying account information (assigning specific features to trigger nodes), and establishing connections with existing entities (creating edges to construct a complete trigger). The heterogeneous nature of these graphs significantly amplifies the complexity of these operations compared to homogeneous settings, making the activation process both inefficient and time-intensive in real-world HeG scenarios.

Furthermore, most existing graph backdoor attacks rely on dynamic subgraph-based triggers, where the features of trigger nodes are adaptively generated based on the features of the target node and its neighboring nodes. This approach faces particular challenges in HeGs, where real-world dynamics cause node features and structural connections to undergo constant evolution. For instance, in heterogeneous social networks, users frequently update profiles, establish new relationships across different entity types, or modify interaction patterns. This leads to our hypothesis that once a trigger is crafted for a HeG at a specific moment, subsequent changes to the target node and its heterogeneous neighborhood during the trigger establishment phase may render the initial trigger ineffective, resulting in backdoor activation failure.

To test this hypothesis within real-world HeG scenarios, we extended the experimental framework from Section~\ref{sec:io1} IO1, fixing the attack budget \(B_a\) at 10 \% of the number of nodes and edges in the training dataset, and evaluated the performance of state-of-the-art GBAs under simulated real-world dynamics. We simulated these dynamics by introducing varying levels of noise to the node features of test nodes and their surrounding neighbors, sampling from the same distribution as the corresponding node features while keeping the graph structure unchanged due to challenges in accurately modeling changes in structural connections. The representative results on the ACM dataset, using HetGNN as the attacked model, are showcased in Figure \ref{fig:io2}, with detailed results of each dataset in Appendix \ref{app:io2}.

The results reveal a critical distinction in backdoor effectiveness across different attacks under the condition of node feature perturbations. Overall, the Attack Success Rate (ASR) of existing graph backdoor attacks decreases as the features of target nodes and their surrounding neighbors change to a greater extent. Specifically, dynamic subgraph-based attacks, except DPGBA, exhibit a significant ASR drop. This decline stems from mismatches between the training and activation phases caused by noise, with interdependencies among different node types in heterogeneous graphs amplifying the impact of feature perturbations. Although DPGBA consistently maintains a high ASR regardless of changes in node features, its clean metrics are notably low and exhibit significant variability. In contrast, fixed subgraph-based attacks (such as SBA-SAMPLE and SBA-GEN) demonstrate only a slight reduction in ASR in most scenarios, as they rely on less adaptive, predefined structures that are less sensitive to such feature variations.

\noindent\fcolorbox{white}{shadecolor}{%
\begin{minipage}{0.978\linewidth}
\textbf{Takeaways:} 
In real-world heterogeneous graph scenarios, it is inefficient to activate the backdoor using a subgraph-based trigger. Moreover, subgraph-based backdoor triggers in GNNs are ineffective for HGNNs due to their sensitivity to node feature changes and the complex dynamics of heterogeneous environments, leading to frequent activation failures. 
\end{minipage}}

\vspace{1mm}
\noindent\textbf{IO3: Evaluating Backdoor - Challenges in Accurately Measuring Attack Effectiveness in HeGs.}
\label{sec:io3} HeGs structures inherently feature interdependent, mutually influential nodes that violate the independence assumption underlying standard ASR measurements. While ASR, defined as the proportion of triggered samples predicted as the target class, works well for independent samples in CV, NLP, and graph classification, it becomes problematic for node classification in HeGs. Some existing works evaluate backdoor attack ASR by simultaneously attaching triggers to multiple test nodes. However, this neglects that a node's classification is influenced not only by its own trigger but also by triggers on neighboring nodes through various metapaths, which leads to an inaccurate evaluation of attack effectiveness.

\begin{figure}
    \centering
    \includegraphics[width=0.85\linewidth]{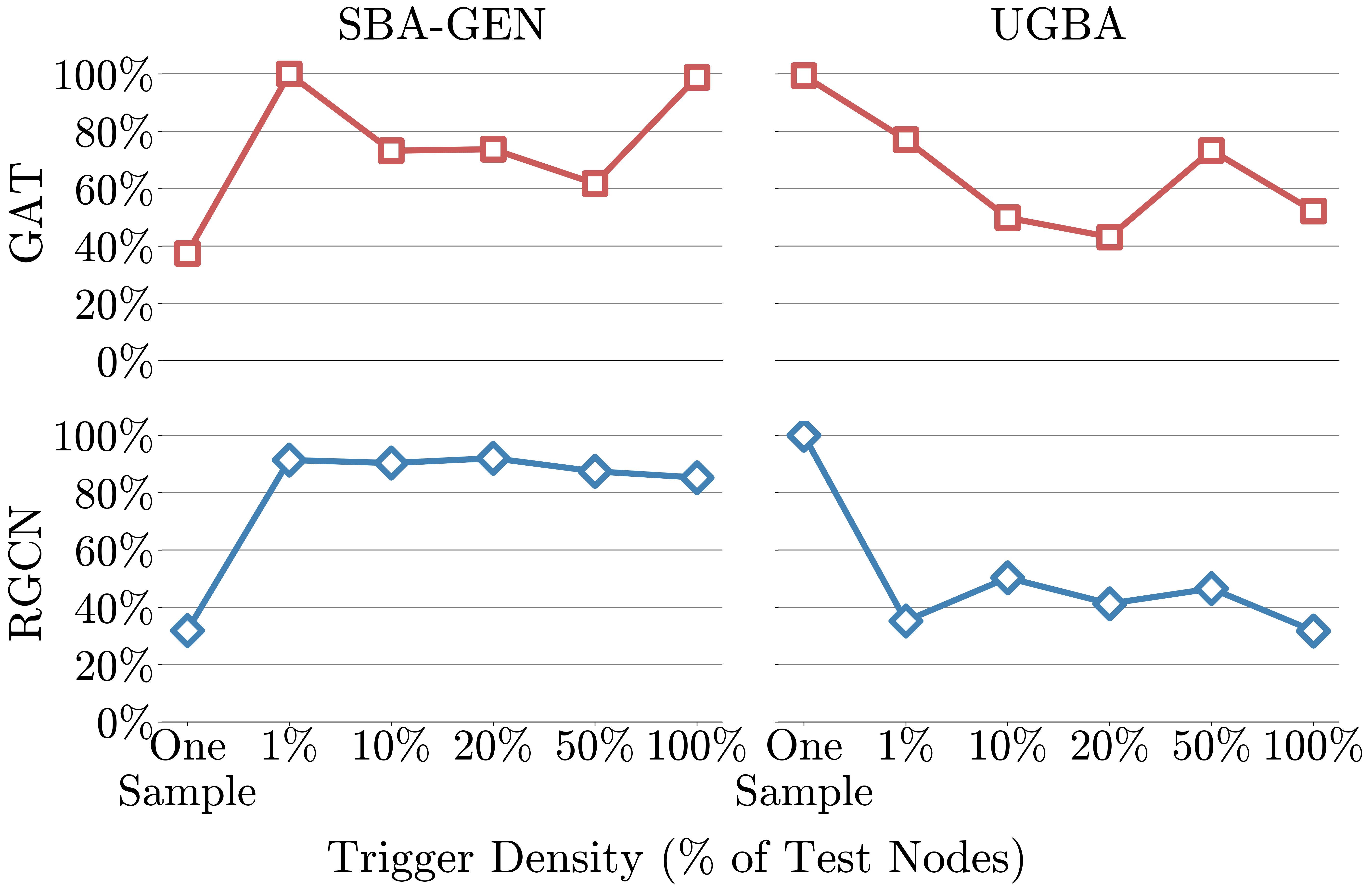}
    \caption{Impact of Trigger Density on Attack Success Rate (ASR) of Graph Backdoor Attacks in Heterogeneous Graphs. Representative Results of SBA-GEN and UGBA Attacking GAT and RGCN on the ACM Dataset.}
    \label{fig:io3}
\end{figure}

To validate this hypothesis, we conducted experiments to investigate the effects of simultaneously assigning triggers to varying proportions of test nodes on ASR measurements based on the experimental framework from Section~\ref{sec:io2} IO2. Specifically, we adjusted the proportion of test nodes that are simultaneously assigned triggers, ranging from individual samples to 1\% up to 100\%. For instance, we select 1\% of the test nodes without replacement from the entire test set, attach triggers, and compute the ASR, repeating this process until all test nodes are covered (100 iterations in this case); the average ASR across all iterations is then taken as the final ASR. Figure \ref{fig:io3} presents the experimental results of SBA-GEN and UGBA attacking GAT and RGCN on the ACM dataset as representative examples, with detailed results of each dataset in Appendix \ref{app:io3}, which exhibit similar patterns.

The results reveal two distinct patterns in HeG environments: as the proportion of test nodes with attached triggers increases, the ASR consistently exhibits either an upward or downward trend. The rising trend likely stems from a cooperative effect, where multiple triggers across different node types in close proximity reinforce each other's influence, amplifying the HGNN's misclassification toward the target label. Conversely, the declining trend may arise from interference, where overlapping triggers introduce conflicting signals across heterogeneous relations, diluting their individual effectiveness and confusing the model.

\noindent\fcolorbox{white}{shadecolor}{%
\begin{minipage}{0.978\linewidth}
\textbf{Takeaways:} Some previous research on evaluating the effectiveness of graph backdoor attacks often overlooks the complex interdependencies among nodes in heterogeneous graphs (HeGs), leading to potentially inflated or distorted Attack Success Rate (ASR) metrics. 

\textbf{In response,} we have improved the ASR measurement protocol in our experiments by attaching triggers to only one sample at a time during testing and repeating this process until all nodes in the test set are evaluated. This approach isolates the impact of individual triggers, enabling a more accurate assessment of attack effectiveness.


\end{minipage}}

\vspace{5mm}
\stepcounter{section} 
\noindent\fcolorbox{sectionborder}{tasktwocolor}{%
  \parbox{\dimexpr\linewidth-2\fboxsep-2\fboxrule\relax}{%
    \Large\textbf{\thesection \hspace{3mm} Research Task 2: Our Attack - HGBA}%
  }%
}\vspace{2mm}

\noindent\textbf{Motivation and Key Intuition.} Our investigation results
translate into critical challenges that must be addressed: \textbf{(1)} how to reduce the attack budget while effectively injecting a backdoor, and \textbf{(2)} how to enable robust backdoor activation despite node feature changes. \ul{\textit{We observe that these challenges originate from a common root: the conventional reliance on subgraph-based triggers, which fundamentally misaligns with the unique properties of heterogeneous graphs.}} 

By reevaluating the characteristics of graph data and analyzing the operational features of HGNNs, we identified that graph data possess a distinctive attribute—inter-node relationships—that has been underexploited in prior work. Moreover, HGNNs uniquely excel at capturing these relationships and the complex dependencies among nodes, making them particularly powerful for heterogeneous graphs. \ul{\textit{This understanding led to our key innovation in HGBA: shifting the trigger design from feature-centric subgraphs to metapath-based connections between nodes of the same type.}}

\vspace{1mm}
\noindent\textbf{Design Concept: Relation-Based Triggers.} 
The relation-based trigger mechanism establishes a specific connection between a poisoned node and a predefined trigger node, anchoring the backdoor in stable structural properties rather than volatile features. By requiring only a single edge for each poisoned node instead of complex subgraphs with multiple nodes and edges, this approach significantly reduces the attack budget while enabling rapid activation with minimal operations and ensuring robustness against feature shifting—directly addressing the identified limitations.

\vspace{1mm}
\noindent\textbf{Goal and Overview.} Building on this foundation, we propose \textbf{H}eterogeneous \textbf{G}raph \textbf{B}ackdoor \textbf{A}ttack (\textbf{HGBA}), whose primary objective is to stealthily and efficiently plant backdoors in HGNNs under a low attack budget, enabling their effective and robust activation in practical scenarios. HGBA addresses the challenges identified in IO1 by significantly reducing the attack budget through its minimal edge-based design. It resolves the activation issues highlighted in IO2 by simplifying the trigger mechanism to operate efficiently and remain robust despite node feature changes in real-world scenarios.
\subsection{HGBA Framework}
\label{sec:hgba_framework}
\begin{figure*}[h]
    \centering
    \includegraphics[width=0.85\linewidth]{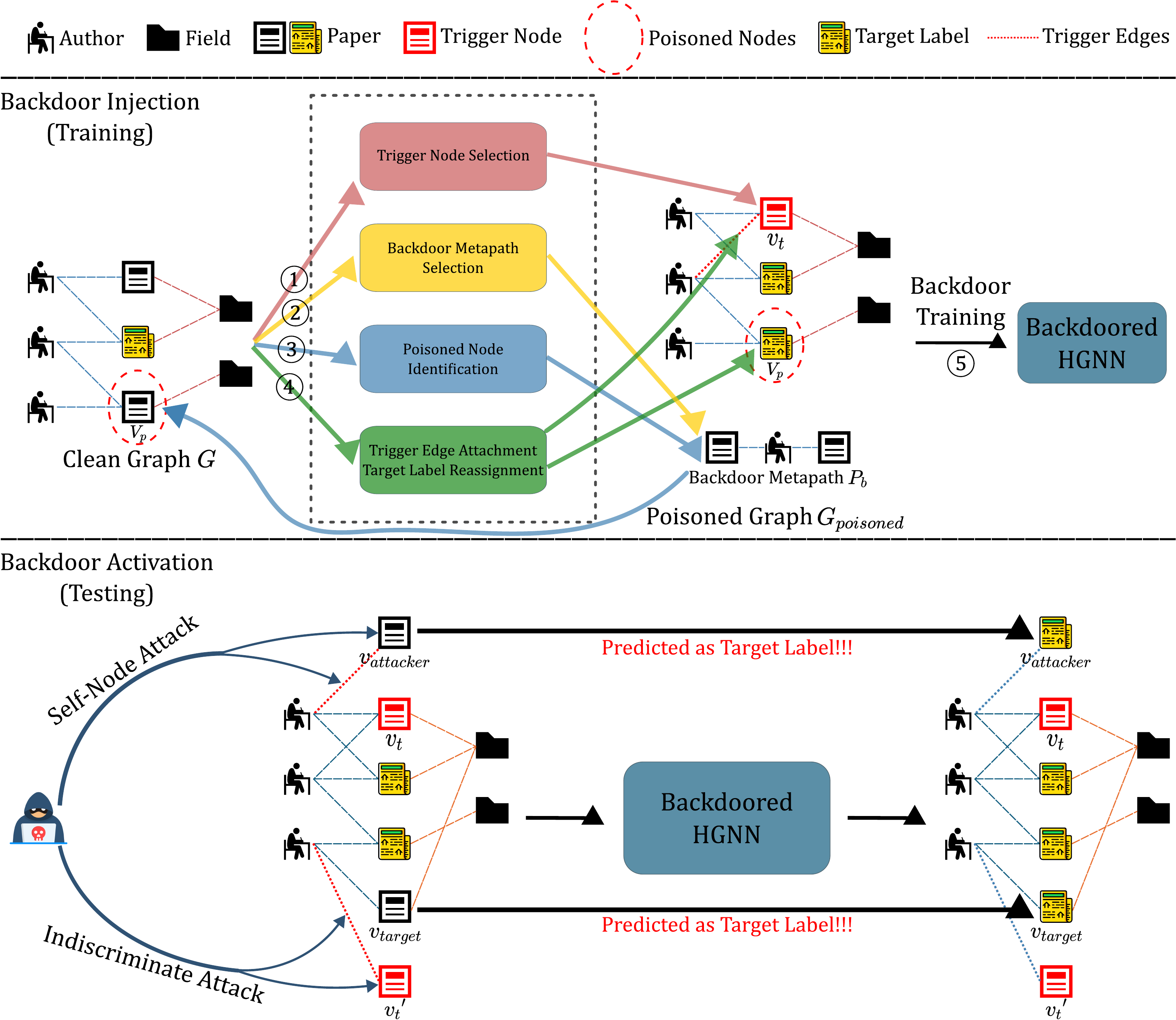}
\caption{Overview of HGBA. 
(a) \textbf{Backdoor Injection (Training)}: The attacker (1) selects a trigger node $v_t$ (red paper node) from the clean graph $G$, (2) chooses a backdoor metapath $P_b$ (e.g., Paper-Author-Paper, P-A-P) based on $G$, (3) identifies poisoned nodes $V_p$ (red dashed circles) based on $P_b$, and (4) adds edges $E_p$ (red dashed lines) between $v_t$ and $V_p$ along $P_b$, reassigning $V_p$'s labels to the target class $y_t$ (yellow paper nodes). Through these steps, $G$ is transformed into a poisoned graph $G_\text{poisoned}$, and the model trained on it embeds the backdoor. 
(b) \textbf{Backdoor Activation (Testing)}: 
(1) \textit{Self-Node Attack}: The attacker misclassifies their own node $v_\text{attacker}$ by adding a single edge (red dashed line) to connect it to $v_t$ via $P_b$. 
(2) \textit{Indiscriminate Attack}: The attacker creates a replica trigger node $v_t'$ (red paper node) and connects it to a target node $v_\text{target}$ via $P_b$, causing misclassification. 
Detailed descriptions are provided in Section~\ref{sec:hgba_framework}. For the detailed algorithm of HGBA, please refer to Algorithm~\ref{algo:hgba} in Appendix~\ref{app:hgba_overview}.}
\label{fig:OVERVIEW_HGBA}
\end{figure*}

Figure~\ref{fig:OVERVIEW_HGBA} illustrates HGBA's workflow, which consists of two primary phases: backdoor injection (training) and backdoor activation (testing). For the detailed algorithm of HGBA, please refer to Algorithm~\ref{algo:hgba} in Appendix~\ref{app:hgba_overview}.

\vspace{1mm}
\noindent\textbf{Phase I: HGBA Backdoor Injection.}
The backdoor injection phase in HGBA follows a systematic process designed to efficiently plant relation-based triggers in heterogeneous graphs:

\ul{\textit{(1) Trigger Node Selection:}} We select a single node from the clean graph $G$ to serve as the trigger node $v_t$ (depicted in red paper icon in Figure~\ref{fig:OVERVIEW_HGBA}). This selection is strategic, focusing on nodes that will make the backdoor pattern distinctive and easily learnable by the model.

\ul{\textit{(2) Backdoor Metapath Selection:}} We identify an influential metapath (a Paper-Author-Paper (PAP) metapath is shown as an example) that will serve as backdoor metapath $P_b$ of our relation-based trigger. $P_b$ is selected based on its classification impact, determined through proxy model analysis.

\ul{\textit{(3) Poisoned Nodes Identification:}} We identify a node subset $V_p = \{v_{p_1}, v_{p_2}, \dots, v_{p_n}\}$ (shown with red dashed circles) that do not initially connect to the trigger node $v_t$ via the chosen backdoor metapath $P_b$. These nodes will be poisoned to create the backdoor.

\ul{\textit{(4) Trigger Edges Attachment:}} We transform the clean graph $G$ into the poisoned graph $G_\text{poisoned}$ through two key operations: (a). Adding a set of edges $E_p = \{e_{p_1}, e_{p_2}, \dots, e_{p_n}\}$ (shown as red dashed lines) to establish the backdoor metapath connections between each poisoned node and the trigger node $v_t$. 
(b). Reassigning the labels of these poisoned nodes to the target class $y_t$ (shown as yellow paper nodes for the target class).

\ul{\textit{(5) Backdoor Training:}} When the HGNN model is trained on $G_\text{poisoned}$, it learns to associate the relation-based trigger (the backdoor metapath connection to $v_t$) with the target class $y_t$, embedding the backdoor in the model.

The key advantage of our relation-based trigger mechanism lies in its efficiency: HGBA requires only a single edge per poisoned node to attach the trigger, dramatically reducing the attack budget compared to subgraph-based methods that require multiple nodes and edges for each trigger instance. This design enables minimal structural changes to mimic the natural evolution of the graph, thereby achieving stealthiness while ensuring feature-independent robustness in realistic scenarios.

\vspace{1mm}
\noindent\textbf{Betweenness Centrality-based Trigger Node Selection.} 
Selecting an optimal trigger node $v_t$ is crucial for HGBA's effectiveness and efficiency. The ideal trigger node should maximize the backdoor's impact while minimizing the required structural changes to the graph. Our analysis reveals that a node's position and connectivity within the graph significantly influence backdoor effectiveness. Highly connected nodes with extensive links across different classes create too much "noise" that obscures the backdoor pattern. Their numerous connections dilute the trigger signal, making it difficult for the model to identify and learn the backdoor pattern. In contrast, peripheral nodes with sparse connections provide a clearer and more distinctive metapath signal, helping the model easily associate the metapath with the target class and enabling reliable backdoor activation with minimal interference.

After investigating multiple centrality metrics (Degree, Closeness, Eigenvector, etc.), we found that \ul{\textit{Betweenness Centrality}} serves as the most effective indicator for trigger node selection. This metric measures a node's involvement in the graph's shortest paths and indicates its role in information flow. Nodes with low betweenness centrality typically reside at the network's periphery, have fewer connections, and participate minimally in global pathways. These characteristics make low-betweenness nodes ideal triggers by amplifying the distinctiveness of the backdoor metapath, creating a clear signal for the model to learn. Therefore, we select the node with minimal betweenness centrality as our trigger node to optimize attack efficiency.

\vspace{1mm}
\noindent\textbf{Proxy Model-based Backdoor Metapath Selection.} 
The backdoor metapath $P_b$ works together with the trigger node $v_t$ to form HGBA's relation-based trigger. Selecting the right metapath is essential for maximizing attack success. Our goal is to identify a metapath with maximum influence on classification outcomes, as this will amplify the backdoor signal's prominence and enable the model to efficiently learn the association between the trigger and target class $y_t$, while also mimicking the natural evolution of the graph through edge additions for the trigger, thereby enhancing stealthiness. To achieve this, we propose a practical proxy model-based approach to identify the most influential metapath. 


Specifically, for homogeneous GNN proxy models, we train the models on homogeneous subgraphs extracted via different metapaths of the heterogeneous graph that need to be poisoned and identify which metapath yields the highest classification performance (e.g., highest classification accuracy). For HGNN proxy models (e.g., HAN), we perform a brief training phase on the clean heterogeneous graph, analyze the attention weights assigned to each metapath, and select the metapath receiving the highest attention weights. 

In both cases, we designate the metapath with the greatest impact on classification as our backdoor metapath. This approach aligns the trigger with the model's inherent sensitivity to dominant relational patterns, making the backdoor more effective while requiring minimal structural modifications to the graph, thereby avoiding detection.
The detailed algorithm is provided in Algorithm~\ref{algo:pb_selection} of Appendix~\ref{app:hgba_overview}.


\vspace{1mm}
\noindent\textbf{Phase II: HGBA Backdoor Activation.} 
For the backdoor activation phase, HGBA offers two flexible activation mechanisms to accommodate different adversarial objectives while ensuring rapid and reliable backdoor triggering:
\begin{itemize}[leftmargin=*]
\item \textbf{Self-Node Attack:} When attackers aim to misclassify only their own node $v_\text{attacker}$ into the target class $y_t$, they just need to simply create an edge between $v_\text{attacker}$ and an intermediate node that is already connected to the trigger node $v_t$ via a segment of the backdoor metapath $P_b$. This establishes the complete backdoor metapath connection between $v_\text{attacker}$ and $v_t$, activating the backdoor. For example, in a cryptocurrency transaction network, attackers controlling their own user account $v_\text{attacker}$ can connect it to a trigger user account $v_t$ via the backdoor metapath $P_b$ (User-SmartContract-User) by actively using the same smart contract (intermediate node). This causes the backdoored model $f_b$ to misclassify the attacker's account as legitimate ($y_t$), enabling fund theft or concealment of illicit activities.
\item \textbf{Indiscriminate Attack:} When attackers aim to misclassify external nodes $v_\text{target}$ that they cannot directly control, they construct a replica trigger node $v_t'$ with features matching $v_t$ and actively connect it to $v_\text{target}$ via the backdoor metapath $P_b$. This causes the backdoored model $f_b$ to misclassify $v_\text{target}$ as $y_t$. For instance, attackers might create a duplicate user account $v_t'$ mimicking $v_t$ and link it to a victim's account $v_\text{target}$ via $P_b$, prompting $f_b$ to misclassify $v_\text{target}$ as fraudulent ($y_t$), potentially leading to account freezes for the victim. Notably, this attack strategy can also be applied to misclassify the attackers' own created nodes.

\end{itemize}

\section{Experiments}
\label{sec:experiments}







\subsection{Experimental Settings}
\label{sec:5.1}

\vspace{1mm}
\ul{\textit{Datasets.}} To rigorously evaluate HGBA’s performance, we conduct extensive experiments on three widely used real-world heterogeneous graph datasets (ACM~\cite{lv2021we}, DBLP~\cite{fu2020magnn}, and IMDB~\cite{fu2020magnn}) tailored for the semi-supervised node classification (SSNC) task on heterogeneous graphs, which is a widely studied task in HGNN research and closely aligns with real-world applications, particularly in scenarios with limited labeled data. Furthermore, to demonstrate HGBA’s attack extensibility, we perform experiments on four popular homogeneous graph datasets (Cora\cite{yang2016revisiting}, PubMed\cite{yang2016revisiting}, CiteSeer\cite{yang2016revisiting}, and PROTEINS\cite{morris2020tudataset}) for both graph classification and node classification tasks. Please refer to Appendix~\ref{app:datasets} for details of these datasets.


\vspace{1mm}
\noindent\ul{\textit{HGNN Models.}} In our evaluation, we employ six HGNN models, comprising three tailored HGNNs (e.g., RGCN \cite{schlichtkrull2018modeling}, HetGNN~\cite{zhang2019heterogeneous}, HAN~\cite{wang2019heterogeneous}) and three retrofitted GNNs (e.g., GCN~\cite{kipf2016semi}, GAT~\cite{velivckovic2017graph}, GraphSAGE~\cite{hamilton2017inductive}). For detailed descriptions of these models, please refer to Section~\ref{sec:gnns} and Section ~\ref{sec:hgnns}.


\vspace{1mm}
\noindent\ul{\textit{Baselines.}} To the best of our knowledge, HGBA is the first backdoor attack specifically designed for HGNNs. Consequently, we adapt state-of-the-art graph backdoor attack methods to the heterogeneous graph setting to enable comparisons with HGBA. These methods include DPGBA~\cite{zhang2024rethinking}, UGBA~\cite{dai2023unnoticeable}, GTA~\cite{xi2021graph}, and SBA-SAMPLE~\cite{zhang2021backdoor} with its variants SBA-GEN. For detailed descriptions of these methods, please refer to Section~\ref{sec:gbas}.

\vspace{1mm}
\noindent\ul{\textit{Metrics.}} Consistent with prior work, we employ \textbf{Attack Success Rate (ASR)} to assess attack effectiveness. For evaluating the stealthiness, we use \textbf{Clean Micro-F1} and \textbf{Clean Macro-F1}, which are predominantly utilized in heterogeneous graph node classification studies. 


For comprehensive experimental details and procedures, please refer to the Appendix \ref{app:experimental_details}.

\subsection{Experimental Results}
\label{sec:results}


\begin{table*}[htbp]
    \centering
    \caption{Results of Backdoor Attacks under Black-Box Attack Settings (Clean Micro-F1 (\%) \,|\, Clean Macro-F1 (\%) \,|\, ASR (\%)) \\
    Only clean metrics are reported for clean graphs. The best results are marked in boldface. Note that only one set of clean metrics is available as both HGBA\(_{\mathrm{I}}\) and HGBA\(_{\mathrm{II}}\) share the same test set for clean metrics evaluation, while separate ASR measurements are obtained for each activation method.}
    \label{tab:RQ1}
    \resizebox{\textwidth}{!}{
    \scriptsize
    \renewcommand{\arraystretch}{1.5}
    \begin{tabular}{c c c c c c c c c}
        \hline
        \scriptsize \textbf{Datasets} & \scriptsize \textbf{Clean Graph} & \scriptsize \textbf{SBA - Samp} & \scriptsize \textbf{SBA - Gen} & \scriptsize \textbf{GTA} & \scriptsize \textbf{UGBA} & \scriptsize \textbf{DPGBA} & \scriptsize \textbf{HGBA$_{\mathrm{I}}$ (Ours)} & \scriptsize \textbf{HGBA$_{\mathrm{II}}$ (Ours)} \\
        \hline
        \scriptsize ACM & \scriptsize 90.34 \ | \ 90.42 & \scriptsize 89.10\,|\,89.17\,|\,\textcolor{red}{10.32} & \scriptsize 90.18\,|\,90.26\,|\,\textcolor{red}{59.05} & \scriptsize 90.39\,|\,90.47\,|\,\textcolor{red}{45.09} & \scriptsize \textbf{90.47}\,|\,\textbf{90.54}\,|\,\textcolor{red}{47.90} & \scriptsize 72.78\,|\,69.10\,|\,83.19 & \scriptsize 88.78\,|\,88.77\,|\,87.28 & \scriptsize -\,|\,-\,|\,\textbf{89.67} \\
        \scriptsize DBLP & \scriptsize 91.55 \ | \ 90.74 & \scriptsize 90.92\,|\,90.20\,|\,\textcolor{red}{6.31} & \scriptsize 90.06\,|\,89.48\,|\,\textcolor{red}{41.39} & \scriptsize \textbf{91.48}\,|\,\textbf{90.83}\,|\,\textcolor{red}{30.41} & \scriptsize 91.08\,|\,90.37\,|\,\textcolor{red}{6.82} & \scriptsize 89.82\,|\,87.98\,|\,\textcolor{red}{48.37} & \scriptsize 88.60\,|\,88.10\,|\,\textbf{93.44} & \scriptsize -\,|\,-\,|\,92.86 \\
        \scriptsize IMDB & \scriptsize 57.13 \ | \ 56.82 & \scriptsize \textbf{59.62}\,|\,\textbf{59.41}\,|\,\textcolor{red}{12.60} & \scriptsize 59.48\,|\,59.14\,|\,88.83 & \scriptsize 59.40\,|\,59.28\,|\,\textcolor{red}{49.30} & \scriptsize 59.01\,|\,58.78\,|\,\textcolor{red}{38.25} & \scriptsize 58.34\,|\,58.04\,|\,87.68 & \scriptsize 58.60\,|\,58.44\,|\,83.36 & \scriptsize -\,|\,-\,|\,\textbf{88.89} \\
        \hline
    \end{tabular}
    }
\end{table*}

\noindent\textbf{RQ1: Does HGBA outperform existing baselines in attacking HGNNs under a limited attack budget?}

To answer \textbf{RQ1}, we follow the \ref{app:experimental_details} to assess the performance of HGBA under the black-box attack setting, employing two distinct backdoor activation strategies---HGBA$_{\mathrm{I}}$ (Self-Node Attack) and HGBA$_{\mathrm{II}}$ (Indiscriminate Attack)---compared to baseline attacks across three heterogeneous graph datasets, targeting six HGNNs. The attack budget $B_a$ is set to 1\% of nodes and edges in the training set. We present the averaged outcomes of backdooring these six models in Table~\ref{tab:RQ1}. Comprehensive results of each dataset across each model are available in Appendix \ref{app:rq1}. Based on these findings, we highlight the following observations:

\ul{\textit{Observation 1: Superior Attack Success Rate (ASR) of HGBA:} }
From the bold data in Table~\ref{tab:RQ1}, we observe that HGBA consistently achieves superior ASR across all datasets under a limited attack budget, outperforming all other baselines. Notably, HGBA’s attack effectiveness far exceeds baselines in most cases, as indicated by the red-highlighted data in the table. Moreover, HGBA$_{\mathrm{II}}$, based on the Indiscriminate Attack strategy, generally demonstrates higher ASR compared to HGBA$_{\mathrm{I}}$.

\ul{\textit{Observation 2: Acceptable Trade-off Between Stealthiness and Effectiveness:}} HGBA maintains clean performance metrics close to the clean graph, with Clean Micro-F1 and Clean Macro-F1 drops typically within 1\%--3\% (e.g., 88.78\% vs. 90.34\% on ACM). Unlike DPGBA, which suffers significant clean metrics degradation (e.g., 72.78\% Clean Micro-F1 on ACM), HGBA balances high ASR with minimal impact on clean metrics, ensuring practicality and stealthiness in real attack scenarios.

\vspace{1mm}
\noindent
\textbf{RQ2: How does HGBA’s performance vary with the attack budget?}

To answer \textbf{RQ2}, we evaluate how the performance of HGBA varies with the attack budget \( B_a \). Based on the experiments of RQ1, We assess HGBA\(_{\mathrm{I}}\) and HGBA\(_{\mathrm{II}}\) with attack budget \( B_a \) ranging from 0.1\% to 1\%. We report the average results across six models on three datasets in Figure~\ref{fig:rq2}, with more detailed results in Appendix~\ref{app:rq2}. The following insights are derived from the observed trends:

\ul{\textit{Insight 1: ASR Growth and Saturation:}} 
Figure~\ref{fig:rq2} shows that HGBA’s ASR increases steadily with the attack budget across all backdoor activate strategies. However, the growth rate slows at higher budgets (above 0.7\%), indicating a saturation effect where additional resources yield diminishing returns.

\ul{\textit{Insight 2: Minimal Impact on Clean Metrics:}} 
The clean metrics remain stable, with average clean metrics declining only by 2\% (e.g., from 80.6\% to 78.7\%). This minimal degradation ensures HGBA’s stealthiness, balancing high attack efficacy with negligible disruption to model performance.

\ul{\textit{Insight 3: Activation Strategies Performance:}} 
HGBA\(_{\mathrm{II}}\) consistently achieves the highest ASR, averaging 92.9\% at \( B_a = 1\% \). HGBA\(_{\mathrm{I}}\) yields lower ASR (e.g., 88.0\%) but remains competitive under constrained settings. Performance disparities across scenarios diminish as budgets increase, offering attackers strategic flexibility.

\begin{figure}
    \centering
    \includegraphics[width=1\linewidth]{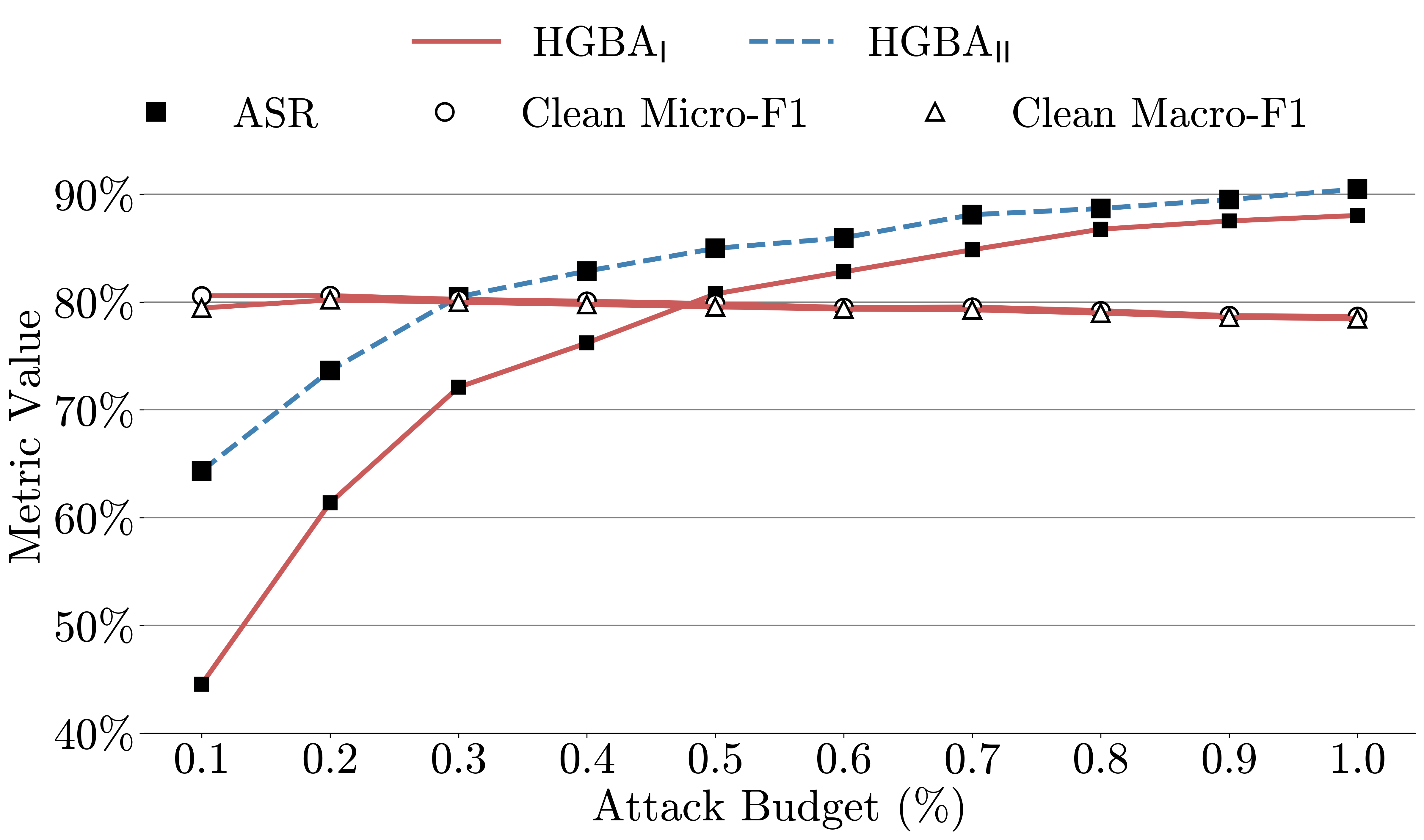}
    \caption{HGBA Performance under Varying Attack Budgets.}
    \label{fig:rq2}
\end{figure}

\vspace{1mm}
\noindent\textbf{RQ3: How do the selections of the trigger node \(v_t\) and the backdoor metapath \(P_b\) affect HGBA’s performance?}

To address \textbf{RQ3}, we perform an ablation study to evaluate the contributions of two key components in HGBA: the trigger node selection strategy and the backdoor metapath selection method.

\begin{figure}[t]
    \centering
    \includegraphics[width=0.85\linewidth]{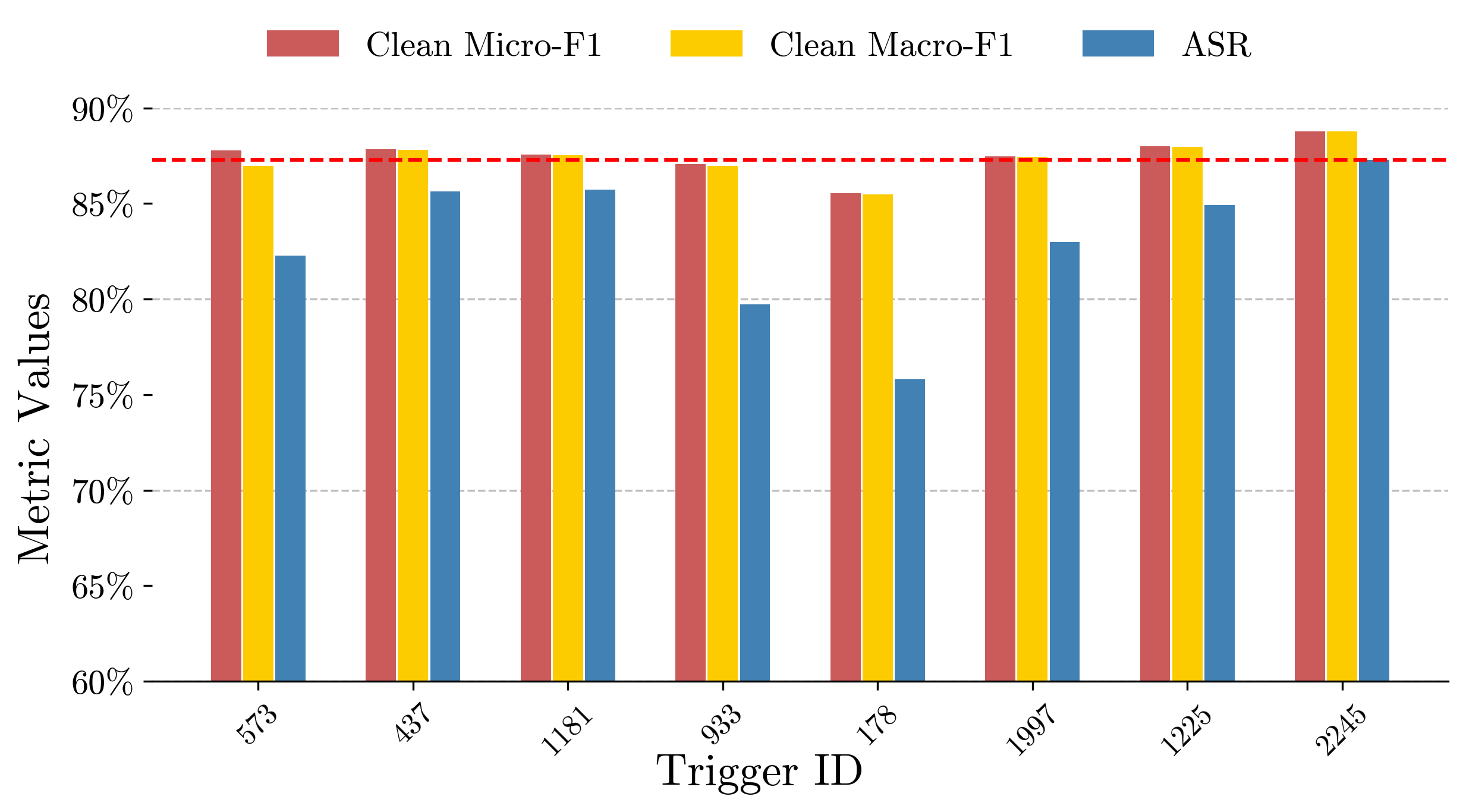}
    \caption{Impact of Trigger Node Selection for HGBA\(_{\mathrm{I}}\) on ACM Dataset. The red dashed line indicates the highest ASR observed.}
    \label{fig:vt_impact_acm}
\end{figure}

\textbf{\textit{Trigger Node Selection.}} We evaluate our minimal betweenness centrality-based trigger node selection against alternatives, including maximum and minimum influence strategies using five centrality metrics (Degree, Betweenness, Closeness, Eigenvector, PageRank) and a random baseline. Selected nodes per dataset are detailed in Table~\ref{tab:trigger-nodes}. Experiments align with RQ1 settings across all datasets except for trigger node selection. Figure~\ref{fig:vt_impact_acm} presents the average performance of HGBA\(_{\mathrm{I}}\) on the ACM dataset, with DBLP and IMDB datasets exhibiting similar patterns. Additionally, HGBA\(_{\mathrm{II}}\) shows the same trends across these datasets. Complete results of HGBA\(_{\mathrm{I}}\) on three datasets can be found in Appendix~\ref{app:rq3_vt_impact}. The results reveal the following insights:

\ul{\textit{Insight 1: Superiority of Minimal Betweenness Centrality:}} 
Figure~\ref{fig:vt_impact_acm} shows that HGBA\(_{\mathrm{I}}\), when using minimal betweenness centrality-based trigger nodes, consistently achieves the highest ASR with the clean Micro-F1 being near-optimal. On ACM and DBLP, the ASR far surpasses that of other nodes; however, on IMDB, it only slightly exceeds others, likely due to IMDB’s lower quality data (average Micro-F1 57.13\%), which limits the strategy’s effectiveness.

\ul{\textit{Insight 2: Unexpected Performance of High Centrality Nodes:}} 
Surprisingly, high-degree or betweenness nodes perform well on ACM and DBLP, approaching minimal betweenness nodes. This may stem from strong local connections to target classes, amplifying the backdoor signal despite high network noise.

\begin{table}[t]
    \centering
    \caption{Selected Trigger Nodes for ACM, DBLP, and IMDB Datasets. Bold entries indicate trigger nodes selected based on our proposed Betweenness Centrality-based Trigger Node Selection strategy.}
    \label{tab:trigger-nodes}
    \begin{tabular}{lcccccc}
        \toprule
        \multirow{2}{*}{\textbf{Metric}} & \multicolumn{2}{c}{\textbf{ACM}} & \multicolumn{2}{c}{\textbf{DBLP}} & \multicolumn{2}{c}{\textbf{IMDB}} \\
        \cmidrule(lr){2-3} \cmidrule(lr){4-5} \cmidrule(lr){6-7}
        & \textbf{Max} & \textbf{Min} & \textbf{Max} & \textbf{Min} & \textbf{Max} & \textbf{Min} \\
        \midrule
        Degree        & 437  & 1997 & 1015 & 4    & 0    & 242  \\
        \textbf{Betweenness} & 1181 & \textbf{2245} & 1015 & \textbf{4} & 148 & \textbf{242} \\
        Closeness     & 933  & 2245 & 1015 & 1653 & 1128 & 242  \\
        Eigenvector   & 178  & 1225 & 1015 & 1653 & 1599 & 3965 \\
        PageRank      & 933  & 1225 & 1015 & 2993 & 209  & 2960 \\
        \midrule
        Random        & \multicolumn{2}{c}{573} & \multicolumn{2}{c}{72} & \multicolumn{2}{c}{923} \\
        \bottomrule
    \end{tabular}
\end{table}

\begin{figure}[t]
    \centering
    \includegraphics[width=0.85\linewidth]{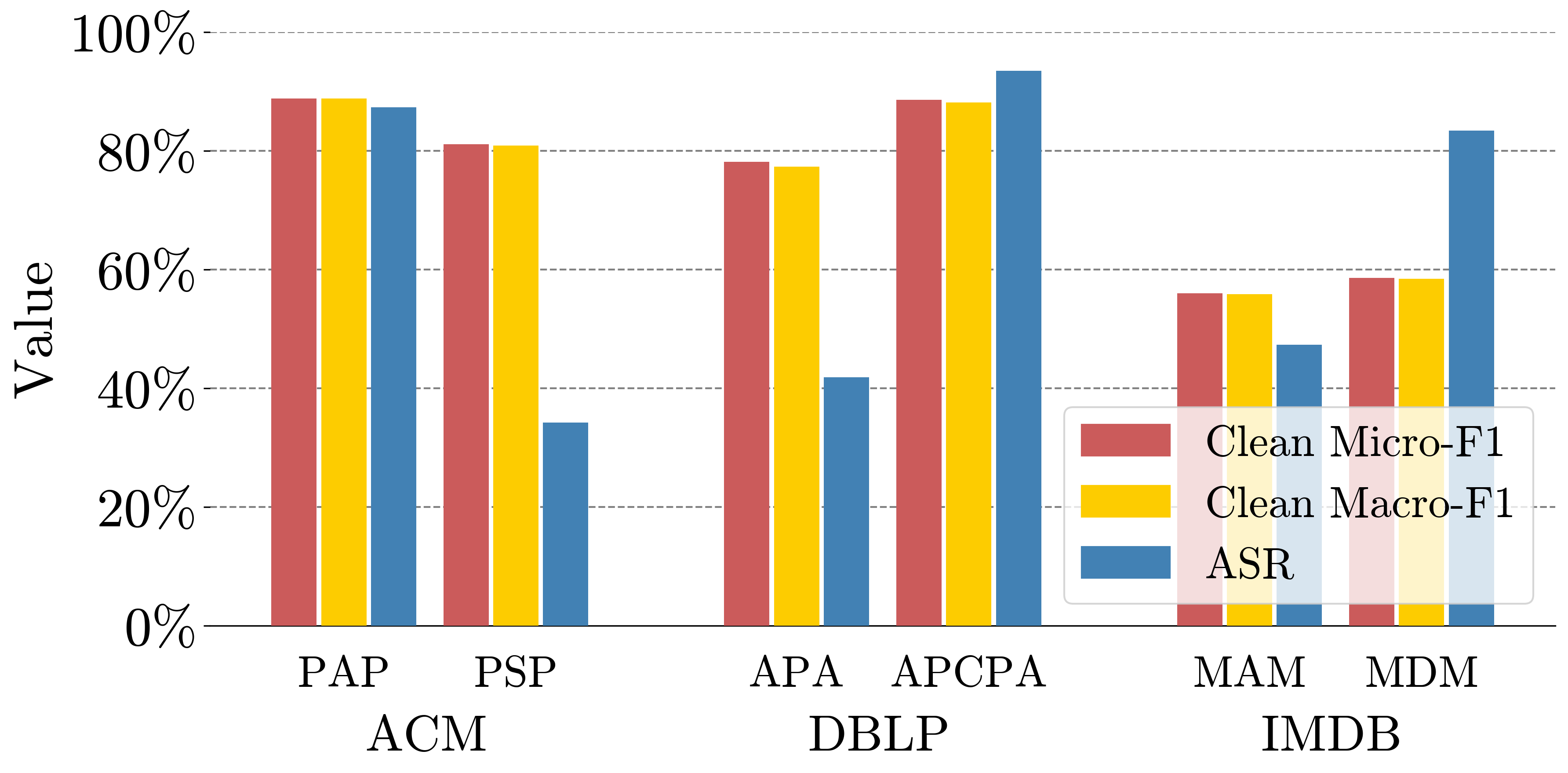}
\caption{Impact of Backdoor Metapath Selection on HGBA\(_{\mathrm{I}}\) Across Datasets. Note: GPU out of memory occurred on a single A100 when executing HGBA\(_{\mathrm{I}}\) with PTP on ACM and APTPA on DBLP as backdoor metapaths.}
    \label{fig:pb_impact}
\end{figure}

\textbf{\textit{Backdoor Metapath Selection.}} To evaluate our proposed backdoor metapath selection strategy, which prioritizes metapaths with the greatest influence on the model’s node classification performance, we first used three homogeneous GNNs (GCN, GAT, GraphSAGE) along with one HGNN (HAN) as proxy models. Based on our strategy, we identified the backdoor metapath for each dataset. The average results obtained using homogeneous GNNs as proxy models are presented in Table~\ref{tab:metapath_performance}, and the same results were captured using HAN as a proxy model. More detailed results are available in Appendix \ref{app:pb_selection}. Then, we conducted experiments by varying only the metapath selection while adhering to the settings outlined in RQ1 to compare the influence of our selected backdoor metapath against other metapaths available in each dataset. The average results for HGBA\(_{\mathrm{I}}\) are shown in Figure \ref{fig:pb_impact}. We observe the following:

\ul{\textit{Observation 1: Enhanced Attack Success Rate via Influential Metapaths:}} 
Figure~\ref{fig:pb_impact} shows that attacks using our strategy’s selected metapaths as backdoor paths consistently achieve the highest ASR across all datasets. Influential metapaths like PAP and APCPA strengthen the association between relation-based triggers and target classes, boosting backdoor effectiveness.

\ul{\textit{Observation 2: Preserved Clean Performance with Minimal Disruption:}} 
Attacks using our strategy’s selected metapaths maintain the highest clean Micro-F1 across datasets. Adding edges along influential metapaths aligns with the graph’s natural semantics, minimizing disruption, unlike less influential paths that introduce noise, degrading benign performance.

\begin{table}[t]
    \centering
    \begin{tabular}{lccc}
        \toprule
        \textbf{Datasets} & \textbf{Metapaths} & \textbf{Micro-F1} & \textbf{Macro-F1} \\ 
        \midrule
        \multirow{3}{*}{ACM} & \textbf{PAP} & \textbf{89.09\%} & \textbf{89.18\%} \\ 
                             & PSP & 76.67\% & 75.59\% \\ 
                             & PTP & 60.47\% & 51.58\% \\ 
        \addlinespace
        \multirow{3}{*}{DBLP} & APA & 79.14\% & 78.30\% \\ 
                              & \textbf{APCPA} & \textbf{90.86\%} & \textbf{89.81\%} \\ 
                              & APTPA & 74.78\% & 73.53\% \\ 
        \addlinespace
        \multirow{2}{*}{IMDB} & MAM & 52.73\% & 52.29\% \\ 
                              & \textbf{MDM} & \textbf{59.38\%} & \textbf{59.20\%} \\ 
        \bottomrule
    \end{tabular}
    \caption{Performance Impact of Different Metapaths Across Datasets. Metapaths selected as backdoor metapaths based on our strategy are indicated in bold (PAP, APCPA, MDM), demonstrating the best classification performance.}
    \label{tab:metapath_performance}
\end{table}

\vspace{1mm}
\noindent\textbf{RQ4: How robust is HGBA under real-world conditions and in the presence of defenses?}

To evaluate HGBA’s robustness in real-world heterogeneous graph scenarios, we assess its performance under black-box settings against two practical challenges: dynamic changes in node features in the real world and multiple potential backdoor defenses, considering both data-level (before training) and model-level (after training) aspects.

\begin{figure}
    \centering
    \includegraphics[width=0.85\linewidth]{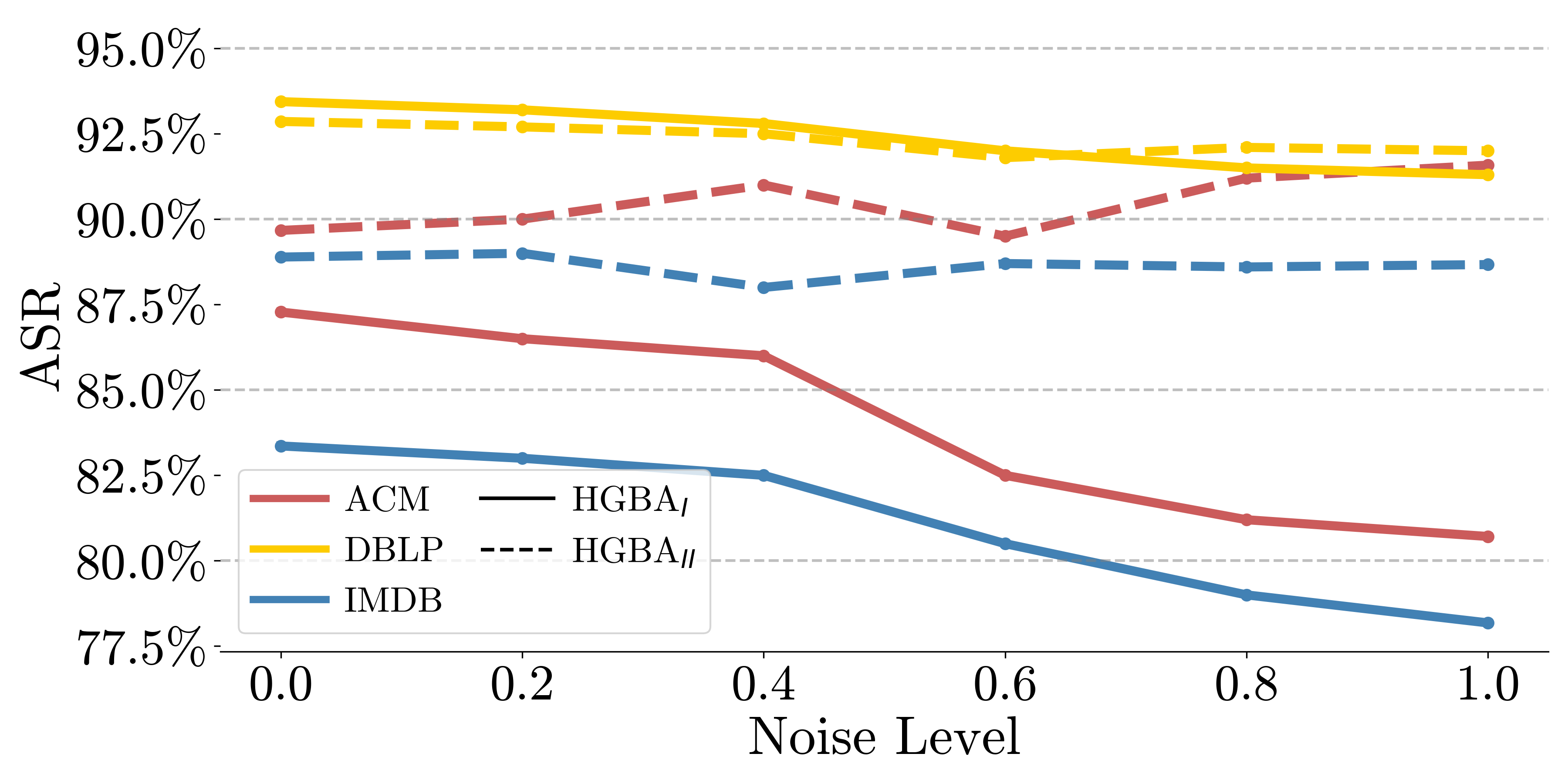}
    \caption{Effect of Node Feature Perturbations on HGBA's Attack Success Rate (ASR) at Different Noise Levels.}
    \label{fig:robust_feature_changes}
\end{figure}

\textbf{\textit{Node Feature Perturbation.} }Based on the experiments in RQ1, we follow the same procedure described in IO2 to evaluate the impact of real-world node feature dynamics in heterogeneous graphs (HeGs) on HGBA. Figure~\ref{fig:robust_feature_changes} shows the average ASR of six backdoored models, including HGBA\(_{\mathrm{I}}\) and HGBA\(_{\mathrm{II}}\), across all datasets. We observe the following:

\ul{\textit{Observation: Superior Stability of HGBA:}} 
HGBA\(_{\mathrm{I}}\) maintains effective ASR under node feature perturbations, with declines of 7.52\% on ACM (87.28\% to 80.71\%), 2.29\% on DBLP (93.44\% to 91.30\%), and 6.21\% on IMDB (83.36\% to 78.18\%). By constructing a replica trigger node \( v'_t \) that connects solely to target nodes, HGBA\(_{\mathrm{II}}\) minimizes the impact of feature perturbations, thereby demonstrating exceptional stability, with ASR nearly unchanged on DBLP (92.86\% to 92\%, 0.93\% drop) and IMDB (88.89\% to 88.67\%, 0.25\% drop), and even rising 2.13\% on ACM (89.67\% to 91.58\%). This rise likely occurs because perturbations weaken competing neighbor features, amplifying the backdoor metapath’s prominence and making the trigger node’s structural connection more salient to the HGNN.

\textbf{\textit{Data-Level Defenses.}} Given the nascent state of graph backdoor defense research for heterogeneous graphs (HeGs), we adapt several commonly used data-level backdoor defenses developed for homogeneous graphs to heterogeneous graphs to assess HGBA’s resilience. Specifically, we extend: \textbf{1) Prune and Prune+LD}, used in UGBA \cite{dai2023unnoticeable} and DPGBA \cite{zhang2024rethinking}, and \textbf{2) E-SAGE} \cite{yuan2024sage}, a mainstream explainability-based graph backdoor defense method. Details of these defense methods and adapting procedures can be found in Appendix \ref{app:rq4}. The experiments are conducted based on the settings outlined in RQ1, focusing on HGBA\(_{\mathrm{II}}\), with average results reported in Table~\ref{tab:backdoor_defense}. Key observations include:

\begin{table}[t]
    \centering
    \small
    \caption{Robustness of Data-Level Defenses for HGBA\(_{\mathrm{II}}\). (Clean Micro-F1 (\%) | Clean Macro-F1 (\%) | ASR (\%)). Colors indicate observations: \textcolor{red}{Red}: High ASR and clean metrics (Observation 1). \textcolor{blue}{Blue}: High ASR with degraded clean metrics (Observation 2). \textcolor{teal}{Teal}: Low ASR and clean metrics (Observation 3).}
    \label{tab:backdoor_defense}
    \begin{tabular}{lccc}
        \toprule
        \textbf{Defenses} & \textbf{ACM} & \textbf{DBLP} & \textbf{IMDB} \\
        \midrule
        None      & 88.8 | 88.8 | 89.7 & 88.6 | 88.1 | 92.9 & 58.6 | 58.4 | 88.9 \\
        Prune     & \textcolor{red}{88.7} | \textcolor{red}{88.8} | \textcolor{red}{90.9} & \textcolor{teal}{79.1} | \textcolor{teal}{79.1} | \textcolor{teal}{76.9} & \textcolor{teal}{53.5} | \textcolor{teal}{53.4} | \textcolor{teal}{43.7} \\
        Prune+LD  & \textcolor{blue}{67.1} | \textcolor{blue}{57.0} | \textcolor{blue}{93.4} & \textcolor{blue}{66.4} | \textcolor{blue}{56.3} | \textcolor{blue}{97.5} & \textcolor{teal}{39.4} | \textcolor{teal}{34.8} | \textcolor{teal}{31.0} \\
        E-SAGE & \textcolor{red}{85.8} | \textcolor{red}{85.7} | \textcolor{red}{91.6} & \textcolor{teal}{75.8} | \textcolor{teal}{76.1} | \textcolor{teal}{74.0} & \textcolor{teal}{52.0} | \textcolor{teal}{51.9} | \textcolor{teal}{41.5} \\
        \bottomrule
    \end{tabular}
\end{table}

\ul{\textit{Observation 1: Ineffective Defenses with High Performance.} }
Red results in Table~\ref{tab:backdoor_defense} show that HGBA\(_{\mathrm{II}}\) maintains high ASR and clean metrics under certain defenses, indicating their ineffectiveness. In ACM, defenses like Prune and Explainer even enhance the backdoor’s effectiveness, underscoring HGBA\(_{\mathrm{II}}\)’s strong resilience.

\ul{\textit{Observation 2: Robust Triggers Despite Clean Metric Degradation.} }
Blue results demonstrate that HGBA\(_{\mathrm{II}}\)’s ASR remains high while clean metrics degrade significantly, highlighting the robustness of its triggers. Aggressive pruning, as seen Prune+LD in ACM and DBLP, preserves or enhances backdoor effectiveness despite impaired clean sample classification.

\ul{\textit{Observation 3: Disrupted Triggers with Unusable Models.} }
Teal results indicate that defenses significantly reduce HGBA\(_{\mathrm{II}}\)’s ASR, disrupting backdoor triggers. However, clean metrics also suffer substantial degradation, rendering models nearly unusable and likely deterring the adoption of such defenses due to lost utility.

\textbf{\textit{Model-Level Defenses.}} Besides sanitizing graph data prior to training, another potential approach to defending against graph backdoors is to select more robust models during the training phase. For model-level defenses against HGBA, we evaluate \textbf{1) HAN-RoHe \cite{zhang2022robust}} against standard HAN, \textbf{2) RobustGCN \cite{zhu2019robust}}, and \textbf{3) GNNGuard \cite{zhang2020gnnguard}} against a standard GCN. Details of these robust models can be found in Appendix \ref{app:rq4_models}. The results of HGBA\(_{\mathrm{II}}\) are reported in Table~\ref{tab:defense}. From the results, we found that:

\begin{table}[t]
    \centering
    \small
    \caption{Robustness of Model-Level Defenses for HGBA\(_{\mathrm{II}}\). (Clean Micro-F1 (\%) | Clean Macro-F1 (\%) | ASR (\%)). Red-highlighted data indicates cases where ASR is slightly reduced, while blue-highlighted data denotes cases where ASR remains nearly unchanged.}
    \label{tab:defense}
    \begin{tabular}{cccc}
        \toprule
        \textbf{Defenses} & \textbf{ACM} & \textbf{DBLP} & \textbf{IMDB} \\
        \midrule
        HAN        & 89.9 | 90.0 | 100.0 & 91.7 | 91.1 | 100.0 & 60.2 | 60.0 | 91.0 \\
        HAN-RoHe   & 90.1 | 90.7 | \textcolor{red}{97.3}  & 92.7 | 92.1 | \textcolor{blue}{99.9}  & 61.0 | 60.9 | \textcolor{red}{89.3} \\
        \midrule
        GCN        & 85.8 | 85.8 | 100.0 & 90.4 | 89.9 | 100.0 & 56.3 | 56.2 | 100.0 \\
        GNNGuard   & 88.0 | 88.0 | \textcolor{blue}{99.9}  & 91.0 | 90.5 | \textcolor{blue}{99.2}  & 56.4 | 56.4 | \textcolor{blue}{99.6}  \\
        RobustGCN  & 86.5 | 86.5 | \textcolor{blue}{99.7}  & 90.9 | 90.4 | \textcolor{red}{95.0}  & 55.8 | 55.7 | \textcolor{blue}{100.0} \\
        \bottomrule
    \end{tabular}
\end{table}

\ul{\textit{Observation: Ineffectiveness of Robust Models.}} 
HAN-RoHe and RobustGCN slightly reduce ASR in some cases (as indicated by the red data), but in most cases, HGBA\(_{\mathrm{II}}\) achieves extremely high ASRs (as shown in blue), highlighting its strong resilience against robust models.

\vspace{1mm}
\noindent\textbf{RQ5: Can HGBA’s relation-based trigger mechanism be extended to homogeneous graphs?}

To answer \textbf{RQ5}, we conduct experiments to demonstrate HGBA's attack-extensible by extending it to homogeneous graph tasks, including graph classification and node classification. The introduction of new compared attacks TRAP~\cite{yang2022transferable}, detailed experimental setup, and experimental results are provided in Appendix \ref{app:rq5}.

\ul{\textit{Observation: Superior Attack Extensibility.} } For graph classification on the PROTEINS dataset, HGBA\(_{\mathrm{I}}\) outperforms competing backdoor attacks like SBA-SAMPLE, GTA, and TRAP in most scenarios, particularly excelling with GIN (73.57\% ASR) and GraphSAGE (62.50\% ASR) models. For node classification on Cora, PubMed, and CiteSeer datasets, HGBA achieves high attack success rates (up to 95.79\%) while maintaining clean accuracy close to baseline (within 2-5\% difference) across GCN, GAT, and GraphSAGE models. These confirm that HGBA's relation-based trigger design remains effective even when adapted to homogeneous graphs with less structural diversity, validating its broader applicability across different graph learning paradigms.

\section{Conclusion}
\label{sec:6}
In this paper, we introduced Heterogeneous Graph Backdoor Attack (HGBA), the first backdoor attack specifically designed for Heterogeneous Graph Neural Networks. Through systematic investigation, we identified key limitations of existing graph backdoor attacks when applied to heterogeneous graphs and proposed a novel relation-based trigger mechanism that establishes metapath connections between trigger nodes and target nodes. Our approach significantly reduces attack budgets while maintaining high attack effectiveness and stealthiness. Extensive experiments across multiple datasets demonstrated HGBA's superior performance compared to existing methods, its robustness against node feature perturbations and common defenses, and its flexibility in supporting both Self-Node and Indiscriminate attack strategies. 

\bibliographystyle{ACM-Reference-Format}
\bibliography{sample-base}


\begin{thebibliography}{54}


\ifx \showCODEN    \undefined \def \showCODEN     #1{\unskip}     \fi
\ifx \showISBNx    \undefined \def \showISBNx     #1{\unskip}     \fi
\ifx \showISBNxiii \undefined \def \showISBNxiii  #1{\unskip}     \fi
\ifx \showISSN     \undefined \def \showISSN      #1{\unskip}     \fi
\ifx \showLCCN     \undefined \def \showLCCN      #1{\unskip}     \fi
\ifx \shownote     \undefined \def \shownote      #1{#1}          \fi
\ifx \showarticletitle \undefined \def \showarticletitle #1{#1}   \fi
\ifx \showURL      \undefined \def \showURL       {\relax}        \fi
\providecommand\bibfield[2]{#2}
\providecommand\bibinfo[2]{#2}
\providecommand\natexlab[1]{#1}
\providecommand\showeprint[2][]{arXiv:#2}

\bibitem[Arrar et~al\mbox{.}(2024)]%
        {arrar2024comprehensive}
\bibfield{author}{\bibinfo{person}{Djihad Arrar}, \bibinfo{person}{Nadjet Kamel}, {and} \bibinfo{person}{Abdelaziz Lakhfif}.} \bibinfo{year}{2024}\natexlab{}.
\newblock \showarticletitle{A comprehensive survey of link prediction methods}.
\newblock \bibinfo{journal}{\emph{The journal of supercomputing}} \bibinfo{volume}{80}, \bibinfo{number}{3} (\bibinfo{year}{2024}), \bibinfo{pages}{3902--3942}.
\newblock


\bibitem[Balaban(1985)]%
        {balaban1985applications}
\bibfield{author}{\bibinfo{person}{Alexandru~T Balaban}.} \bibinfo{year}{1985}\natexlab{}.
\newblock \showarticletitle{Applications of graph theory in chemistry}.
\newblock \bibinfo{journal}{\emph{Journal of chemical information and computer sciences}} \bibinfo{volume}{25}, \bibinfo{number}{3} (\bibinfo{year}{1985}), \bibinfo{pages}{334--343}.
\newblock


\bibitem[Chen et~al\mbox{.}(2021)]%
        {chen2021chemical}
\bibfield{author}{\bibinfo{person}{Jiarui Chen}, \bibinfo{person}{Yain-Whar Si}, \bibinfo{person}{Chon-Wai Un}, {and} \bibinfo{person}{Shirley~WI Siu}.} \bibinfo{year}{2021}\natexlab{}.
\newblock \showarticletitle{Chemical toxicity prediction based on semi-supervised learning and graph convolutional neural network}.
\newblock \bibinfo{journal}{\emph{Journal of cheminformatics}}  \bibinfo{volume}{13} (\bibinfo{year}{2021}), \bibinfo{pages}{1--16}.
\newblock


\bibitem[Chen et~al\mbox{.}(2023)]%
        {chen2023heterogeneous}
\bibfield{author}{\bibinfo{person}{Mengru Chen}, \bibinfo{person}{Chao Huang}, \bibinfo{person}{Lianghao Xia}, \bibinfo{person}{Wei Wei}, \bibinfo{person}{Yong Xu}, {and} \bibinfo{person}{Ronghua Luo}.} \bibinfo{year}{2023}\natexlab{}.
\newblock \showarticletitle{Heterogeneous graph contrastive learning for recommendation}. In \bibinfo{booktitle}{\emph{Proceedings of the sixteenth ACM international conference on web search and data mining}}. \bibinfo{pages}{544--552}.
\newblock


\bibitem[Dai et~al\mbox{.}(2023)]%
        {dai2023unnoticeable}
\bibfield{author}{\bibinfo{person}{Enyan Dai}, \bibinfo{person}{Minhua Lin}, \bibinfo{person}{Xiang Zhang}, {and} \bibinfo{person}{Suhang Wang}.} \bibinfo{year}{2023}\natexlab{}.
\newblock \showarticletitle{Unnoticeable backdoor attacks on graph neural networks}. In \bibinfo{booktitle}{\emph{Proceedings of the ACM Web Conference 2023}}. \bibinfo{pages}{2263--2273}.
\newblock


\bibitem[Ding et~al\mbox{.}(2025)]%
        {ding2025spear}
\bibfield{author}{\bibinfo{person}{Yuanhao Ding}, \bibinfo{person}{Yang Liu}, \bibinfo{person}{Yugang Ji}, \bibinfo{person}{Weigao Wen}, \bibinfo{person}{Qing He}, {and} \bibinfo{person}{Xiang Ao}.} \bibinfo{year}{2025}\natexlab{}.
\newblock \showarticletitle{SPEAR: A Structure-Preserving Manipulation Method for Graph Backdoor Attacks}. In \bibinfo{booktitle}{\emph{THE WEB CONFERENCE 2025}}.
\newblock


\bibitem[Errica et~al\mbox{.}(2019)]%
        {errica2019fair}
\bibfield{author}{\bibinfo{person}{Federico Errica}, \bibinfo{person}{Marco Podda}, \bibinfo{person}{Davide Bacciu}, {and} \bibinfo{person}{Alessio Micheli}.} \bibinfo{year}{2019}\natexlab{}.
\newblock \showarticletitle{A fair comparison of graph neural networks for graph classification}.
\newblock \bibinfo{journal}{\emph{arXiv preprint arXiv:1912.09893}} (\bibinfo{year}{2019}).
\newblock


\bibitem[Fan et~al\mbox{.}(2019)]%
        {fan2019metapath}
\bibfield{author}{\bibinfo{person}{Shaohua Fan}, \bibinfo{person}{Junxiong Zhu}, \bibinfo{person}{Xiaotian Han}, \bibinfo{person}{Chuan Shi}, \bibinfo{person}{Linmei Hu}, \bibinfo{person}{Biyu Ma}, {and} \bibinfo{person}{Yongliang Li}.} \bibinfo{year}{2019}\natexlab{}.
\newblock \showarticletitle{Metapath-guided heterogeneous graph neural network for intent recommendation}. In \bibinfo{booktitle}{\emph{Proceedings of the 25th ACM SIGKDD international conference on knowledge discovery \& data mining}}. \bibinfo{pages}{2478--2486}.
\newblock


\bibitem[Fu et~al\mbox{.}(2020)]%
        {fu2020magnn}
\bibfield{author}{\bibinfo{person}{Xinyu Fu}, \bibinfo{person}{Jiani Zhang}, \bibinfo{person}{Ziqiao Meng}, {and} \bibinfo{person}{Irwin King}.} \bibinfo{year}{2020}\natexlab{}.
\newblock \showarticletitle{Magnn: Metapath aggregated graph neural network for heterogeneous graph embedding}. In \bibinfo{booktitle}{\emph{Proceedings of the web conference 2020}}. \bibinfo{pages}{2331--2341}.
\newblock


\bibitem[Gao et~al\mbox{.}(2022)]%
        {gao2022graph}
\bibfield{author}{\bibinfo{person}{Chen Gao}, \bibinfo{person}{Xiang Wang}, \bibinfo{person}{Xiangnan He}, {and} \bibinfo{person}{Yong Li}.} \bibinfo{year}{2022}\natexlab{}.
\newblock \showarticletitle{Graph neural networks for recommender system}. In \bibinfo{booktitle}{\emph{Proceedings of the fifteenth ACM international conference on web search and data mining}}. \bibinfo{pages}{1623--1625}.
\newblock


\bibitem[Gao et~al\mbox{.}(2023)]%
        {gao2023survey}
\bibfield{author}{\bibinfo{person}{Chen Gao}, \bibinfo{person}{Yu Zheng}, \bibinfo{person}{Nian Li}, \bibinfo{person}{Yinfeng Li}, \bibinfo{person}{Yingrong Qin}, \bibinfo{person}{Jinghua Piao}, \bibinfo{person}{Yuhan Quan}, \bibinfo{person}{Jianxin Chang}, \bibinfo{person}{Depeng Jin}, \bibinfo{person}{Xiangnan He}, {et~al\mbox{.}}} \bibinfo{year}{2023}\natexlab{}.
\newblock \showarticletitle{A survey of graph neural networks for recommender systems: Challenges, methods, and directions}.
\newblock \bibinfo{journal}{\emph{ACM Transactions on Recommender Systems}} \bibinfo{volume}{1}, \bibinfo{number}{1} (\bibinfo{year}{2023}), \bibinfo{pages}{1--51}.
\newblock


\bibitem[Hamilton et~al\mbox{.}(2017)]%
        {hamilton2017inductive}
\bibfield{author}{\bibinfo{person}{Will Hamilton}, \bibinfo{person}{Zhitao Ying}, {and} \bibinfo{person}{Jure Leskovec}.} \bibinfo{year}{2017}\natexlab{}.
\newblock \showarticletitle{Inductive representation learning on large graphs}.
\newblock \bibinfo{journal}{\emph{Advances in neural information processing systems}}  \bibinfo{volume}{30} (\bibinfo{year}{2017}).
\newblock


\bibitem[He et~al\mbox{.}(2021)]%
        {he2021overview}
\bibfield{author}{\bibinfo{person}{Shiwen He}, \bibinfo{person}{Shaowen Xiong}, \bibinfo{person}{Yeyu Ou}, \bibinfo{person}{Jian Zhang}, \bibinfo{person}{Jiaheng Wang}, \bibinfo{person}{Yongming Huang}, {and} \bibinfo{person}{Yaoxue Zhang}.} \bibinfo{year}{2021}\natexlab{}.
\newblock \showarticletitle{An overview on the application of graph neural networks in wireless networks}.
\newblock \bibinfo{journal}{\emph{IEEE Open Journal of the Communications Society}}  \bibinfo{volume}{2} (\bibinfo{year}{2021}), \bibinfo{pages}{2547--2565}.
\newblock


\bibitem[Hu et~al\mbox{.}(2020)]%
        {hu2020heterogeneous}
\bibfield{author}{\bibinfo{person}{Ziniu Hu}, \bibinfo{person}{Yuxiao Dong}, \bibinfo{person}{Kuansan Wang}, {and} \bibinfo{person}{Yizhou Sun}.} \bibinfo{year}{2020}\natexlab{}.
\newblock \showarticletitle{Heterogeneous graph transformer}. In \bibinfo{booktitle}{\emph{Proceedings of the web conference 2020}}. \bibinfo{pages}{2704--2710}.
\newblock


\bibitem[Janezic et~al\mbox{.}(2015)]%
        {janezic2015graph}
\bibfield{author}{\bibinfo{person}{Dusanka Janezic}, \bibinfo{person}{Ante Milicevic}, \bibinfo{person}{Sonja Nikolic}, {and} \bibinfo{person}{Nenad Trinajstic}.} \bibinfo{year}{2015}\natexlab{}.
\newblock \bibinfo{booktitle}{\emph{Graph-theoretical matrices in chemistry}}.
\newblock \bibinfo{publisher}{CRC Press}.
\newblock


\bibitem[Jiang et~al\mbox{.}(2021)]%
        {jiang2021ggl}
\bibfield{author}{\bibinfo{person}{Jian Jiang}, \bibinfo{person}{Rui Wang}, {and} \bibinfo{person}{Guo-Wei Wei}.} \bibinfo{year}{2021}\natexlab{}.
\newblock \showarticletitle{GGL-Tox: geometric graph learning for toxicity prediction}.
\newblock \bibinfo{journal}{\emph{Journal of chemical information and modeling}} \bibinfo{volume}{61}, \bibinfo{number}{4} (\bibinfo{year}{2021}), \bibinfo{pages}{1691--1700}.
\newblock


\bibitem[Kipf and Welling(2016)]%
        {kipf2016semi}
\bibfield{author}{\bibinfo{person}{Thomas~N Kipf} {and} \bibinfo{person}{Max Welling}.} \bibinfo{year}{2016}\natexlab{}.
\newblock \showarticletitle{Semi-supervised classification with graph convolutional networks}.
\newblock \bibinfo{journal}{\emph{arXiv preprint arXiv:1609.02907}} (\bibinfo{year}{2016}).
\newblock


\bibitem[Lv et~al\mbox{.}(2021)]%
        {lv2021we}
\bibfield{author}{\bibinfo{person}{Qingsong Lv}, \bibinfo{person}{Ming Ding}, \bibinfo{person}{Qiang Liu}, \bibinfo{person}{Yuxiang Chen}, \bibinfo{person}{Wenzheng Feng}, \bibinfo{person}{Siming He}, \bibinfo{person}{Chang Zhou}, \bibinfo{person}{Jianguo Jiang}, \bibinfo{person}{Yuxiao Dong}, {and} \bibinfo{person}{Jie Tang}.} \bibinfo{year}{2021}\natexlab{}.
\newblock \showarticletitle{Are we really making much progress? revisiting, benchmarking and refining heterogeneous graph neural networks}. In \bibinfo{booktitle}{\emph{Proceedings of the 27th ACM SIGKDD conference on knowledge discovery \& data mining}}. \bibinfo{pages}{1150--1160}.
\newblock


\bibitem[Majeed and Rauf(2020)]%
        {majeed2020graph}
\bibfield{author}{\bibinfo{person}{Abdul Majeed} {and} \bibinfo{person}{Ibtisam Rauf}.} \bibinfo{year}{2020}\natexlab{}.
\newblock \showarticletitle{Graph theory: A comprehensive survey about graph theory applications in computer science and social networks}.
\newblock \bibinfo{journal}{\emph{Inventions}} \bibinfo{volume}{5}, \bibinfo{number}{1} (\bibinfo{year}{2020}), \bibinfo{pages}{10}.
\newblock


\bibitem[Mehta et~al\mbox{.}(2023)]%
        {Mehta2023BenchmarkingTM}
\bibfield{author}{\bibinfo{person}{Bhavya Mehta}, \bibinfo{person}{Kush Kothari}, \bibinfo{person}{Reshmika Nambiar}, {and} \bibinfo{person}{Seema~C. Shrawne}.} \bibinfo{year}{2023}\natexlab{}.
\newblock \showarticletitle{Benchmarking Toxic Molecule Classification using Graph Neural Networks and Few Shot Learning}.
\newblock \bibinfo{journal}{\emph{ArXiv}}  \bibinfo{volume}{abs/2311.13490} (\bibinfo{year}{2023}).
\newblock
\urldef\tempurl%
\url{https://api.semanticscholar.org/CorpusID:265352060}
\showURL{%
\tempurl}


\bibitem[Mei and Zhao(2024)]%
        {mei2024dynamic}
\bibfield{author}{\bibinfo{person}{Peng Mei} {and} \bibinfo{person}{Yu~Hong Zhao}.} \bibinfo{year}{2024}\natexlab{}.
\newblock \showarticletitle{Dynamic network link prediction with node representation learning from graph convolutional networks}.
\newblock \bibinfo{journal}{\emph{Scientific Reports}} \bibinfo{volume}{14}, \bibinfo{number}{1} (\bibinfo{year}{2024}), \bibinfo{pages}{538}.
\newblock


\bibitem[Morris et~al\mbox{.}(2020)]%
        {morris2020tudataset}
\bibfield{author}{\bibinfo{person}{Christopher Morris}, \bibinfo{person}{Nils~M Kriege}, \bibinfo{person}{Franka Bause}, \bibinfo{person}{Kristian Kersting}, \bibinfo{person}{Petra Mutzel}, {and} \bibinfo{person}{Marion Neumann}.} \bibinfo{year}{2020}\natexlab{}.
\newblock \showarticletitle{Tudataset: A collection of benchmark datasets for learning with graphs}.
\newblock \bibinfo{journal}{\emph{arXiv preprint arXiv:2007.08663}} (\bibinfo{year}{2020}).
\newblock


\bibitem[Myers et~al\mbox{.}(2014)]%
        {myers2014information}
\bibfield{author}{\bibinfo{person}{Seth~A Myers}, \bibinfo{person}{Aneesh Sharma}, \bibinfo{person}{Pankaj Gupta}, {and} \bibinfo{person}{Jimmy Lin}.} \bibinfo{year}{2014}\natexlab{}.
\newblock \showarticletitle{Information network or social network? The structure of the Twitter follow graph}. In \bibinfo{booktitle}{\emph{Proceedings of the 23rd international conference on world wide web}}. \bibinfo{pages}{493--498}.
\newblock


\bibitem[Newman et~al\mbox{.}(2002)]%
        {newman2002random}
\bibfield{author}{\bibinfo{person}{Mark~EJ Newman}, \bibinfo{person}{Duncan~J Watts}, {and} \bibinfo{person}{Steven~H Strogatz}.} \bibinfo{year}{2002}\natexlab{}.
\newblock \showarticletitle{Random graph models of social networks}.
\newblock \bibinfo{journal}{\emph{Proceedings of the national academy of sciences}} \bibinfo{volume}{99}, \bibinfo{number}{suppl\_1} (\bibinfo{year}{2002}), \bibinfo{pages}{2566--2572}.
\newblock


\bibitem[Salamat et~al\mbox{.}(2021)]%
        {salamat2021heterographrec}
\bibfield{author}{\bibinfo{person}{Amirreza Salamat}, \bibinfo{person}{Xiao Luo}, {and} \bibinfo{person}{Ali Jafari}.} \bibinfo{year}{2021}\natexlab{}.
\newblock \showarticletitle{HeteroGraphRec: A heterogeneous graph-based neural networks for social recommendations}.
\newblock \bibinfo{journal}{\emph{Knowledge-Based Systems}}  \bibinfo{volume}{217} (\bibinfo{year}{2021}), \bibinfo{pages}{106817}.
\newblock


\bibitem[Schlichtkrull et~al\mbox{.}(2018)]%
        {schlichtkrull2018modeling}
\bibfield{author}{\bibinfo{person}{Michael Schlichtkrull}, \bibinfo{person}{Thomas~N Kipf}, \bibinfo{person}{Peter Bloem}, \bibinfo{person}{Rianne Van Den~Berg}, \bibinfo{person}{Ivan Titov}, {and} \bibinfo{person}{Max Welling}.} \bibinfo{year}{2018}\natexlab{}.
\newblock \showarticletitle{Modeling relational data with graph convolutional networks}. In \bibinfo{booktitle}{\emph{The semantic web: 15th international conference, ESWC 2018, Heraklion, Crete, Greece, June 3--7, 2018, proceedings 15}}. Springer, \bibinfo{pages}{593--607}.
\newblock


\bibitem[Sozol et~al\mbox{.}(2024)]%
        {sozol2024anomaly}
\bibfield{author}{\bibinfo{person}{Md~Shariar Sozol}, \bibinfo{person}{Golam~Mostafa Saki}, {and} \bibinfo{person}{Md~Mostafizur Rahman}.} \bibinfo{year}{2024}\natexlab{}.
\newblock \showarticletitle{Anomaly Detection in Cybersecurity with Graph-Based Approaches}.
\newblock \bibinfo{journal}{\emph{International Journal of Scientific Research in Engineering and Management (IJSREM)}} \bibinfo{volume}{8}, \bibinfo{number}{8} (\bibinfo{year}{2024}), \bibinfo{pages}{1--7}.
\newblock


\bibitem[Tang and Liu(2010)]%
        {tang2010graph}
\bibfield{author}{\bibinfo{person}{Lei Tang} {and} \bibinfo{person}{Huan Liu}.} \bibinfo{year}{2010}\natexlab{}.
\newblock \showarticletitle{Graph mining applications to social network analysis}.
\newblock \bibinfo{journal}{\emph{Managing and mining graph data}} (\bibinfo{year}{2010}), \bibinfo{pages}{487--513}.
\newblock


\bibitem[Trinajstic(2018)]%
        {trinajstic2018chemical}
\bibfield{author}{\bibinfo{person}{Nenad Trinajstic}.} \bibinfo{year}{2018}\natexlab{}.
\newblock \bibinfo{booktitle}{\emph{Chemical graph theory}}.
\newblock \bibinfo{publisher}{CRC press}.
\newblock


\bibitem[Veli{\v{c}}kovi{\'c} et~al\mbox{.}(2017)]%
        {velivckovic2017graph}
\bibfield{author}{\bibinfo{person}{Petar Veli{\v{c}}kovi{\'c}}, \bibinfo{person}{Guillem Cucurull}, \bibinfo{person}{Arantxa Casanova}, \bibinfo{person}{Adriana Romero}, \bibinfo{person}{Pietro Lio}, {and} \bibinfo{person}{Yoshua Bengio}.} \bibinfo{year}{2017}\natexlab{}.
\newblock \showarticletitle{Graph attention networks}.
\newblock \bibinfo{journal}{\emph{arXiv preprint arXiv:1710.10903}} (\bibinfo{year}{2017}).
\newblock


\bibitem[Wang et~al\mbox{.}(2021b)]%
        {wang2021review}
\bibfield{author}{\bibinfo{person}{Jianian Wang}, \bibinfo{person}{Sheng Zhang}, \bibinfo{person}{Yanghua Xiao}, {and} \bibinfo{person}{Rui Song}.} \bibinfo{year}{2021}\natexlab{b}.
\newblock \showarticletitle{A review on graph neural network methods in financial applications}.
\newblock \bibinfo{journal}{\emph{arXiv preprint arXiv:2111.15367}} (\bibinfo{year}{2021}).
\newblock


\bibitem[Wang et~al\mbox{.}(2021a)]%
        {wang2021graph}
\bibfield{author}{\bibinfo{person}{Shoujin Wang}, \bibinfo{person}{Liang Hu}, \bibinfo{person}{Yan Wang}, \bibinfo{person}{Xiangnan He}, \bibinfo{person}{Quan~Z Sheng}, \bibinfo{person}{Mehmet~A Orgun}, \bibinfo{person}{Longbing Cao}, \bibinfo{person}{Francesco Ricci}, {and} \bibinfo{person}{Philip~S Yu}.} \bibinfo{year}{2021}\natexlab{a}.
\newblock \showarticletitle{Graph learning based recommender systems: A review}.
\newblock \bibinfo{journal}{\emph{arXiv preprint arXiv:2105.06339}} (\bibinfo{year}{2021}).
\newblock


\bibitem[Wang et~al\mbox{.}(2019)]%
        {wang2019heterogeneous}
\bibfield{author}{\bibinfo{person}{Xiao Wang}, \bibinfo{person}{Houye Ji}, \bibinfo{person}{Chuan Shi}, \bibinfo{person}{Bai Wang}, \bibinfo{person}{Yanfang Ye}, \bibinfo{person}{Peng Cui}, {and} \bibinfo{person}{Philip~S Yu}.} \bibinfo{year}{2019}\natexlab{}.
\newblock \showarticletitle{Heterogeneous graph attention network}. In \bibinfo{booktitle}{\emph{The world wide web conference}}. \bibinfo{pages}{2022--2032}.
\newblock


\bibitem[Wu et~al\mbox{.}(2019)]%
        {wu2019net}
\bibfield{author}{\bibinfo{person}{Jun Wu}, \bibinfo{person}{Jingrui He}, {and} \bibinfo{person}{Jiejun Xu}.} \bibinfo{year}{2019}\natexlab{}.
\newblock \showarticletitle{Net: Degree-specific graph neural networks for node and graph classification}. In \bibinfo{booktitle}{\emph{Proceedings of the 25th ACM SIGKDD international conference on knowledge discovery \& data mining}}. \bibinfo{pages}{406--415}.
\newblock


\bibitem[Wu et~al\mbox{.}(2022)]%
        {wu2022graph}
\bibfield{author}{\bibinfo{person}{Shiwen Wu}, \bibinfo{person}{Fei Sun}, \bibinfo{person}{Wentao Zhang}, \bibinfo{person}{Xu Xie}, {and} \bibinfo{person}{Bin Cui}.} \bibinfo{year}{2022}\natexlab{}.
\newblock \showarticletitle{Graph neural networks in recommender systems: a survey}.
\newblock \bibinfo{journal}{\emph{Comput. Surveys}} \bibinfo{volume}{55}, \bibinfo{number}{5} (\bibinfo{year}{2022}), \bibinfo{pages}{1--37}.
\newblock


\bibitem[Xi et~al\mbox{.}(2021)]%
        {xi2021graph}
\bibfield{author}{\bibinfo{person}{Zhaohan Xi}, \bibinfo{person}{Ren Pang}, \bibinfo{person}{Shouling Ji}, {and} \bibinfo{person}{Ting Wang}.} \bibinfo{year}{2021}\natexlab{}.
\newblock \showarticletitle{Graph backdoor}. In \bibinfo{booktitle}{\emph{30th USENIX security symposium (USENIX Security 21)}}. \bibinfo{pages}{1523--1540}.
\newblock


\bibitem[Xiao et~al\mbox{.}(2022)]%
        {xiao2022graph}
\bibfield{author}{\bibinfo{person}{Shunxin Xiao}, \bibinfo{person}{Shiping Wang}, \bibinfo{person}{Yuanfei Dai}, {and} \bibinfo{person}{Wenzhong Guo}.} \bibinfo{year}{2022}\natexlab{}.
\newblock \showarticletitle{Graph neural networks in node classification: survey and evaluation}.
\newblock \bibinfo{journal}{\emph{Machine Vision and Applications}} \bibinfo{volume}{33}, \bibinfo{number}{1} (\bibinfo{year}{2022}), \bibinfo{pages}{4}.
\newblock


\bibitem[Yan et~al\mbox{.}(2023)]%
        {yan2023graph}
\bibfield{author}{\bibinfo{person}{Bo Yan}, \bibinfo{person}{Cheng Yang}, \bibinfo{person}{Chuan Shi}, \bibinfo{person}{Yong Fang}, \bibinfo{person}{Qi Li}, \bibinfo{person}{Yanfang Ye}, {and} \bibinfo{person}{Junping Du}.} \bibinfo{year}{2023}\natexlab{}.
\newblock \showarticletitle{Graph mining for cybersecurity: A survey}.
\newblock \bibinfo{journal}{\emph{ACM Transactions on Knowledge Discovery from Data}} \bibinfo{volume}{18}, \bibinfo{number}{2} (\bibinfo{year}{2023}), \bibinfo{pages}{1--52}.
\newblock


\bibitem[Yang et~al\mbox{.}(2022)]%
        {yang2022transferable}
\bibfield{author}{\bibinfo{person}{Shuiqiao Yang}, \bibinfo{person}{Bao~Gia Doan}, \bibinfo{person}{Paul Montague}, \bibinfo{person}{Olivier De~Vel}, \bibinfo{person}{Tamas Abraham}, \bibinfo{person}{Seyit Camtepe}, \bibinfo{person}{Damith~C Ranasinghe}, {and} \bibinfo{person}{Salil~S Kanhere}.} \bibinfo{year}{2022}\natexlab{}.
\newblock \showarticletitle{Transferable graph backdoor attack}. In \bibinfo{booktitle}{\emph{Proceedings of the 25th international symposium on research in attacks, intrusions and defenses}}. \bibinfo{pages}{321--332}.
\newblock


\bibitem[Yang et~al\mbox{.}(2016)]%
        {yang2016revisiting}
\bibfield{author}{\bibinfo{person}{Zhilin Yang}, \bibinfo{person}{William Cohen}, {and} \bibinfo{person}{Ruslan Salakhudinov}.} \bibinfo{year}{2016}\natexlab{}.
\newblock \showarticletitle{Revisiting semi-supervised learning with graph embeddings}. In \bibinfo{booktitle}{\emph{International conference on machine learning}}. PMLR, \bibinfo{pages}{40--48}.
\newblock


\bibitem[Yao et~al\mbox{.}(2025)]%
        {yao2025mecon}
\bibfield{author}{\bibinfo{person}{Zihao Yao}, \bibinfo{person}{Fanding Huang}, \bibinfo{person}{Yannan Li}, \bibinfo{person}{Wei Duan}, \bibinfo{person}{Peng Qian}, \bibinfo{person}{Nan Yang}, {and} \bibinfo{person}{Willy Susilo}.} \bibinfo{year}{2025}\natexlab{}.
\newblock \showarticletitle{Mecon: A GNN-based graph classification framework for MEV activity detection}.
\newblock \bibinfo{journal}{\emph{Expert Systems with Applications}}  \bibinfo{volume}{269} (\bibinfo{year}{2025}), \bibinfo{pages}{126486}.
\newblock


\bibitem[Yuan et~al\mbox{.}(2024)]%
        {yuan2024sage}
\bibfield{author}{\bibinfo{person}{Dingqiang Yuan}, \bibinfo{person}{Xiaohua Xu}, \bibinfo{person}{Lei Yu}, \bibinfo{person}{Tongchang Han}, \bibinfo{person}{Rongchang Li}, {and} \bibinfo{person}{Meng Han}.} \bibinfo{year}{2024}\natexlab{}.
\newblock \showarticletitle{E-SAGE: Explainability-Based Defense Against Backdoor Attacks on Graph Neural Networks}. In \bibinfo{booktitle}{\emph{International Conference on Wireless Artificial Intelligent Computing Systems and Applications}}. Springer, \bibinfo{pages}{402--414}.
\newblock


\bibitem[Zeng et~al\mbox{.}(2019)]%
        {zeng2019graphsaint}
\bibfield{author}{\bibinfo{person}{Hanqing Zeng}, \bibinfo{person}{Hongkuan Zhou}, \bibinfo{person}{Ajitesh Srivastava}, \bibinfo{person}{Rajgopal Kannan}, {and} \bibinfo{person}{Viktor Prasanna}.} \bibinfo{year}{2019}\natexlab{}.
\newblock \showarticletitle{Graphsaint: Graph sampling based inductive learning method}.
\newblock \bibinfo{journal}{\emph{arXiv preprint arXiv:1907.04931}} (\bibinfo{year}{2019}).
\newblock


\bibitem[Zhang et~al\mbox{.}(2019)]%
        {zhang2019heterogeneous}
\bibfield{author}{\bibinfo{person}{Chuxu Zhang}, \bibinfo{person}{Dongjin Song}, \bibinfo{person}{Chao Huang}, \bibinfo{person}{Ananthram Swami}, {and} \bibinfo{person}{Nitesh~V Chawla}.} \bibinfo{year}{2019}\natexlab{}.
\newblock \showarticletitle{Heterogeneous graph neural network}. In \bibinfo{booktitle}{\emph{Proceedings of the 25th ACM SIGKDD international conference on knowledge discovery \& data mining}}. \bibinfo{pages}{793--803}.
\newblock


\bibitem[Zhang et~al\mbox{.}(2023)]%
        {zhang2023graph}
\bibfield{author}{\bibinfo{person}{Hangfan Zhang}, \bibinfo{person}{Jinghui Chen}, \bibinfo{person}{Lu Lin}, \bibinfo{person}{Jinyuan Jia}, {and} \bibinfo{person}{Dinghao Wu}.} \bibinfo{year}{2023}\natexlab{}.
\newblock \showarticletitle{Graph contrastive backdoor attacks}. In \bibinfo{booktitle}{\emph{International Conference on Machine Learning}}. PMLR, \bibinfo{pages}{40888--40910}.
\newblock


\bibitem[Zhang and Chen(2018)]%
        {zhang2018link}
\bibfield{author}{\bibinfo{person}{Muhan Zhang} {and} \bibinfo{person}{Yixin Chen}.} \bibinfo{year}{2018}\natexlab{}.
\newblock \showarticletitle{Link prediction based on graph neural networks}.
\newblock \bibinfo{journal}{\emph{Advances in neural information processing systems}}  \bibinfo{volume}{31} (\bibinfo{year}{2018}).
\newblock


\bibitem[Zhang et~al\mbox{.}(2022)]%
        {zhang2022robust}
\bibfield{author}{\bibinfo{person}{Mengmei Zhang}, \bibinfo{person}{Xiao Wang}, \bibinfo{person}{Meiqi Zhu}, \bibinfo{person}{Chuan Shi}, \bibinfo{person}{Zhiqiang Zhang}, {and} \bibinfo{person}{Jun Zhou}.} \bibinfo{year}{2022}\natexlab{}.
\newblock \showarticletitle{Robust heterogeneous graph neural networks against adversarial attacks}. In \bibinfo{booktitle}{\emph{Proceedings of the AAAI conference on artificial intelligence}}, Vol.~\bibinfo{volume}{36}. \bibinfo{pages}{4363--4370}.
\newblock


\bibitem[Zhang and Zitnik(2020)]%
        {zhang2020gnnguard}
\bibfield{author}{\bibinfo{person}{Xiang Zhang} {and} \bibinfo{person}{Marinka Zitnik}.} \bibinfo{year}{2020}\natexlab{}.
\newblock \showarticletitle{Gnnguard: Defending graph neural networks against adversarial attacks}.
\newblock \bibinfo{journal}{\emph{Advances in neural information processing systems}}  \bibinfo{volume}{33} (\bibinfo{year}{2020}), \bibinfo{pages}{9263--9275}.
\newblock


\bibitem[Zhang et~al\mbox{.}(2021b)]%
        {zhang2021graph}
\bibfield{author}{\bibinfo{person}{Xiao-Meng Zhang}, \bibinfo{person}{Li Liang}, \bibinfo{person}{Lin Liu}, {and} \bibinfo{person}{Ming-Jing Tang}.} \bibinfo{year}{2021}\natexlab{b}.
\newblock \showarticletitle{Graph neural networks and their current applications in bioinformatics}.
\newblock \bibinfo{journal}{\emph{Frontiers in genetics}}  \bibinfo{volume}{12} (\bibinfo{year}{2021}), \bibinfo{pages}{690049}.
\newblock


\bibitem[Zhang et~al\mbox{.}(2021a)]%
        {zhang2021backdoor}
\bibfield{author}{\bibinfo{person}{Zaixi Zhang}, \bibinfo{person}{Jinyuan Jia}, \bibinfo{person}{Binghui Wang}, {and} \bibinfo{person}{Neil~Zhenqiang Gong}.} \bibinfo{year}{2021}\natexlab{a}.
\newblock \showarticletitle{Backdoor attacks to graph neural networks}. In \bibinfo{booktitle}{\emph{Proceedings of the 26th ACM Symposium on Access Control Models and Technologies}}. \bibinfo{pages}{15--26}.
\newblock


\bibitem[Zhang et~al\mbox{.}(2024)]%
        {zhang2024rethinking}
\bibfield{author}{\bibinfo{person}{Zhiwei Zhang}, \bibinfo{person}{Minhua Lin}, \bibinfo{person}{Enyan Dai}, {and} \bibinfo{person}{Suhang Wang}.} \bibinfo{year}{2024}\natexlab{}.
\newblock \showarticletitle{Rethinking graph backdoor attacks: A distribution-preserving perspective}. In \bibinfo{booktitle}{\emph{Proceedings of the 30th ACM SIGKDD Conference on Knowledge Discovery and Data Mining}}. \bibinfo{pages}{4386--4397}.
\newblock


\bibitem[Zhao et~al\mbox{.}(2021)]%
        {zhao2021graphsmote}
\bibfield{author}{\bibinfo{person}{Tianxiang Zhao}, \bibinfo{person}{Xiang Zhang}, {and} \bibinfo{person}{Suhang Wang}.} \bibinfo{year}{2021}\natexlab{}.
\newblock \showarticletitle{Graphsmote: Imbalanced node classification on graphs with graph neural networks}. In \bibinfo{booktitle}{\emph{Proceedings of the 14th ACM international conference on web search and data mining}}. \bibinfo{pages}{833--841}.
\newblock


\bibitem[Zhao et~al\mbox{.}(2024)]%
        {zhao2024disambiguated}
\bibfield{author}{\bibinfo{person}{Tianxiang Zhao}, \bibinfo{person}{Xiang Zhang}, {and} \bibinfo{person}{Suhang Wang}.} \bibinfo{year}{2024}\natexlab{}.
\newblock \showarticletitle{Disambiguated node classification with graph neural networks}. In \bibinfo{booktitle}{\emph{Proceedings of the ACM Web Conference 2024}}. \bibinfo{pages}{914--923}.
\newblock


\bibitem[Zhu et~al\mbox{.}(2019)]%
        {zhu2019robust}
\bibfield{author}{\bibinfo{person}{Dingyuan Zhu}, \bibinfo{person}{Ziwei Zhang}, \bibinfo{person}{Peng Cui}, {and} \bibinfo{person}{Wenwu Zhu}.} \bibinfo{year}{2019}\natexlab{}.
\newblock \showarticletitle{Robust graph convolutional networks against adversarial attacks}. In \bibinfo{booktitle}{\emph{Proceedings of the 25th ACM SIGKDD international conference on knowledge discovery \& data mining}}. \bibinfo{pages}{1399--1407}.
\newblock


\end{thebibliography}
\appendix

\section{Details of Datasets}
\label{app:datasets}

\subsection{Heterogeneous Graph Datasets}
\label{app:hetdatasets}
In our work, we evaluate the performance of HGBA and baseline attacks on heterogeneous graphs based on ACM, DBLP, and IMDB Datasets. These datasets are the most commonly used real-world heterogeneous graph datasets in the HGNN research field for semi-supervised node classification (SSNC) tasks. Detailed statistics for these datasets are summarized in Table~\ref{tab:datasets}.

\textit{(i)}~ACM~\cite{lv2021we} – a citation network extracted from the ACM digital library{\footnote{\href{https://dl.acm.org/}{https://dl.acm.org/}}}, refined in subsequent work~\cite{lv2021we} to incorporate all paper citation and reference edges for enhanced relational complexity;

\textit{(ii)}~DBLP~\cite{fu2020magnn} – a bibliographic network of computer science literature sourced from DBLP{\footnote{\href{https://dblp.org/}{https://dblp.org/}}}, 

\textit{(iii)}~IMDB~\cite{fu2020magnn} – a movie network from the IMDB database, capturing movies, directors, and actors across multiple genres{\footnote{\href{https://www.imdb.com/}{https://www.imdb.com/}}}.

\begin{table*}[t]
\centering
\caption{Statistics of Heterogeneous Graph Datasets Used in Experiments}
\label{tab:datasets}
\begin{tabular}{c c c c c c c c c c c}
\toprule
\textbf{Dataset} & \textbf{Node Type} & \textbf{\# Nodes} & \textbf{Feat. Dim.} & \textbf{Edge Types (\# Edges)} & \textbf{Target} & \textbf{\#Classes} & \textbf{Train} & \textbf{Val} & \textbf{Test} & \textbf{Metapaths} \\
\midrule
\multirow{4}{*}{ACM} 
  & Paper (P) & 3,025 & 1,902 & \multirow{4}{*}{\centering \parbox{2cm}{\centering P-P (5,343) \\[0em] P-S (3,025) \\[0em] P-T (255,619) \\[0em] P-A (9,949)}} & \multirow{4}{*}{\centering Paper} & \multirow{4}{*}{\centering 3} & \multirow{4}{*}{\centering 605} & \multirow{4}{*}{\centering 303} & \multirow{4}{*}{\centering 2,117} & \multirow{4}{*}{\centering \parbox{1cm}{\centering PAP \\[0.5em] PSP \\[0.5em] PTP}} \\
  & Author (A) & 5,959 & 1,902 & & & & & & & \\
  & Subject (S) & 56 & 1,902 & & & & & & & \\
  & Term (T) & 1,902 & - & & & & & & & \\
\midrule
\multirow{4}{*}{DBLP} 
  & Author (A) & 4,057 & 334 & \multirow{4}{*}{\centering \parbox{2cm}{\centering A-P (19,645) \\[0.5em] P-T (85,810) \\[0.5em] P-C (14,328)}} & \multirow{4}{*}{\centering Author} & \multirow{4}{*}{\centering 4} & \multirow{4}{*}{\centering 811} & \multirow{4}{*}{\centering 406} & \multirow{4}{*}{\centering 2,840} & \multirow{4}{*}{\centering \parbox{1cm}{\centering APA \\[0.5em] APTPA \\[0.5em] APCPA}} \\
  & Paper (P) & 14,328 & 4,231 & & & & & & & \\
  & Term (T) & 7,723 & 50 & & & & & & & \\
  & Conference (C) & 20 & - & & & & & & & \\
\midrule
\multirow{3}{*}{IMDB} 
  & Movie (M) & 4,278 & 3,066 & \multirow{3}{*}{\centering \parbox{2cm}{\centering M-D (4,278) \\[0.5em] M-A (12,828)}} & \multirow{3}{*}{\centering Movie} & \multirow{3}{*}{\centering 5} & \multirow{3}{*}{\centering 856} & \multirow{3}{*}{\centering 428} & \multirow{3}{*}{\centering 2,994} & \multirow{3}{*}{\centering \parbox{1cm}{\centering MDM \\[0.5em] MAM}} \\
  & Director (D) & 2,081 & 3,066 & & & & & & & \\
  & Actor (A) & 5,257 & 3,066 & & & & & & & \\
\bottomrule
\end{tabular}
\end{table*}

\subsection{Homogeneous Graph Datasets}

We utilize five homogeneous graph datasets, including Cora, PubMed, and CiteSeer ~\cite{yang2016revisiting} for node classification, PROTEINS \cite{morris2020tudataset} for graph classification, to evaluate HGBA's attack-extensibility, and Flickr \cite{zeng2019graphsaint} as used in IO1 to explore the impact of the attack budget. Detailed statistics for these datasets are summarized in Table~\ref{tab:homo_datasets_nc} \& Table~\ref{tab:homo_datasets_gc}.

\textit{(i)} ~Cora, PubMed, and CiterSeer~\cite{yang2016revisiting} - These datasets are citation networks in which nodes represent papers and edges indicate citation links. For Cora and CiteSeer, nodes are characterized by binary word vectors that reflect the presence or absence of words from a fixed dictionary. PubMed, however, uses TF/IDF-weighted word vectors to describe each node. Across all three datasets, nodes are classified according to their research domains.

\textit{(ii)}~Flickr~\cite{zeng2019graphsaint} - In this network, each node represents a single image uploaded to Flickr. Edges connect pairs of images that share specific attributes, such as location, gallery, or comments. Node features are captured by a 500-dimensional bag-of-words model from NUS-wide. For labels, the images are grouped manually into 7 unique categories.

\textit{(iii)}~PROTEINS~\cite{morris2020tudataset} - A collection of proteins categorized as either enzymes or non-enzymes. Nodes denote amino acids, with edges linking pairs that are within 6 Angstroms of each other.

\begin{table}[htbp]
    \centering
    \caption{Statistics of Homogeneous Graph Datasets Used in Experiments on the Node Classification Task.}
    \label{tab:homo_datasets_nc}
    \begin{tabular}{lcccc}
        \toprule
        Datasets & \#Nodes & \#Edges & \#Feature & \#Classes \\
        \midrule
        Cora & 2,708 & 5,429 & 1,443 & 7 \\
        Pubmed & 19,717 & 44,338 & 500 & 3 \\
        CiteSeer & 3,327 & 4,732 & 3,703 & 6 \\
        Flickr & 89,250 & 899,756 & 500 & 7 \\
        \bottomrule
    \end{tabular}
\end{table}

\begin{table}[htbp]
    \centering
    \caption{Statistics of Homogeneous Graph Datasets Used in Experiments on the Graph Classification Task.}
    \label{tab:homo_datasets_gc}
    \small 
    \begin{tabular}{lccccc}
        \toprule
        Datasets & \#Graphs & \#Nodes & \#Avg Nodes & \#Avg Edges & \#Classes \\
        \midrule
        PROTEINS & 1,113 & 43,471 & 39.06 & 72.82 & 2 \\
        \bottomrule
    \end{tabular}
\end{table}

\section{Experimental Details}
\label{app:experimental_details}

\subsection{Dataset Splitting}

We conduct experiments on transductive semi-supervised node classification tasks over heterogeneous graphs, where node labels are predicted using the entire graph structure, the features of all nodes, and the labels observed only for training nodes. Following the common practice in most studies, for each dataset, we randomly partition nodes into 20\% for training, 10\% for validation, and 70\% for testing, as detailed in Table~\ref{tab:datasets}.

\subsection{Dataset Poisoning}

For the fixed-subgraph-based SBA-SAMPLE and SBA-GEN, we generate a homogeneous trigger structure of size 3 using the Erdős-Rényi (ER) model, which is proved to can produce more effective trigger structures in prior work. For SBA-SAMPLE, node features are randomly sampled from the training graph’s node feature matrix, while for SBA-GEN, they are drawn from a Gaussian distribution matching the mean and variance of the training graph’s node feature matrix. 

For the adaptive subgraph-based GTA, UGBA, and DPGBA, we implement these methods on subgraphs extracted via the most influential meta-path to obtain corresponding homogeneous triggers. 

For all resulting homogeneous triggers, we introduce intermediate nodes between trigger nodes along the most influential meta-path, transforming them into heterogeneous triggers connected to poisoned nodes, as illustrated in Figure~\ref{fig:io1}. Notably, in each dataset, we designate the class with the fewest instances as the adversary-specified target class\(y_t\) to mitigate the impact of imbalanced data distributions (class 0 for ACM and DBLP, class 1 for IMDB).

\subsection{Training and Evaluation}

In simulating the role of standard researchers and developers operating in real-world scenarios, we train a clean model on the clean heterogeneous graph dataset by optimizing the cross-entropy loss. We then select the model with the lowest validation loss via early stopping and calculate its Micro-F1 and Macro-F1 scores on the test nodes \(V_{\text{test}}\).

For clean models, we train the clean model on the clean heterogeneous graph dataset by optimizing the cross-entropy loss. Then, we select the model with the lowest validation loss via early stopping and calculate its Micro-F1 and Macro-F1 scores on the test nodes \(V_{test}\).

For backdoored models under black-box settings, to closely simulate a realistic black-box attack, we shift roles from attacker to regular user after poisoning the dataset. We follow the same steps as training a clean model to obtain the backdoored model. The only difference is that we additionally compute the ASR by poisoning the test set post-evaluation using the improved strategy from the takeaway of Section ~\ref{sec:io3} IO3. It is worth noting that, for HGBA, since we have two different backdoor activation strategies, we will obtain two sets of ASR values.

\subsection{Other Details} 

We repeat each experiment five times, reporting average metrics. Consistent settings apply across all attacks, with trigger size limited to 3 nodes; additional parameters are summarized in Table~\ref{tab:parameters}.

\begin{table*}[t]
\centering
\caption{Default Parameter Settings}
\label{tab:parameters}
\begin{tabular}{c c c c c} 
\toprule
\multicolumn{5}{c}{\textit{Model Architecture}} \\
\midrule
\textbf{Models} & \textbf{Architecture} & \textbf{Aggregator} & \textbf{Dropout} & \textbf{\#Attention Heads} \\
\midrule
GCN & 2 AL & Weighted Sum & 0.0 & -  \\
GAT & 2 AL & Attention - weighted Sum & 0.6 & 8  \\
GraphSAGE & 2 AL & Mean & 0.5 & -  \\
RGCN & 2 AL & Relation - weighted Sum & 0.0 & -  \\
HetGNN & 2 AL & Heterogeneous Mean & 0.0 & -  \\
HGT & 2 AL & Heterogeneous Multi - head Attention & 0.0 & 8  \\
HAN & 1 AL & Metapath - based Attention & 0.6 & 8  \\
\midrule
\multicolumn{5}{c}{\textit{Other Settings of Model Architecture}} \\
\midrule
\multicolumn{2}{c}{Hidden Channels} & \multicolumn{3}{c}{128} \\ 
\multicolumn{2}{c}{Classifier Architecture} & \multicolumn{3}{c}{FCN (1 FC + 1 SM)} \\ 
\midrule
\multicolumn{5}{c}{\textit{Model Training}} \\
\midrule
\multicolumn{2}{c}{Optimizer} & \multicolumn{3}{c}{Adam} \\ 
\multicolumn{2}{c}{Learning Rate} & \multicolumn{3}{c}{0.003} \\ 
\multicolumn{2}{c}{Weight Decay} & \multicolumn{3}{c}{0.0001} \\ 
\multicolumn{2}{c}{Epochs} & \multicolumn{3}{c}{200} \\ 
\multicolumn{2}{c}{Batch Size} & \multicolumn{3}{c}{Full graph} \\ 
\multicolumn{2}{c}{Patience (Early Stopping)} & \multicolumn{3}{c}{30} \\ 
\multicolumn{2}{c}{Delta (Early Stopping)} & \multicolumn{3}{c}{0.001} \\ 
\bottomrule
\multicolumn{5}{l}{\footnotesize FC: fully - connected layer, SM: softmax layer.} \\
\multicolumn{5}{l}{\footnotesize Delta: Minimum improvement threshold for validation loss. }
\end{tabular}
\end{table*}

\section{Detailed Experimental Results of IO1}
\label{app:io1}

Table~\ref{tab:gnn_results} shows the performance of homogeneous GNNs (GCN, GAT, GraphSAGE) in terms of accuracy on clean graphs across three datasets (Cora, Pubmed, Flickr). 

Table~\ref{tab:hgnn_performance} details the performance of HGNNs on clean graphs, reporting both micro-F1 and macro-F1 scores for six models (GCN, GAT, GraphSAGE, RGCN, HetGNN, HAN) across ACM, DBLP, and IMDB datasets. 

Table~\ref{tab:attack_performance_hogs} then focuses on the average performance of graph backdoor attacks on three homogeneous GNNs under different attack budgets (1\%, 3\%, 5\%, 10\%), presenting clean accuracy and attack success rate (ASR). 

Table~\ref{tab:attack_performance_hegs} for HGNNs further demonstrates the impact of these attacks on six HGNN models, providing clean micro-F1, clean macro-F1, and ASR metrics across the same range of attack budgets.

\begin{table}[htbp]
    \centering
    \caption{Performance of GNNs on Clean Graph. (Accuracy(\%))}
    \label{tab:gnn_results}
    \begin{tabular}{l c c c}
        \toprule
        Dataset & GCN & GAT & GraphSAGE \\
        \midrule
        Cora & 82.9 & 84.5 & 81.8 \\
        Pubmed & 85.1 & 83.9 & 85.7 \\
        Flickr & 45.5 & 46.5 & 47 \\
        \bottomrule
    \end{tabular}
\end{table}

\begin{table}[htbp]
    \centering
    \caption{Performance of HGNNs on Clean Graph.  (Clean Micro-F1 (\%) | Clean Macro-F1 (\%))}
    \label{tab:hgnn_performance}
    \begin{tabular}{c c c c c c c}
        \toprule
        \multirow{2}{*}[1.5ex]{\centering Models} & \multicolumn{2}{c}{\centering ACM} & \multicolumn{2}{c}{\centering DBLP} & \multicolumn{2}{c}{\centering IMDB} \\
        \midrule
        GCN & $87.57$ | $87.69$ &  & $92.09$ | $91.48$ &  & $50.32$ | $49.78$ &  \\
        GAT & $88.90$ | $89.01$ &  & $88.02$ | $86.09$ &  & $49.53$ | $48.79$ &  \\
        GraphSAGE & $90.79$ | $90.85$ &  & $92.48$ | $91.87$ &  & $58.35$ | $58.29$ &  \\
        RGCN & $91.89$ | $91.96$ &  & $92.54$ | $91.96$ &  & $62.68$ | $62.62$ &  \\
        HetGNN & $92.13$ | $92.19$ &  & $91.45$ | $90.86$ &  & $60.83$ | $60.57$ &  \\
        HAN & $90.75$ | $90.82$ &  & $92.72$ | $92.14$ &  & $61.03$ | $60.89$ &  \\
        \bottomrule
    \end{tabular}
\end{table}

\begin{table}[htbp]
    \centering
    \caption{Average Performance of Graph Backdoor Attacks on Three GNNs across Three Datasets under Different Attack Budgets. (Clean Acc (\%) | ASR (\%))}
    \label{tab:attack_performance_hogs}
    \footnotesize
    \begin{tabular}{c c c c c}
        \toprule
        \multirow{2}{*}[1.2ex]{\centering Attacks} & \multicolumn{1}{c}{\centering 1\%} & \multicolumn{1}{c}{\centering 3\%} & \multicolumn{1}{c}{\centering 5\%} & \multicolumn{1}{c}{\centering 10\%} \\
        \midrule
        SBA-SAMPLE & 71.69 | 35.20 & 70.91 | 81.58 & 70.53 | 88.74 & 69.73 | 93.80 \\
        SBA-GEN & 71.44 | 31.47 & 70.72 | 85.42 & 70.48 | 90.45 & 69.57 | 94.40 \\
        GTA & 71.19 | 83.06 & 70.65 | 78.26 & 69.57 | 77.97 & 65.88 | 81.64 \\
        UGBA & 70.55 | 85.03 & 70.37 | 94.29 & 70.16 | 96.16 & 69.85 | 95.29 \\
        DPGBA & 70.27 | 95.63 & 70.25 | 97.11 & 69.92 | 97.49 & 69.15 | 98.54 \\
        \bottomrule
    \end{tabular}
\end{table}

\begin{table}[htbp]
    \centering
    \caption{Average Performance of Graph Backdoor Attacks on Six HGNNs across Three Datasets under Different Attack Budgets. (Clean Micro-F1 (\%) | Clean Macro-F1 (\%) | ASR(\%))}
    \label{tab:attack_performance_hegs}
    \scriptsize
    \begin{tabular}{c c c c c}
        \toprule
        \multirow{2}{*}[1.2ex]{\centering Attacks} & \multicolumn{1}{c}{\centering 1\%} & \multicolumn{1}{c}{\centering 3\%} & \multicolumn{1}{c}{\centering 5\%} & \multicolumn{1}{c}{\centering 10\%} \\
        \midrule
        SBA-SAMPLE & 79.88 | 79.59 | 9.74 & 78.16 | 77.65 | 19.53 & 79.45 | 79.14 | 28.28 & 78.44 | 78.16 | 34.29 \\
        SBA-GEN & 79.91 | 79.63 | 63.42 & 80.72 | 80.51 | 72.30 & 79.48 | 79.11 | 76.95 & 77.69 | 76.83 | 81.15 \\
        GTA & 80.42 | 80.19 | 41.60 & 80.16 | 79.92 | 59.30 & 79.75 | 79.48 | 65.14 & 78.60 | 77.79 | 71.74 \\
        UGBA & 80.19 | 79.89 | 30.99 & 79.97 | 79.75 | 34.91 & 79.72 | 79.47 | 38.64 & 78.05 | 77.50 | 41.77 \\
        DPGBA & 73.65 | 71.71 | 73.08 & 73.26 | 71.87 | 75.80 & 72.06 | 69.88 | 74.74 & 69.29 | 66.35 | 76.84 \\
        \bottomrule
    \end{tabular}
\end{table}

\section{Detailed Experimental Results for IO2}
\label{app:io2}

\begin{figure}
    \centering
    \includegraphics[width=1\linewidth]{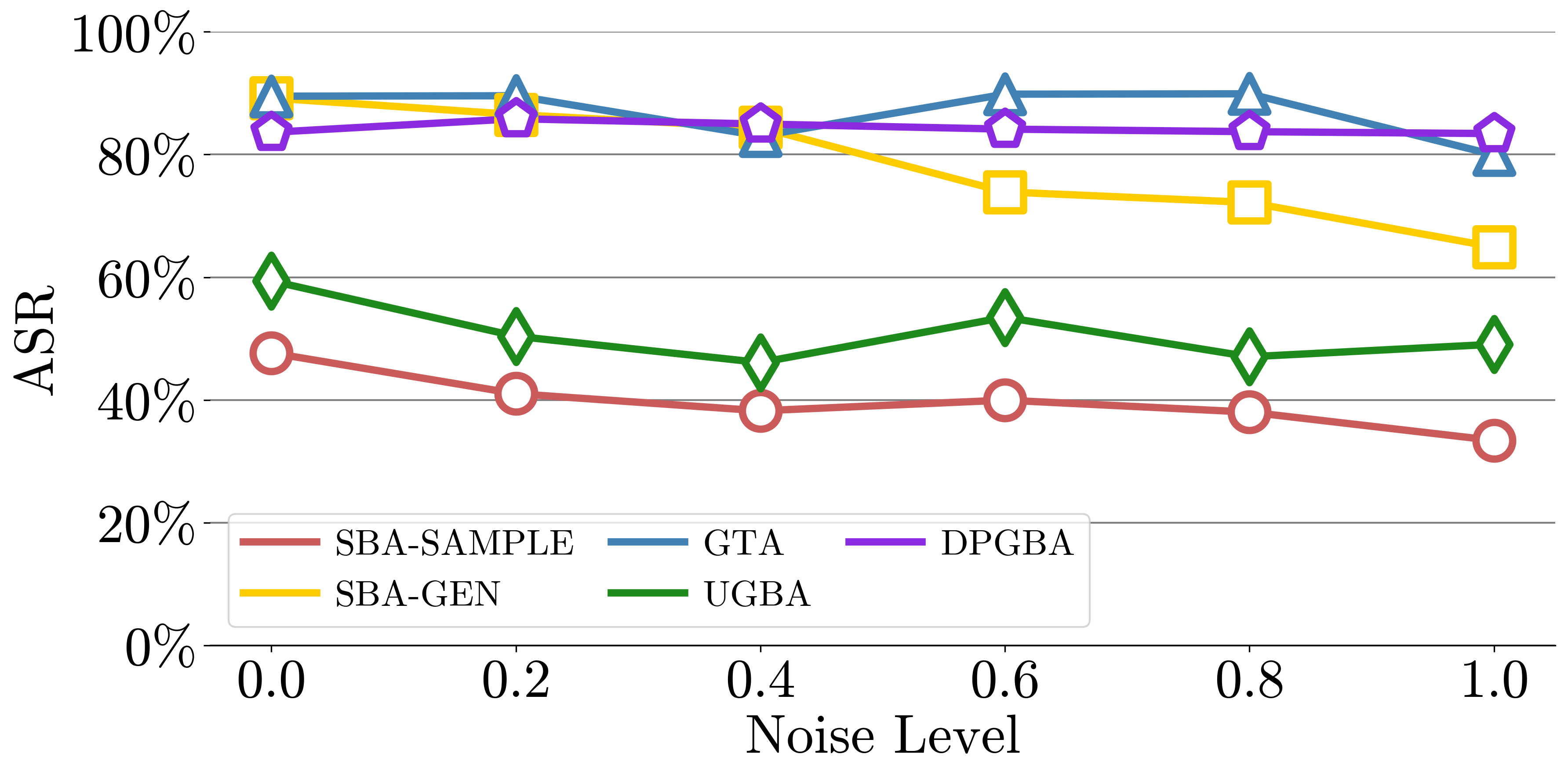}
    \caption{Impact of Node Perturbations on Backdoor Activation for Existing Graph Backdoor Attacks on HGNNs. Average results across six HGNNs on the ACM dataset.}
    \label{fig:acm_io2}
\end{figure}

\begin{figure}
    \centering
    \includegraphics[width=0.85\linewidth]{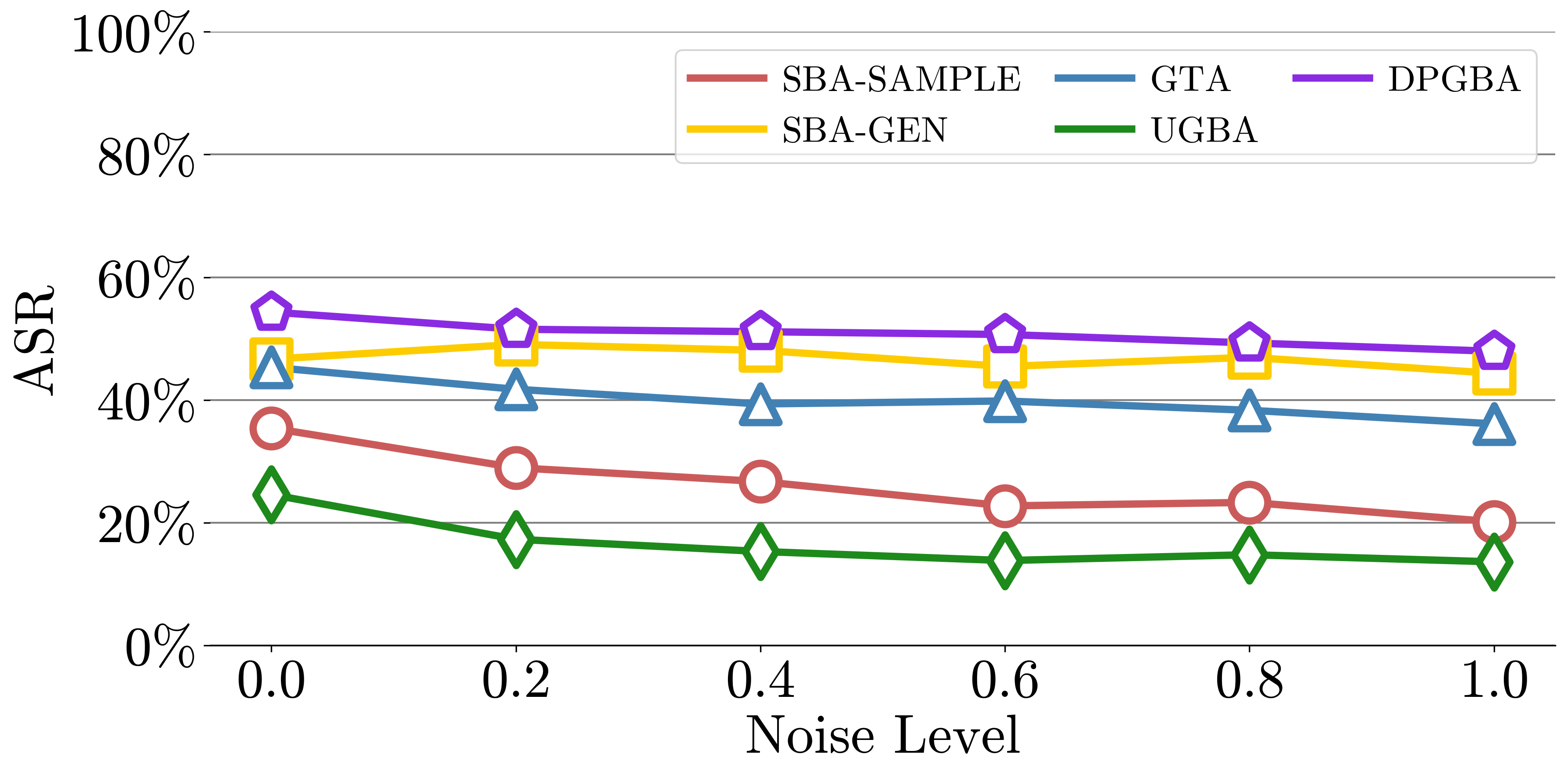}
    \caption{Impact of Node Perturbations on Backdoor Activation for Existing Graph Backdoor Attacks on HGNNs. Average results across six HGNNs on the DBLP dataset.}
    \label{fig:dblp_io2}
\end{figure}

\begin{figure}
    \centering
    \includegraphics[width=0.85\linewidth]{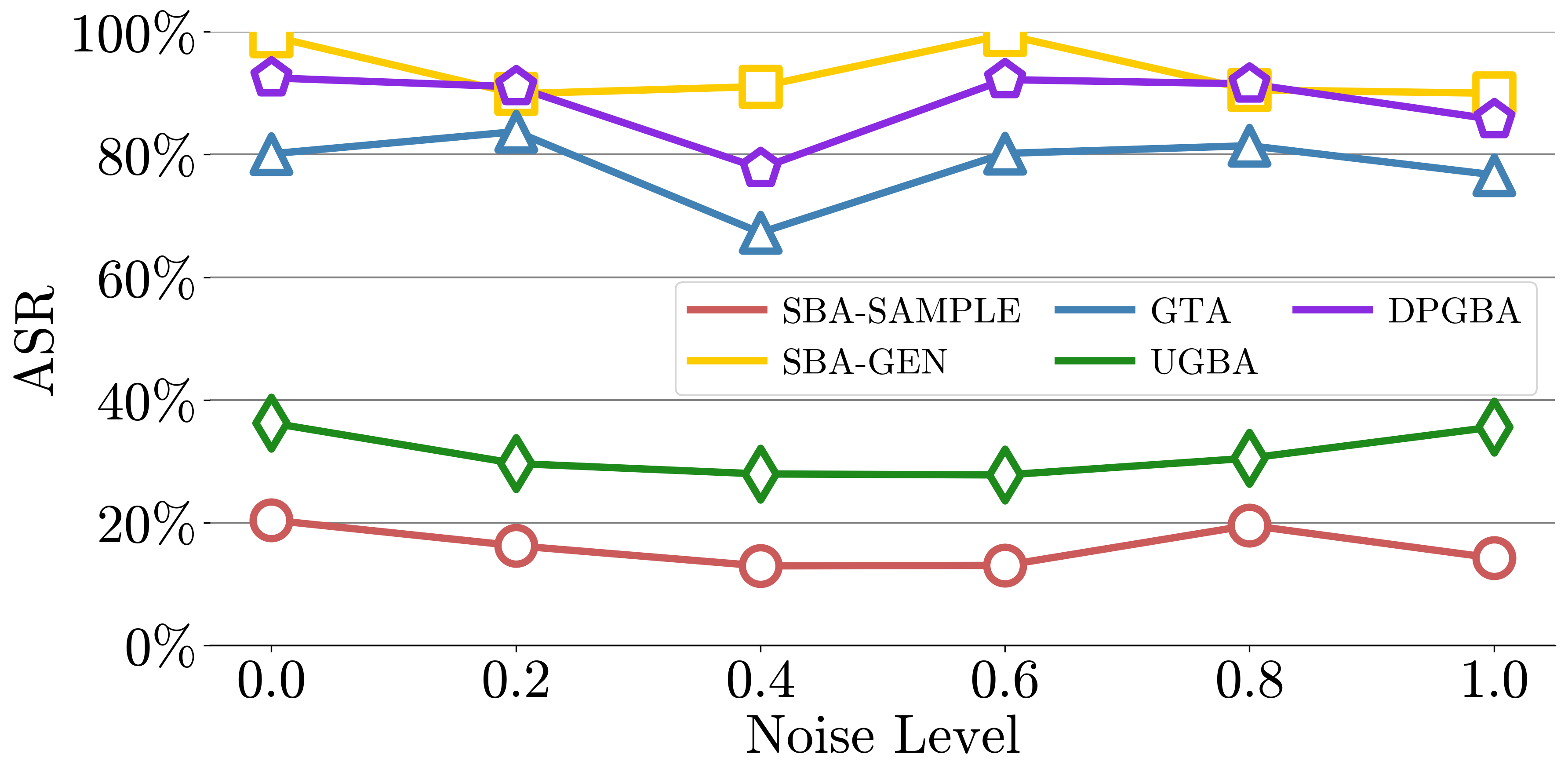}
    \caption{Impact of Node Perturbations on Backdoor Activation for Existing Graph Backdoor Attacks on HGNNs. Average results across six HGNNs on the IMDB dataset.}
    \label{fig:imdb_io2}
\end{figure}

Fig.~\ref{fig:acm_io2}, Fig.~\ref{fig:dblp_io2}, and Fig.~\ref{fig:imdb_io2} illustrate the average impact of node perturbations on backdoor activation for existing graph backdoor attacks on six HGNNs, with each figure corresponding to the ACM, DBLP, and IMDB datasets, respectively.

\section{Detailed Experimental Results for IO3}
\label{app:io3}
\begin{figure*}
    \centering
    \includegraphics[width=0.85\textwidth]{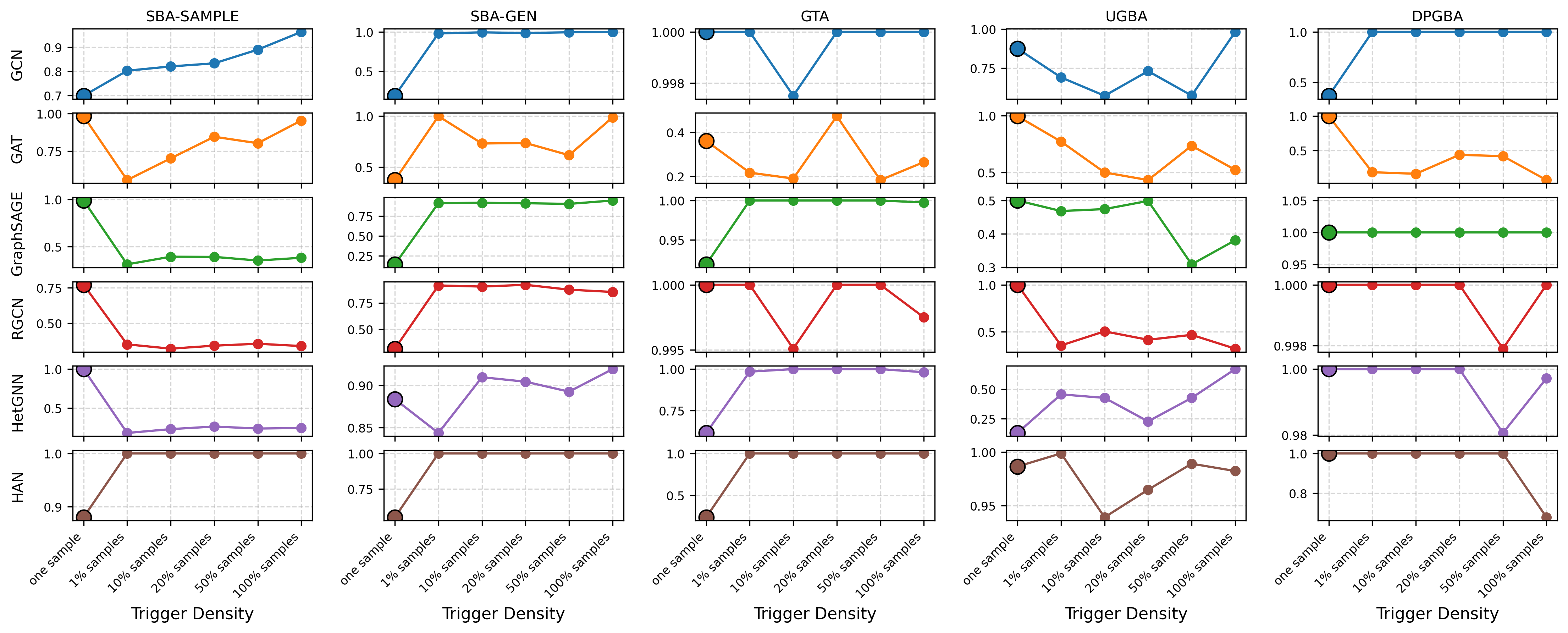}
    \caption{Impact of Trigger Density on Attack Success Rate (ASR) of Graph Backdoor Attacks in Heterogeneous Graphs (ACM Dataset).}
    \label{fig:unaccurate_asr_acm}
\end{figure*}

\begin{figure*}
    \centering
    \includegraphics[width=0.85\textwidth]{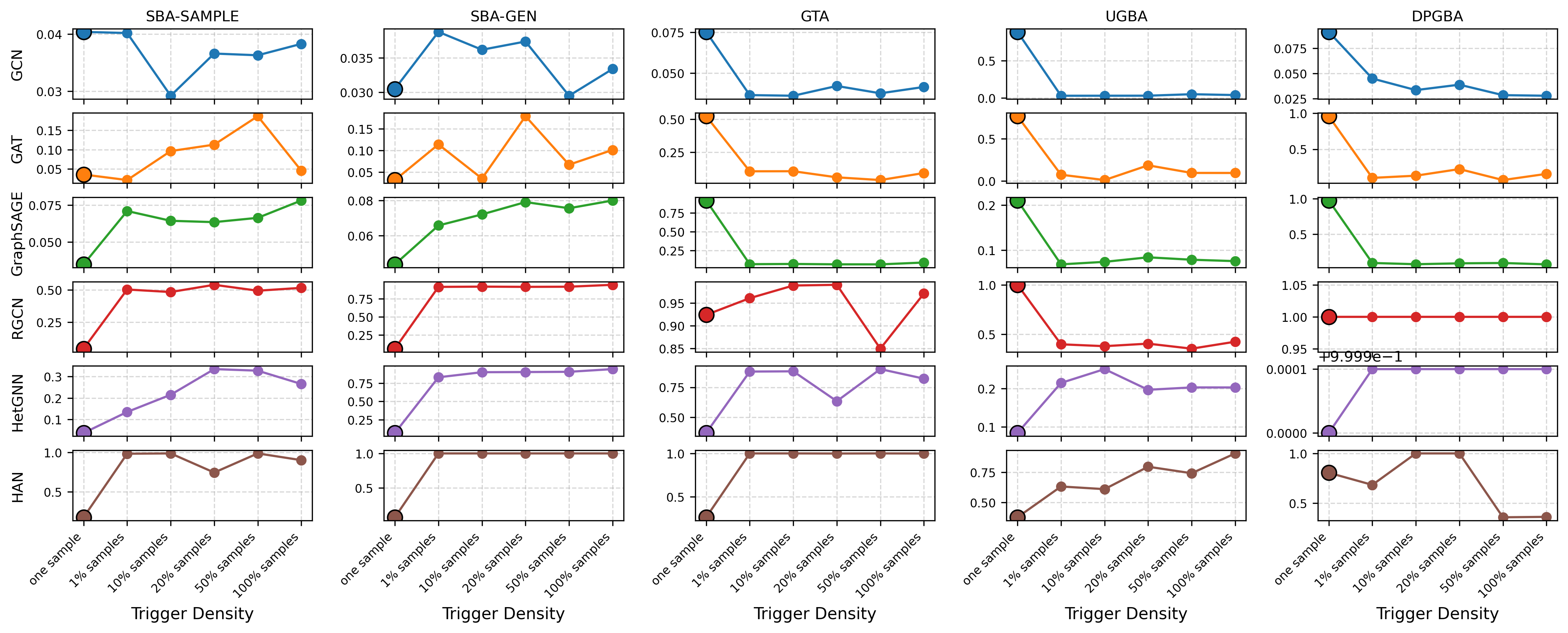}
    \caption{Impact of Trigger Density on Attack Success Rate (ASR) of Graph Backdoor Attacks in Heterogeneous Graphs (DBLP Dataset).}
    \label{fig:unaccurate_asr_dblp}
\end{figure*}

\begin{figure*}
    \centering
    \includegraphics[width=0.85\textwidth]{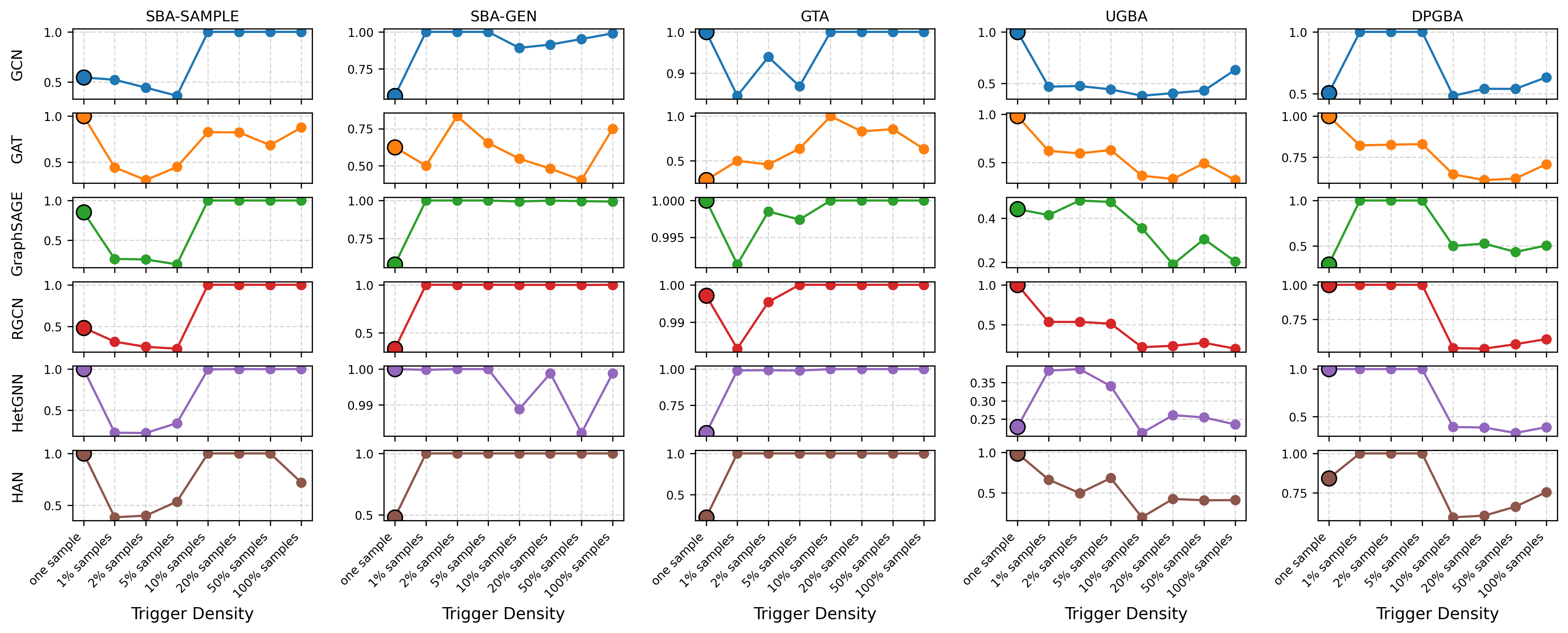}
    \caption{Impact of Trigger Density on Attack Success Rate (ASR) of Graph Backdoor Attacks in Heterogeneous Graphs (IMDB Dataset).}
    \label{fig:unaccurate_asr_imdb}
\end{figure*}

Fig.~\ref{fig:unaccurate_asr_acm}, Fig.~\ref{fig:unaccurate_asr_dblp}, and Fig.~\ref{fig:unaccurate_asr_imdb} show the impact of trigger density on the ASR of graph backdoor attacks in heterogeneous graphs, corresponding to the ACM, DBLP, and IMDB datasets, respectively.

\section{Algorithmic Details of HGBA and Backdoor Metapath Selection}
\label{app:hgba_overview}

Algorithm~\ref{algo:hgba} outlines the overall HGBA process and Algorithm~\ref{algo:pb_selection} focuses on backdoor metapath selection.

\begin{algorithm}
\caption{HGBA: Heterogeneous Graph Backdoor Attack}
\label{algo:hgba}
\begin{algorithmic}[1]
\REQUIRE Clean graph $G = (V, E, X, R, Y)$, target class $y_t$
\ENSURE  Poisoned graph $G_{poisoned}$, trigger node $v_t$, backdoor metapath $P_{b}$
\STATE $v_t \leftarrow \text{SelectTriggerNode}(G)$ \COMMENT{Based on Betweenness Centrality-based Trigger Node Selection Strategy}
\STATE $P_{b} \leftarrow \text{SelectBackdoorMetapath}(G)$ \COMMENT{Based on Algorithm \ref{algo:pb_selection}}
\STATE $V_{poisoned} \leftarrow \text{IdentifyPoisonedNodes}(G, v_t, P_{b})$
\FOR{each $v_{p_i} \in V_p$}
    \STATE $e_{p_i} \leftarrow \text{AddEdges}(v_{p_i}, v_t, P_{b})$
    \STATE $Y(v_{p_i}) \leftarrow y_t$
\ENDFOR
\STATE $E_{p} \leftarrow \bigcup_{i=1}^n e_{p_i}$
\STATE $G_{poisoned} \leftarrow (V, E \cup E_{p}, X, R, Y)$ 
\RETURN $G_{poisoned}, v_t, P_{b}$
\end{algorithmic}
\end{algorithm}

\begin{algorithm}
\caption{Backdoor Metapath Selection}
\label{algo:pb_selection}
\begin{algorithmic}[1]
\REQUIRE
\STATE Clean graph $G$
\STATE Set of candidate metapaths $\mathcal{P} = \{P_1, P_2, \ldots, P_n\}$
\STATE Proxy model $f$
\ENSURE Selected backdoor metapath $P_b$
\IF{$ \text{type} \mathcal(f) = \text{HomoGNN}$}
    \STATE $acc^* \leftarrow 0$ \COMMENT{Best accuracy}
    \STATE $P_b \leftarrow \emptyset$ \COMMENT{Backdoor metapath}
    
    \FORALL{$P_i \in \mathcal{P}$}
        \STATE $G_{P_i} \leftarrow\text{ExtractHomogeneousSubgraph}(G, P_i)$
        \STATE $f_{P_i} \leftarrow \text{TrainHomogeneousGNN}(G_{P_i})$
        \STATE $acc_{P_i} \leftarrow \text{EvaluateModel}(f_{P_i}, G_{P_i})$ 
        
        \IF{$acc_{P_i} > acc^*$}
            \STATE $acc^* \leftarrow acc_{P_i}$
            \STATE $P_b \leftarrow P_i$
        \ENDIF
    \ENDFOR
    
\ELSIF{$\text{type}\mathcal(f) = \text{HGNN}$}
    \STATE $f_H \leftarrow \text{TrainHGNN}(G, \mathcal{P})$ 
    \STATE $W \leftarrow \text{ExtractMetapathWeight}(f_H)$ 
    
    \STATE $w^* \leftarrow 0$
    \STATE $P_b \leftarrow \emptyset$
    
    \FORALL{$P_i \in \mathcal{P}$}
        \IF{$W(P_i) > w^*$}
            \STATE $w^* \leftarrow W(P_i)$
            \STATE $P_b \leftarrow P_i$
        \ENDIF
    \ENDFOR
\ENDIF
\RETURN $P_b$
\end{algorithmic}
\end{algorithm}

\section{DETAILED EXPERIMENTAL RESULTS}
\label{app:detailed_experimenal_results}

\subsection{Comprehensive Experimental Results for RQ1}
\label{app:rq1}

Table~\ref{tab:RQ1_acm}, Table~\ref{tab:RQ1_dblp}, and Table~\ref{tab:RQ1_imdb} report the results of backdoor attacks under black-box settings on ACM, DBLP, and IMDB datasets respectively, presenting clean micro-F1, clean macro-F1, and attack success rate (ASR) metrics for various graph neural network models and attack methods, with the best ASR results highlighted in bold.

\begin{table*}[htbp]
    \centering
    \caption{Results of Backdoor Attacks under Black-Box Attack Settings on ACM Dataset. (Clean Micro-F1 (\%) \,|\, Clean Macro-F1 (\%) \,|\, ASR (\%)). The best ASR results are marked in boldface. The red-highlighted data reflects the low stealthiness of DPGBA. Note that only one set of clean metrics is available as both HGBA\(_{\mathrm{1}}\) and HGBA\(_{\mathrm{2}}\) share the same test set for clean metrics evaluation, while separate ASR measurements are obtained for each activation method.}
    \label{tab:RQ1_acm}
    \resizebox{\textwidth}{!}{
    \scriptsize
    \renewcommand{\arraystretch}{1.5}
    \begin{tabular}{c c c c c c c c}
        \hline
        \scriptsize \textbf{Models} & \scriptsize \textbf{SBA-SAMPLE} & \scriptsize \textbf{SBA-GEN} & \scriptsize \textbf{GTA} & \scriptsize \textbf{UGBA} & \scriptsize \textbf{DPGBA} & \scriptsize \textbf{HGBA$_{\mathrm{1}}$} & \scriptsize \textbf{HGBA$_{\mathrm{2}}$} \\
        \hline
        \scriptsize GCN & \scriptsize 89.09\,|\,89.17\,|\,13.50 & \scriptsize 88.29\,|\,88.37\,|\,78.64 & \scriptsize 87.77\,|\,87.91\,|\,41.25 & \scriptsize 87.02\,|\,87.01\,|\,56.84 & \scriptsize 89.57\,|\,89.57\,|\,\textbf{100.00} & \scriptsize 86.20\,|\,86.23\,|\,98.77 & \scriptsize -\,|\,-\,|\,99.82 \\
        \scriptsize GAT & \scriptsize 87.30\,|\,87.36\,|\,14.88 & \scriptsize 88.43\,|\,88.45\,|\,32.94 & \scriptsize 88.95\,|\,89.06\,|\,12.11 & \scriptsize 88.15\,|\,88.28\,|\,24.67 & \scriptsize 88.15\,|\,88.25\,|\,36.51 & \scriptsize 88.52\,|\,88.56\,|\,88.05 & \scriptsize -\,|\,-\,|\,\textbf{97.06} \\
        \scriptsize GraphSAGE & \scriptsize 90.18\,|\,90.32\,|\,4.41 & \scriptsize 91.22\,|\,91.31\,|\,57.11 & \scriptsize 91.45\,|\,91.47\,|\,64.19 & \scriptsize 92.16\,|\,92.16\,|\,53.98 & \scriptsize 90.79\,|\,90.82\,|\,\textbf{100.00} & \scriptsize 89.30\,|\,89.27\,|\,92.55 & \scriptsize -\,|\,-\,|\,95.10 \\
        \scriptsize RGCN & \scriptsize 90.75\,|\,90.76\,|\,2.44 & \scriptsize 92.21\,|\,92.29\,|\,50.77 & \scriptsize 91.36\,|\,91.46\,|\,42.07 & \scriptsize 91.83\,|\,91.96\,|\,33.24 & \scriptsize \textcolor{red}{52.46}\,|\,\textcolor{red}{43.64}\,|\,\textbf{100.00} & \scriptsize 91.03\,|\,90.47\,|\,96.94 & \scriptsize -\,|\,-\,|\,90.53 \\
        \scriptsize HetGNN & \scriptsize 85.84\,|\,85.94\,|\,1.46 & \scriptsize 90.98\,|\,91.07\,|\,38.95 & \scriptsize 92.40\,|\,92.40\,|\,32.50 & \scriptsize 92.26\,|\,92.35\,|\,18.95 & \scriptsize \textcolor{red}{61.38}\,|\,\textcolor{red}{51.15}\,|\,\textbf{100.00} & \scriptsize 87.93\,|\,87.36\,|\,71.90 & \scriptsize -\,|\,-\,|\,75.87 \\
        \scriptsize HAN & \scriptsize 91.45\,|\,91.48\,|\,25.24 & \scriptsize 89.94\,|\,90.06\,|\,95.89 & \scriptsize 90.42\,|\,90.54\,|\,78.45 & \scriptsize 91.41\,|\,91.46\,|\,99.70 & \scriptsize \textcolor{red}{54.34}\,|\,\textcolor{red}{51.19}\,|\,62.62 & \scriptsize 91.26\,|\,90.71\,|\,\textbf{100.00} & \scriptsize -\,|\,-\,|\,\textbf{100.00} \\
        \hline
    \end{tabular}
    }
\end{table*}

\begin{table*}[htbp]
   \centering
   \caption{Results of Backdoor Attacks under Black-Box Attack Settings on DBLP Dataset. (Clean Micro-F1 (\%) \,|\, Clean Macro-F1 (\%) \,|\, ASR (\%)). The best results are marked in boldface. Note that only one set of clean metrics is available as both HGBA\(_{\mathrm{1}}\) and HGBA\(_{\mathrm{2}}\) share the same test set for clean metrics evaluation, while separate ASR measurements are obtained for each activation method.}
   \label{tab:RQ1_dblp}
   \resizebox{\textwidth}{!}{
   \scriptsize
   \renewcommand{\arraystretch}{1.5}
   \begin{tabular}{c c c c c c c c}
       \hline
       \scriptsize \textbf{Models} & \scriptsize \textbf{SBA-SAMPLE} & \scriptsize \textbf{SBA-GEN} & \scriptsize \textbf{GTA} & \scriptsize \textbf{UGBA} & \scriptsize \textbf{DPGBA} & \scriptsize \textbf{HGBA$_{\mathrm{1}}$} & \scriptsize \textbf{HGBA$_{\mathrm{2}}$} \\
       \hline
       \scriptsize GCN & \scriptsize 92.22\,|\,91.63\,|\,2.36 & \scriptsize 91.41\,|\,90.92\,|\,2.62 & \scriptsize 91.02\,|\,90.30\,|\,2.87 & \scriptsize 91.48\,|\,90.84\,|\,2.60 & \scriptsize 91.73\,|\,91.13\,|\,3.23 & \scriptsize 90.58\,|\,90.10\,|\,\textbf{100.00} & \scriptsize -\,|\,-\,|\,\textbf{100.00} \\
       \scriptsize GAT & \scriptsize 85.96\,|\,84.56\,|\,2.67 & \scriptsize 80.89\,|\,80.19\,|\,4.69 & \scriptsize 91.66\,|\,91.07\,|\,3.84 & \scriptsize 90.11\,|\,89.33\,|\,3.88 & \scriptsize 91.34\,|\,90.54\,|\,7.26 & \scriptsize 82.01\,|\,82.01\,|\,\textbf{78.05} & \scriptsize -\,|\,-\,|\,67.32 \\
       \scriptsize GraphSAGE & \scriptsize 91.62\,|\,91.07\,|\,3.74 & \scriptsize 91.62\,|\,91.00\,|\,3.87 & \scriptsize 91.38\,|\,90.84\,|\,3.61 & \scriptsize 91.80\,|\,91.17\,|\,3.88 & \scriptsize 91.83\,|\,91.17\,|\,4.02 & \scriptsize 86.92\,|\,86.45\,|\,72.48 & \scriptsize -\,|\,-\,|\,\textbf{73.83} \\
       \scriptsize RGCN & \scriptsize 91.94\,|\,91.44\,|\,13.28 & \scriptsize 92.22\,|\,91.72\,|\,72.37 & \scriptsize 91.97\,|\,91.23\,|\,54.31 & \scriptsize 92.33\,|\,91.81\,|\,12.10 & \scriptsize 92.57\,|\,91.96\,|\,\textbf{100.00} & \scriptsize 88.60\,|\,88.09\,|\,99.52 & \scriptsize -\,|\,-\,|\,99.63 \\
       \scriptsize HetGNN & \scriptsize 91.27\,|\,90.54\,|\,7.24 & \scriptsize 91.34\,|\,90.80\,|\,67.03 & \scriptsize 90.50\,|\,89.81\,|\,39.60 & \scriptsize 89.51\,|\,88.65\,|\,9.64 & \scriptsize 91.09\,|\,90.43\,|\,\textbf{100.00} & \scriptsize 83.29\,|\,82.97\,|\,90.00 & \scriptsize -\,|\,-\,|\,91.80 \\
       \scriptsize HAN & \scriptsize 92.50\,|\,91.93\,|\,8.55 & \scriptsize 92.89\,|\,92.27\,|\,97.75 & \scriptsize 92.33\,|\,91.71\,|\,78.20 & \scriptsize 91.24\,|\,90.39\,|\,8.82 & \scriptsize 80.36\,|\,72.65\,|\,75.74 & \scriptsize 91.67\,|\,91.13\,|\,\textbf{100.00} & \scriptsize -\,|\,-\,|\,\textbf{100.00} \\
       \hline
   \end{tabular}
   }
\end{table*}

\begin{table*}[htbp]
   \centering
   \caption{Results of Backdoor Attacks under Black-Box Attack Settings on IMDB Dataset. (Clean Micro-F1 (\%) \,|\, Clean Macro-F1 (\%) \,|\, ASR (\%)). The best results are marked in boldface. Note that only one set of clean metrics is available as both HGBA\(_{\mathrm{1}}\) and HGBA\(_{\mathrm{2}}\) share the same test set for clean metrics evaluation, while separate ASR measurements are obtained for each activation method.}
   \label{tab:RQ1_imdb}
   \resizebox{\textwidth}{!}{
   \scriptsize
   \renewcommand{\arraystretch}{1.5}
   \begin{tabular}{c c c c c c c c}
       \hline
       \scriptsize \textbf{Models} & \scriptsize \textbf{SBA-SAMPLE} & \scriptsize \textbf{SBA-GEN} & \scriptsize \textbf{GTA} & \scriptsize \textbf{UGBA} & \scriptsize \textbf{DPGBA} & \scriptsize \textbf{HGBA$_{\mathrm{1}}$} & \scriptsize \textbf{HGBA$_{\mathrm{2}}$} \\
       \hline
       \scriptsize GCN & \scriptsize 57.51\,|\,57.32\,|\,12.87 & \scriptsize 58.19\,|\,57.91\,|\,\textbf{100.00} & \scriptsize 56.63\,|\,56.41\,|\,51.56 & \scriptsize 57.11\,|\,56.87\,|\,30.13 & \scriptsize 59.06\,|\,58.65\,|\,\textbf{100.00} & \scriptsize 55.38\,|\,55.23\,|\,94.51 & \scriptsize -\,|\,-\,|\,97.78 \\
       \scriptsize GAT & \scriptsize 57.41\,|\,57.26\,|\,14.43 & \scriptsize 57.03\,|\,56.83\,|\,43.35 & \scriptsize 57.28\,|\,57.00\,|\,59.07 & \scriptsize 56.37\,|\,56.20\,|\,64.29 & \scriptsize 58.41\,|\,58.13\,|\,49.35 & \scriptsize 56.16\,|\,56.01\,|\,80.07 & \scriptsize -\,|\,-\,|\,\textbf{98.84} \\
       \scriptsize GraphSAGE & \scriptsize 59.49\,|\,59.51\,|\,13.12 & \scriptsize 59.17\,|\,59.22\,|\,99.88 & \scriptsize 59.83\,|\,59.83\,|\,51.97 & \scriptsize 59.70\,|\,59.76\,|\,32.54 & \scriptsize 59.75\,|\,59.74\,|\,\textbf{100.00} & \scriptsize 59.23\,|\,59.17\,|\,59.55 & \scriptsize -\,|\,-\,|\,75.97 \\
       \scriptsize RGCN & \scriptsize 63.36\,|\,63.16\,|\,11.24 & \scriptsize 63.11\,|\,62.93\,|\,99.79 & \scriptsize 62.14\,|\,62.04\,|\,43.14 & \scriptsize 62.84\,|\,62.72\,|\,31.77 & \scriptsize 58.97\,|\,58.86\,|\,\textbf{100.00} & \scriptsize 62.12\,|\,61.87\,|\,40.04 & \scriptsize -\,|\,-\,|\,71.18 \\
       \scriptsize HetGNN & \scriptsize 59.00\,|\,58.53\,|\,12.30 & \scriptsize 58.21\,|\,56.94\,|\,95.95 & \scriptsize 59.65\,|\,59.71\,|\,36.55 & \scriptsize 58.64\,|\,57.98\,|\,16.00 & \scriptsize 58.48\,|\,58.22\,|\,\textbf{100.00} & \scriptsize 58.42\,|\,58.23\,|\,45.88 & \scriptsize -\,|\,-\,|\,75.50 \\
       \scriptsize HAN & \scriptsize 60.97\,|\,60.70\,|\,11.66 & \scriptsize 61.20\,|\,61.02\,|\,\textbf{100.00} & \scriptsize 60.88\,|\,60.70\,|\,53.50 & \scriptsize 59.38\,|\,59.15\,|\,54.79 & \scriptsize 55.34\,|\,54.63\,|\,76.71 & \scriptsize 60.27\,|\,60.13\,|\,83.11 & \scriptsize -\,|\,-\,|\,91.04 \\
       \hline
   \end{tabular}
   }
\end{table*}

\subsection{Detailed Experimental Results for RQ2}
\label{app:rq2}

\begin{figure}[htbp]
    \centering
    \includegraphics[width=1\linewidth]{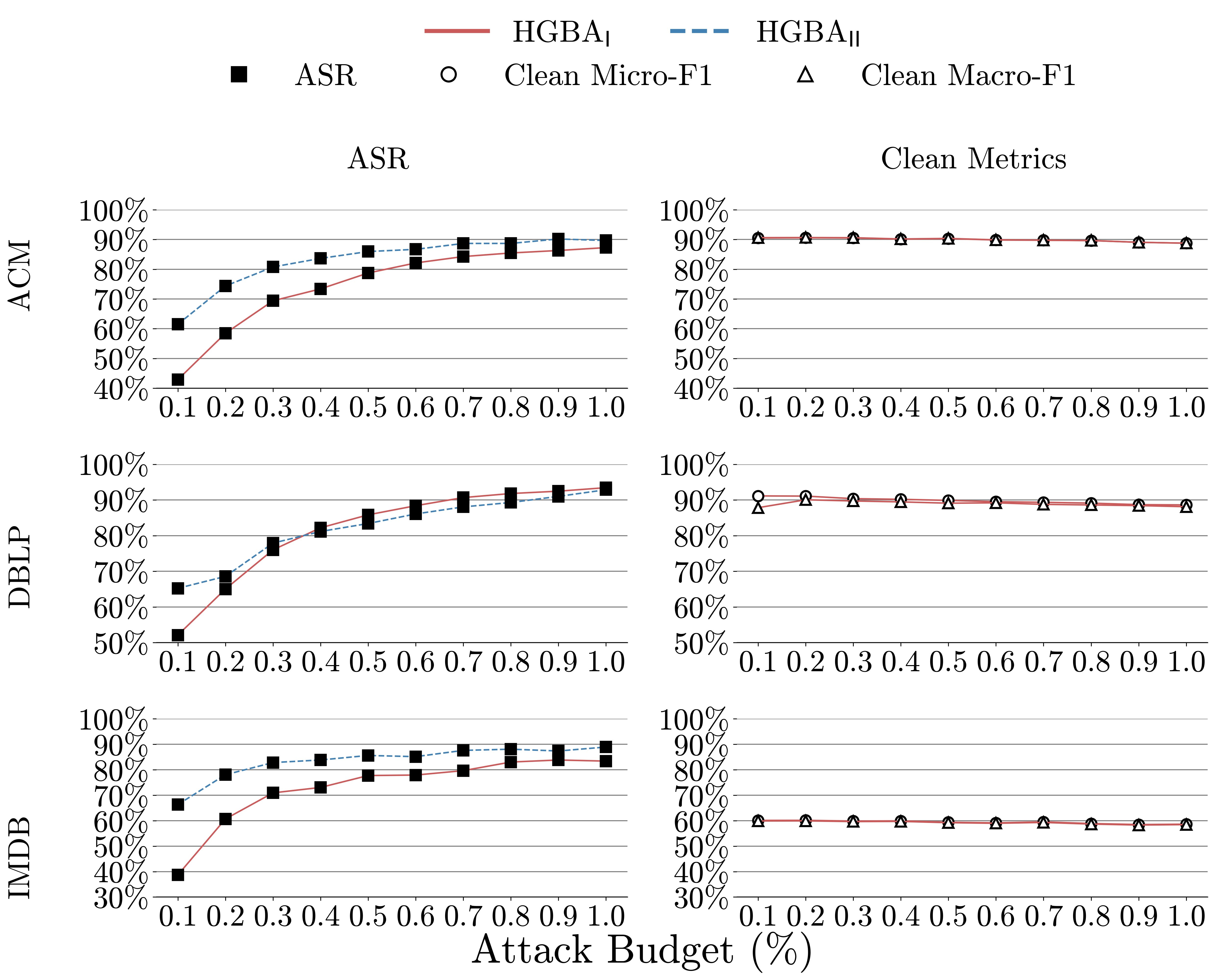}
    \caption{The Average Performance of HGBA Attacking Six HGNNs on Three Datasets under Varying Attack Budgets.}
    \label{fig:rq2_all_datasets}
\end{figure}

Fig.~\ref{fig:rq2_all_datasets} shows the average performance of HGBA when attacking six HGNNs across three datasets under different attack budgets.

\subsection{Detailed Experimental Results for RQ3}
\label{app:rq3}

\subsubsection{Trigger Node Selection Impact on HGBA\(_{\mathrm{I}}\)}
\label{app:rq3_vt_impact}

\begin{figure}[htbp]
    \centering
    \includegraphics[width=1\linewidth]{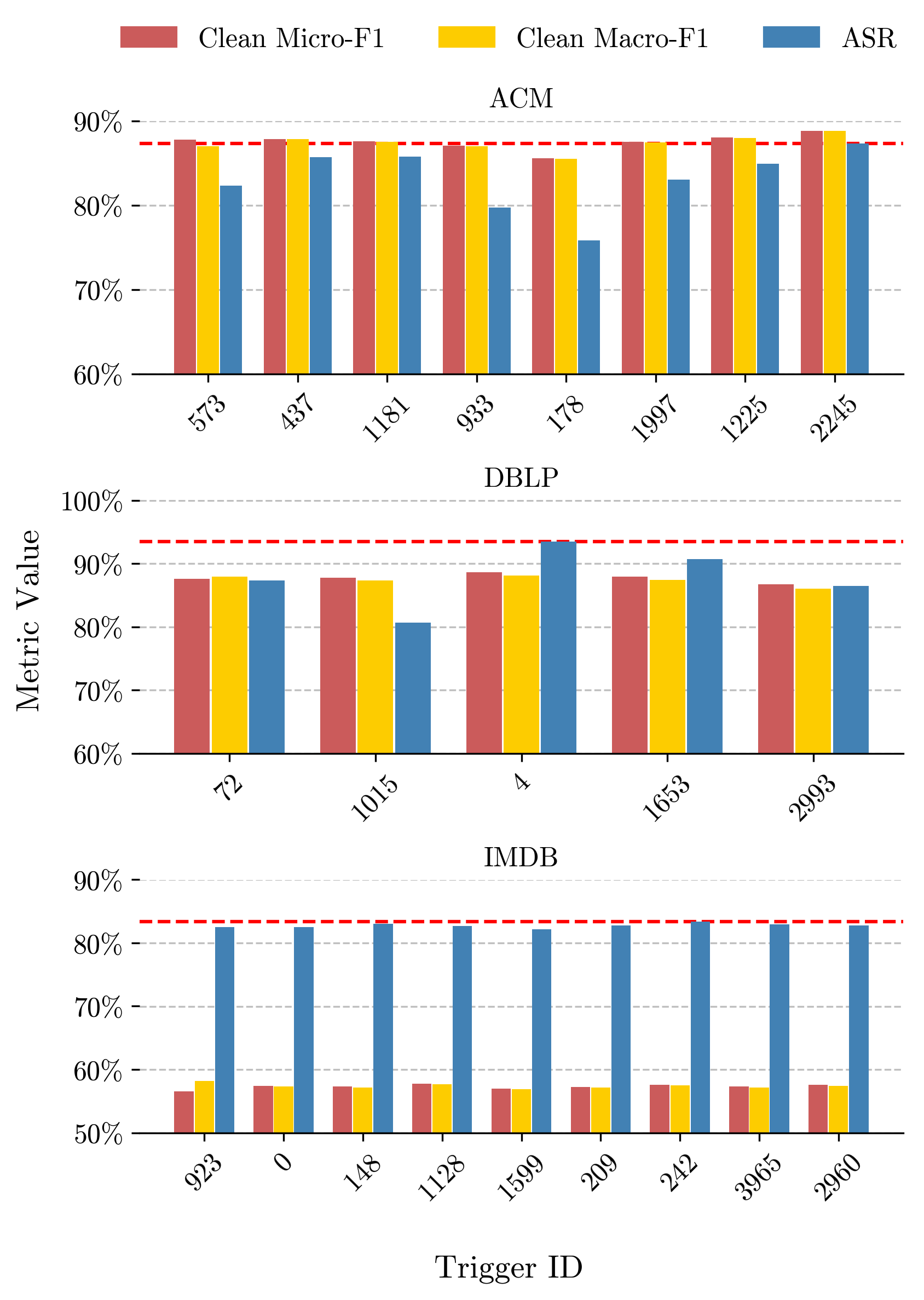}
    \caption{Impact of Trigger Node Selection for HGBA\(_{\mathrm{I}}\) on Three Datasets. The red dashed line indicates the highest ASR observed.  2245, 4, 242 respectively represent the trigger nodes selected for each dataset based on the strategy we proposed.}
    \label{fig:rq3_vt_impact}
\end{figure}

Fig.~\ref{fig:rq3_vt_impact} illustrates the impact of trigger node selection for HGBA\(_{\mathrm{I}}\) across three datasets.

\subsubsection{Results of Backdoor Metapath Selection}
\label{app:pb_selection}
Table~\ref{tab:acm_metapath}, Table~\ref{tab:dblp_metapath}, and Table~\ref{tab:imdb_metapath} display the results of backdoor metapath selection using homogeneous GNNs as proxy models for ACM, DBLP, and IMDB datasets, respectively.

\begin{figure}[t]
    \centering
    \includegraphics[width=0.85\linewidth]{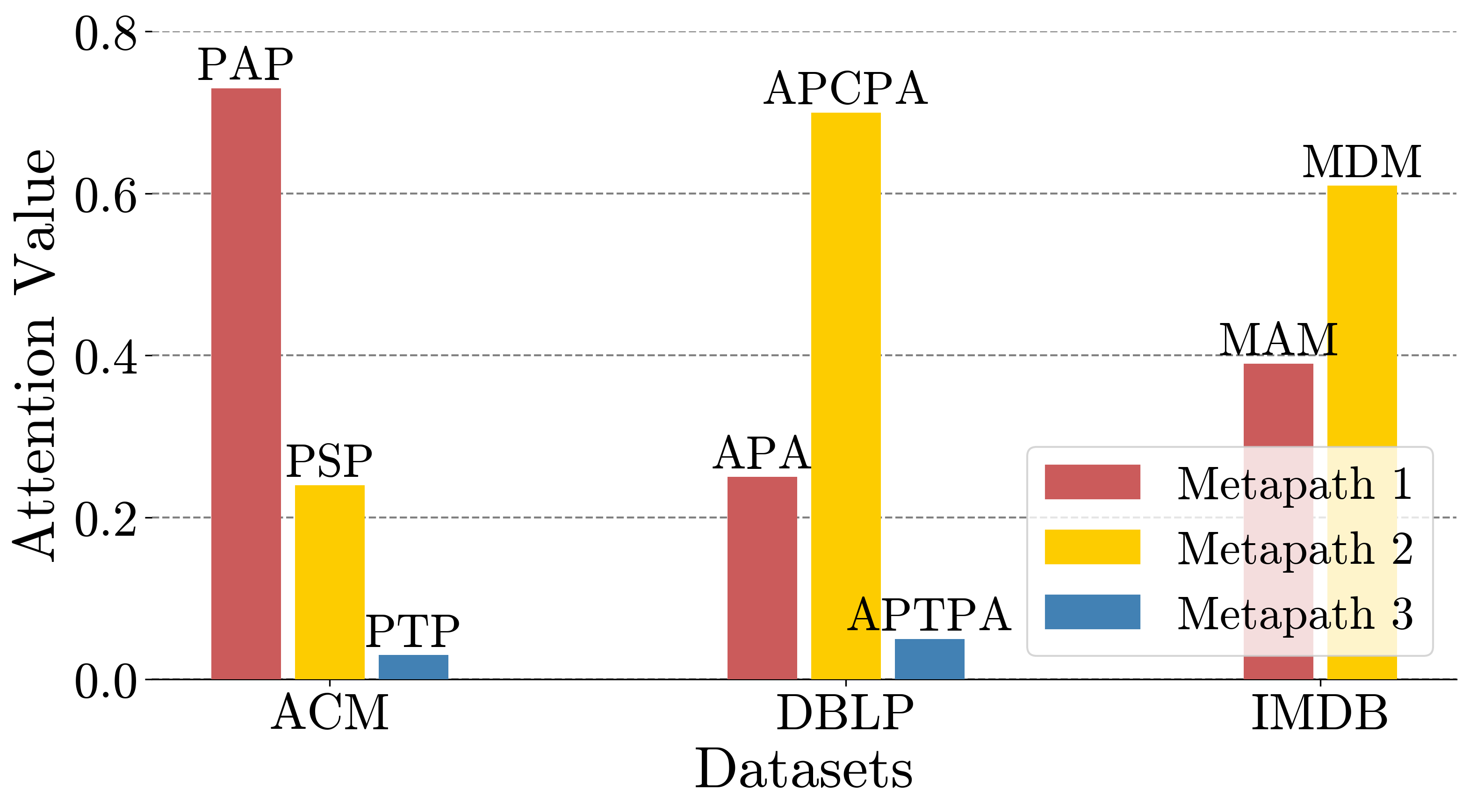}
    \caption{Attention values of metapaths in HAN, illustrating the influence of selected metapaths (PAP, APCPA, MDM) on node classification.}
    \label{fig:han_att_value}
\end{figure}

Fig.~\ref{fig:han_att_value} shows the attention values of metapaths in HAN, highlighting the influence of selected metapaths (PAP, APCPA, MDM) on node classification.

\begin{table}[h]
    \renewcommand{\arraystretch}{1} 
    \centering
    \begin{tabular}{lccc}
        \toprule 
        & \textbf{GCN} & \textbf{GAT} & \textbf{GraphSAGE} \\
        \midrule 
        \textbf{PAP} & \textbf{87.57\% }| \textbf{87.69\%} & \textbf{88.90\% }| \textbf{89.01\% }& \textbf{90.79\% }| \textbf{90.85\%} \\
        \textbf{PSP} & 70.59\% | 68.91\% & 70.79\% | 69.17\% & 88.62\% | 88.69\% \\
        \textbf{PTP} & 35.06\% | 17.31\% & - & 85.87\% | 85.84\% \\
        \bottomrule 
    \end{tabular}
    \caption{Backdoor Metapath Selection Using Homogeneous GNNs as Proxy Models (ACM). "-" means out of GPU memory.}
    \label{tab:acm_metapath}
\end{table}

\begin{table}[h]
    \renewcommand{\arraystretch}{1} 
    \centering
    \begin{tabular}{lccc}
        \toprule 
        & \textbf{GCN} & \textbf{GAT} & \textbf{GraphSAGE} \\
        \midrule 
        \textbf{APA} & 79.10\% | 77.52\% & 78.30\% | 79.96\%  & 78.36\% | 79.26\% \\
        \textbf{APCPA} & \textbf{92.09\% }|\textbf{ 91.48\%} & \textbf{88.02\% }| \textbf{86.09\%}  & \textbf{92.48\%} |\textbf{ 91.87\%} \\
        \textbf{APTPA} & 72.99\% | 71.29\% & -  & 76.57\% | 75.77\% \\
        \bottomrule 
    \end{tabular}
    \caption{Backdoor Metapath Selection Using Homogeneous GNNs as Proxy Models (DBLP). "-" means out of GPU memory.}
    \label{tab:dblp_metapath}
\end{table}

\begin{table}[h]
    \renewcommand{\arraystretch}{1} 
    \centering
    \begin{tabular}{lccc}
        \toprule 
        & \textbf{GCN} & \textbf{GAT} & \textbf{GraphSAGE} \\
        \midrule 
        \textbf{MAM} & 50.32\% | 48.79\% & 49.78\% | 49.53\%  & 58.35\% | 58.29\% \\
        \textbf{MDM} & \textbf{58.99\%} |\textbf{ 58.73\%} & \textbf{59.00\% }| \textbf{58.78\%}  &\textbf{ 60.16\%} | \textbf{60.09\%} \\
        \bottomrule 
    \end{tabular}
    \caption{Backdoor Metapath Selection Using Homogeneous GNNs as Proxy Models (IMDB)}
    \label{tab:imdb_metapath}
\end{table}

\subsection{More Information for RQ4}
\label{app:rq4}

\subsubsection{Introduction of Defense Methods and Robust Models}
\label{app:rq4_models}

\begin{itemize}[leftmargin=*]
    \item \textit{Prune} \cite{dai2023unnoticeable}. It operates by eliminating the edges that connect nodes with a low cosine similarity. Since the edges added by backdoor attackers often link nodes that are not similar to each other, this pruning strategy can effectively disrupt the trigger structure and the attachment edges established by the attackers.

    \item \textit{Prune+LD (Label Discarding)} \cite{dai2023unnoticeable}. Based on pruning the edges between dissimilar nodes as in the Prune method, it also discards the labels of the nodes that are connected by these dissimilar edges. This is aimed at reducing the impact of the dirty labels associated with the poisoned nodes, which can otherwise negatively affect the performance and security of the GNNs.

    \item \textit{ESAGE} \cite{yuan2024sage}. E-SAGE is a proposed approach for defending against GNN backdoor attacks, leveraging explainability. Based on the finding that malicious and benign edges exhibit significant differences in importance scores for explainability evaluation, E-SAGE adaptively conducts an iterative edge pruning process on the graph using edge scores. 

    \item \textit{HAN-RoHe} \cite{zhang2022robust}. RoHe is a robust HGNN framework designed to defend against topology adversarial attacks. It equips an attention purifier that prunes malicious neighbors based on topology and features. Introducing metapath-based transiting probability as a prior criterion mitigates the perturbation enlargement effect, restraining the influence of adversarial hubs. The purifier then masks out neighbors with low confidence, alleviating the negative impact of unreliable neighbors in the soft attention mechanism.
    
    \item \textit{RobustGCN} \cite{zhu2019robust}. Instead of representing nodes as vectors, RobustGCN uses Gaussian distributions as hidden node representations in each convolutional layer, enabling it to absorb adversarial changes through variance adjustments. To address the propagation of attacks, it employs a variance-based attention mechanism, assigning weights to node neighborhoods according to their variances during convolutions. This approach enhances the robustness of GCNs against adversarial attacks.
    
    \item \textit{GNNGuard} \cite{zhang2020gnnguard}. GNNGuard is a general algorithm for defending against training-time attacks on graph structure. It can be integrated into any GNN by detecting and quantifying the relationship between graph structure and node features. It assigns higher weights to edges connecting similar nodes and prunes edges between unrelated nodes. With components like neighbor importance estimation and layer-wise graph memory, it restores GNN performance against various adversarial attacks, including targeted and non-targeted ones.
\end{itemize}

\subsubsection{Defense Implementation Details of RQ4}
\label{app:rq4_ex_details}

For Prune, Prune+LD, and E-SAGE, which sanitize graphs through graph pruning techniques before training, since they were originally proposed for homogeneous graphs, to implement them on heterogeneous graph data, we first extract subgraphs on heterogeneous graphs based on backdoor metapaths. Then, we apply these three methods, record the two-end nodes of the edges that need to be pruned, and return to the heterogeneous graph to delete all the edges between these nodes based on the backdoor metapaths using masks. After that, normal training and evaluation are conducted.

For HAN-RoHe, RobustGCN, and GNNGuard, we simply replace the normal models with these robust models without any additional processing.

\subsection{Details about RQ5}
\label{app:rq5}

\subsubsection{New Compared Method: TRAP}
\label{app:rq5_TRAP}

\begin{itemize}
    \item \textit{TRAP.} It poisons the training dataset with perturbation-based triggers generated through a gradient-based score matrix from a surrogate GCN model. 
\end{itemize}

\subsubsection{Defense Implementation Details of RQ5}
\label{app:rq5_implement_details}

Here, we only describe how to extend the relation-based trigger mechanism of HGBA to homogeneous graph classification and node classification tasks on heterogeneous graph datasets. For the implementation details of other comparison methods, please refer to the related works.

For classification tasks on homogeneous graphs, we randomly select a certain percentage of samples from the training set as poisoned samples according to the poisoning rate. For each poisoned sample, we choose the three nodes with the lowest Betweenness Centrality and connect them to construct a trigger of size 3. The subsequent operations are consistent with those of other methods.

For node classification tasks on homogeneous graphs, we first select the node with the lowest Betweenness Centrality as the trigger node. Then, among the nodes that are not connected to the trigger node, we select some nodes according to the attack budget and establish edges to set up the trigger. The subsequent operations follow the same procedures as other methods.
\subsubsection{Results of RQ5}

Table~\ref{tab:homo_graph} presents the performance comparison of HGBA against baseline backdoor attacks (SBA-SAMPLE, GTA, TRAP) in graph classification tasks using the PROTEINS dataset. 

Table~\ref{tab:homo_node} illustrates the performance of HGBA in node classification tasks under homogeneous graphs.

\label{app:rq5_results}
\begin{table}[t]
    \centering
    \small
    \caption{Performance of HGBA compared to baseline backdoor attacks on graph classification tasks using the PROTEINS dataset. Results report Accuracy on Clean Graph) and paired Clean Accuracy/Attack Success Rate (Clean Acc/ASR) for each method as percentages, averaged across GCN, GIN, and GraphSAGE.}
    \label{tab:homo_graph}
    \begin{tabular}{lccc} 
        \toprule
        \textbf{Attacks} & \textbf{GCN} & \textbf{GIN} & \textbf{GraphSAGE} \\
        \midrule
        Clean Graph & 71.95 & 70.14 & 69.95 \\
        SBA-SAMPLE  & \textbf{71.92}\ |\ 33.33 & \textbf{70.88}\ |\ 46.97 & 69.02\ |\ 33.33 \\
        GTA         & 69.55\ |\ 41.37 & 57.61\ |\ 65.52 & 64.09\ |\ 55.17 \\
        TRAP        & 68.38\ |\ \textbf{77.78} & 68.38\ |\ 68.89 & \textbf{69.32}\ |\ 55.56 \\
        HGBA        & 70.61\ |\ 65.18 & 69.46\ |\ \textbf{73.57} & 68.86\ |\ \textbf{62.50} \\
        \bottomrule
    \end{tabular}
\end{table}
\begin{table}[t]
    \centering
    \small
    \caption{Performance of HGBA on node classification tasks under homogeneous graphs. Results report Accuracy on Clean Graph, Clean Accuracy under backdoor attack (Clean Acc), and Attack Success Rate (ASR) as percentages, averaged across GCN, GAT, and GraphSAGE.}
    \label{tab:homo_node}
    \begin{tabular}{lccc}
        \toprule
        \textbf{Dataset} & \textbf{Clean Graph} & \textbf{Clean Acc} & \textbf{ASR} \\
        \midrule
        Cora     & 87.99 & 85.64 & 85.58 \\
        PubMed   & 87.32 & 84.94 & 95.51 \\
        CiteSeer & 74.94 & 70.18 & 95.79 \\
        \bottomrule
    \end{tabular}
\end{table}








\end{document}